\documentclass[iop, revtex4]{emulateapj}
\usepackage{multirow}
\usepackage{color}
\usepackage[normalem]{ulem}
\usepackage{amssymb}
\usepackage{lscape}

\begin{document}

\title{High-Resolution Near-infrared Spectroscopy of Diffuse Sources around MWC 1080}

\author{Il-Joong Kim\altaffilmark{1}}
\author{Heeyoung Oh\altaffilmark{1}}
\author{Woong-Seob Jeong\altaffilmark{1,2}}
\author{Kwang-Il Seon\altaffilmark{1,2}}
\author{Tae-Soo Pyo\altaffilmark{3,4}}
\author{Jae-Joon Lee\altaffilmark{1}}

\altaffiltext{1}{Korea Astronomy and Space Science Institute, Daejeon, 34055, Republic of Korea; ijkim@kasi.re.kr}
\altaffiltext{2}{Astronomy and Space Science Major, University of Science and Technology, Daejeon, 34113, Republic of Korea}
\altaffiltext{3}{Subaru Telescope, National Astronomical Observatory of Japan, National Institutes of Natural Sciences (NINS), 650 North A'ohoku Place, Hilo, HI 96720, USA}
\altaffiltext{4}{School of Mathematical and Physical Science, SOKENDAI (The Graduate University for Advanced Studies), Hayama, Kanagawa 240-0193, Japan}

\begin{abstract}
To reveal the origins of diffuse H$\alpha$ emissions observed around the Herbig star MWC 1080, we have performed a high-resolution near-infrared (NIR) spectroscopic observation using the Immersion GRating INfrared Spectrograph (IGRINS). In the NIR H and K bands, we detected various emission lines (six hydrogen Brackett lines, seven H$_{2}$ lines, and an {[}\ion{Fe}{2}{]} line) and compared their spatial locations with the optical (H$\alpha$ and {[}\ion{S}{2}{]}) and radio ($^{13}$CO and CS) line maps. The shock-induced H$_{2}$ and {[}\ion{Fe}{2}{]} lines indicate the presence of multiple outflows, consisting of at least three, associated young stars in this region. The kinematics of H$_{2}$ and {[}\ion{Fe}{2}{]} near the northeast (NE) cavity edge supports that the NE main outflow from MWC 1080A is the blueshifted one with a low inclination angle. The H$_{2}$ and {[}\ion{Fe}{2}{]} lines near the southeast molecular region newly reveal that additional highly-blueshifted outflows originate from other young stars. The fluorescent H$_{2}$ lines were found to trace photodissociation regions formed on the cylindrical surfaces of the main outflow cavity, which are expanding outward with a velocity of about 10--15 km s$^{-1}$. For the H$\alpha$ emission, we identify its components associated with two stellar outflows and two young stars in addition to the dominant component of MWC 1080A scattered by dust. We also report a few faint H$\alpha$ features located $\sim$0.4 pc away in the southwest direction from MWC 1080A, which lie near the axes of the NE main outflow and one of the newly-identified outflows.
\end{abstract}

\keywords{infrared: ISM --- ISM: Herbig-Haro objects --- ISM: jets and outflows --- ISM: molecules --- stars: formation --- stars: individual (MWC 1080) --- techniques: spectroscopic}

\section{Introduction} \label{sec:intro}

The MIRIS Pa$\alpha$ Galactic Plane Survey detected a lot of new Pa$\alpha$-emitting sources \citep{2018ApJS..238...28K}, which are expected to be associated with star-forming activities in our Galaxy. The Pa$\alpha$ sources would give important clues to understand the star forming history of our Galaxy. In the Cepheus region, we found 127 Pa$\alpha$ sources (109 extended sources and 18 point-like sources). The extended Pa$\alpha$ sources are mostly \ion{H}{2} regions, while the point-like sources are planetary nebulae or emission-line stars, such as Wolf-Rayet and Herbig Ae/Be stars. For three sources among the point-like Pa$\alpha$ sources, we also found diffuse extended H$\alpha$ features surrounding H$\alpha$ point sources using the INT/WFC Photometric H$\alpha$ Survey (IPHAS) data with a higher spatial resolution. The nebulosity surrounding stars would be caused by the leftover material and partially ionized gas ejected by stars in the process of star formation, colliding with nearby clouds of gas and dust. The extended H$\alpha$ emission originates from recombination in ionized gas or the dust scattering of stellar photons, or both. However, it is not clear which mechanism (ionized gas or dust scattering) is more plausible in many cases. To better understand the origin of the extended H$\alpha$ emission around the emission-line stars (shown as point-like Pa$\alpha$ sources by MIRIS), we have carried out high-resolution near-infrared (NIR) spectroscopic follow-up observations toward them.

MWC 1080, the first follow-up target, is one of the brightest Pa$\alpha$ sources detected in Cepheus through the survey. MWC 1080 is known to be a Herbig Ae/Be star with a B0 spectral type according to the SIMBAD database. \citet{1984ApJ...282..631C} suggested a kinematic distance of 2.5 kpc based on the  $^{13}$CO emission from the parent cloud. On the other hand, \citet{1992ApJ...397..613H} and \citet{2003ApJ...598L..39F} applied a distance of 1 kpc to circumstellar disks around MWC 1080. Its parallax distance has been recently measured to be 1.373 $\pm$ 0.174 kpc \citep{2018yCat.1345....0G}. MWC 1080 is, in fact, a multiple system consisting of a primary star (MWC 1080A), a close (0.$\arcsec$8) companion (MWC 1080B), and a third companion (MWC 1080E) located $\sim$5.$\arcsec$2 east from MWC 1080A \citep{1997A&A...318..472L,2002AJ....124.2207P}. \citet{1997A&A...318..472L} classified MWC 1080B as another Herbig Ae/Be star.

\citet{2018ApJS..238...28K} proposed that the extended H$\alpha$ features around MWC 1080 might be a young \ion{H}{2} region. The previous studies for MWC 1080, however, are likely to suggest that the extended H$\alpha$ features are not a simple \ion{H}{2} region. Diffuse sources around MWC 1080 have been observed at various wavelengths since \citet{1960ApJS....4..337H} included MWC 1080 in his list of stars with nebulosity. \citet{1992PASJ...44...77Y} performed optical spectroscopic observations of the primary star MWC 1080A and its associated nebulosity, and found the existence of a strong H$\alpha$ emission line in the optical spectrum of MWC 1080A, but no evidence of shock or ultraviolet (UV) excitation in the nebular spectra. They concluded, based on the existence of the H$\alpha$ recombination lines and the absence of any forbidden line, that the extended H$\alpha$ features are attributable to scattering of the stellar H$\alpha$ line from MWC 1080A by dusty materials rather than an ionized medium, as in bright reflection nebulae observed at the optical and infrared (IR) wavelengths. On the other hand, the existence of the main outflow cavity discovered later by \citet{2008ApJ...673..315W} and the {[}\ion{S}{2}{]} Herbig-Haro (HH) objects by \citet{1992A&A...262..229P} supports the possibility that, at least, some portions of the extended H$\alpha$ emission are caused by outflows from MWC 1080A and/or other young stars. Because no strong {[}\ion{S}{2}{]} forbidden line was detected in the stellar spectrum of MWC 1080A \citep{1992PASJ...44...77Y}, the {[}\ion{S}{2}{]} HH objects are likely caused by shock excitation. \citet{1989ApL&C..27..299C} reported weak extended 6-cm radio continuums, which come from ionized gas, in association with MWC 1080. \citet{2011ApJ...727...26S} presented the 850 $\micron$ dust continuum, of which the morphology seems to be similar to that of the 6-cm radio continuum. This similarity in morphology suggests that the 850 $\micron$ continuum originates from warm dust in the ionized gas. \citet{1979ApJ...231..115H} also presented evidence for thermal emissions from circumstellar dust, leftover from the star formation process, by far-IR (40--160 $\micron$) observations. \citet{2003ApJ...598L..39F} and \citet{2009A&A...497..117A} studied millimeter continuum images to find evidence for a circumstellar disk around MWC 1080. \citet{2009A&A...497..117A} found several clumps in their 1.3 and 2.7 mm images and proposed, based on the millimeter morphology and spectral energy distribution (SED) from the NIR to the centimeter wavelengths, that the clumps can be a compact disk close to MWC 1080 plus a large toroidal envelope with an inner radius of 6000 AU. \citet{2002RMxAA..38..169G} detected three resolved radio continuum sources around MWC 1080 using the Very Large Array (VLA) observations at 6 cm. The central source (VLA 4) among the three sources seems to coincide with MWC 1080A (see Figure 26 of \citet{2011ApJ...727...26S}). \citet{2002RMxAA..38..169G} suggested that the other two VLA sources may trace embedded young stellar objects (YSOs).

\citet{1992A&A...262..229P} reported the discovery of HH objects associated with MWC 1080 in a continuum-subtracted {[}\ion{S}{2}{]} $\lambda$6730\AA{} image. They suggested the direction of the bipolar outflow from MWC 1080A to have position angles (PAs) of $\simeq$ 60$\arcdeg$ and 240$\arcdeg$ (east from the north) and interpreted all of the HH objects as the rim-brightened edges of the outflow cavity. However, \citet{2008ApJ...673..315W} found that the bipolar outflow cavity has PAs of $\simeq$ 45$\arcdeg$ and 225$\arcdeg$ using molecular maps. They investigated the morphologies of the molecular clouds around MWC 1080 by observing the CS($J$ = 2 $\rightarrow$ 1), $^{13}$CO($J$ = 1 $\rightarrow$ 0), and C$^{18}$O($J$ = 1 $\rightarrow$ 0) lines and identified a cavity elongated along the northeast (NE)-southwest (SW) direction. Particularly, the $^{13}$CO map revealed an almost perfect biconical cavity. They noticed that the size of the cavity is much smaller than those in older Herbig Ae/Be star systems and suggested that a previous bipolar outflow from MWC 1080 (hereafter main outflow) has formed the cavity, and the main outflow may be still active. \citet{2008ApJ...673..315W} also presented a NIR $K'$-band image with $\sim$0.$\arcsec$1 resolution, which shows $\sim$50 young, low-mass stars and a reflection nebula with an hourglass shape in the direction perpendicular to the main outflow axis. Because most of the young stars are located inside the main outflow cavity, \citet{2008ApJ...673..315W} suggested that they were likely formed together with MWC 1080 and then have been exposed to the cavity. Recently, \citet{2014ApJ...796...74L} obtained mid-IR (11.2, 11.6, and 18.5 $\micron$) images for the MWC 1080 region with a sub-arcsecond resolution. They demonstrated that the diffuse mid-IR emissions observed around MWC 1080 trace neither a disk nor an envelope, but the internal surfaces of the main outflow cavity revealed by the $^{13}$CO map. Hence, they concluded that the observed mid-IR emission would originate from warm dust grains, heated by the stellar radiation, located close to the main outflow cavity walls.

This paper presents the results of a high-resolution NIR spectroscopic observation for the MWC 1080 region using the Immersion GRating INfrared Spectrograph (IGRINS). This is the first high-resolution NIR spectroscopy for this region. Because IGRINS has a broad wavelength coverage of the whole H and K bands, we expect to detect various hydrogen Brackett lines related to the Pa$\alpha$ and H$\alpha$ emissions, which would enable us to reveal the origin of the hydrogen recombination line emissions. Also, IGRINS has detected the H$_{2}$ and {[}\ion{Fe}{2}{]} lines induced by stellar outflows \citep[e.g.,][]{2016ApJ...833..275O,2016ApJ...817..148O}, which provides an opportunity to find stellar outflows in this region. Section \ref{sec:data} describes the IGRINS NIR spectroscopy and the IPHAS H$\alpha$ data acquisition. In Section \ref{sec:result}, we present the results of the IGRINS spectroscopy. We report diffuse sources identified in the position-velocity diagrams (PVDs) of the observed emission lines, and measure the line profiles and line fluxes for the individual sources. Also, we classify the diffuse sources into several types according to their origins revealed from the spectroscopic analysis and the spatial correlations with the H$\alpha$, {[}\ion{S}{2}{]}, and $^{13}$CO maps. In Section \ref{sec:discussion}, we discuss the stellar outflows identified in this study and the possible origins of the atomic hydrogen line emission extended widely in the MWC 1080 region. We also discuss the plausible origins of the compact hydrogen line emission detected in three young stars. Finally, Section \ref{sec:conclusions} contains the summary and conclusions of our main results.

\section{Observations and Data Acquisition} \label{sec:data}

\subsection{IGRINS NIR Spectroscopy} \label{subsec:igrins}

\begin{figure*}
\centering
\includegraphics[scale=0.4]{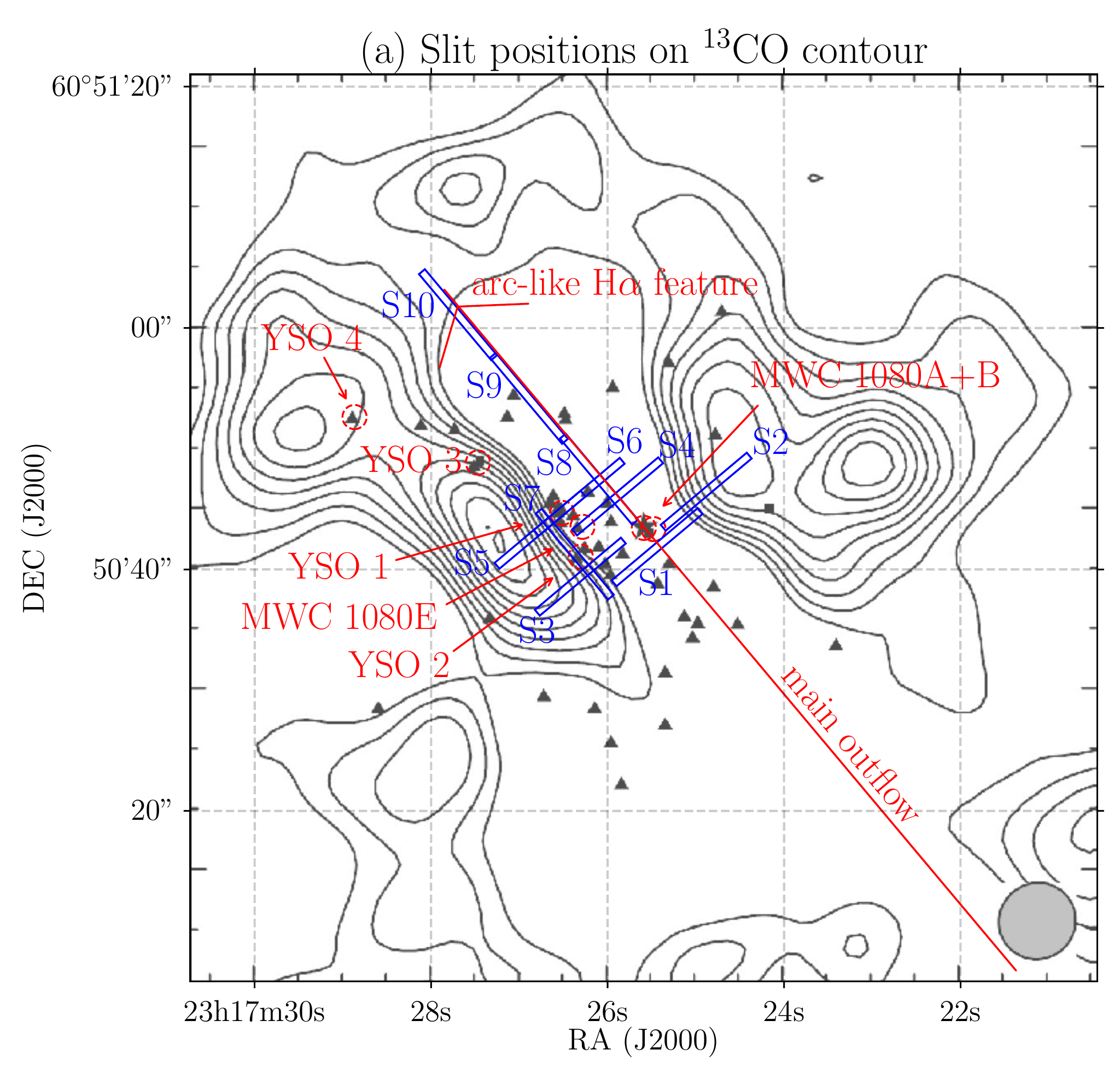}\includegraphics[scale=0.4]{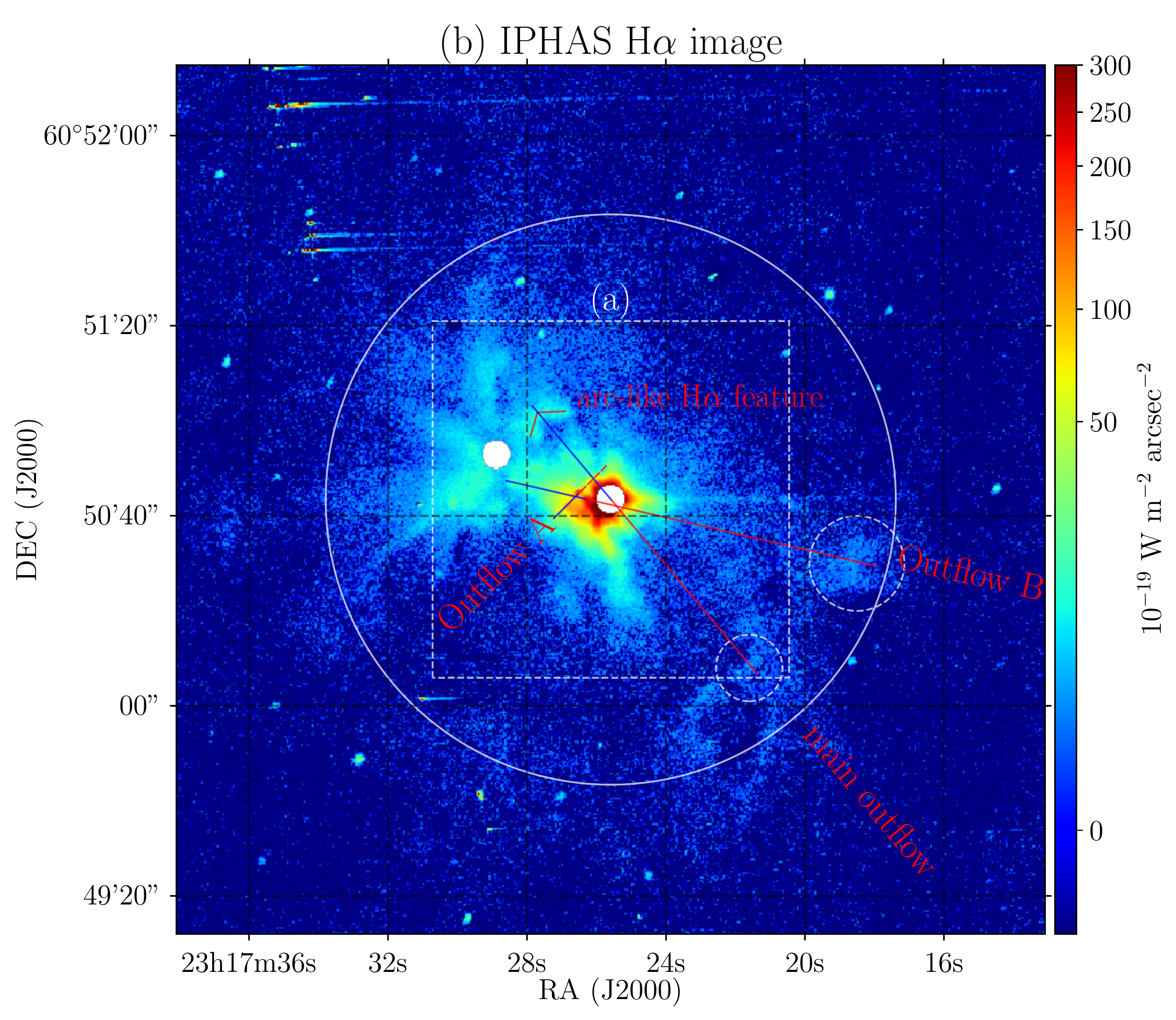}
\caption{(a) IGRINS slit positions (blue 9.$\arcsec$3 $\times$ 0.$\arcsec$63 rectangles) overlaid on the $^{13}$CO contour map, and (b) IPHAS H$\alpha$ image (continuum-subtracted) of the whole MWC 1080 region. In panel (a), the $^{13}$CO map was adapted from Figure 2 of \citet{2008ApJ...673..315W}, and its beam size is shown in the lower right corner of the figure. Triangles indicate the NIR-identified young, low-mass stars \citep{2008ApJ...673..315W}, out of which seven stars are also denoted in red dashed circles. The red diagonal line passing through MWC 1080A at PA = 40$\arcdeg$ is the central axis of the main outflow cavity. In panel (b), the circle with a radius of 1$\arcmin$ represents the whole area of the H$\alpha$ features detected in this paper, and the dashed square indicates the area displayed in (a). The blue+red line, representing the main outflow, with PA = 40$\arcdeg$ is the same as the diagonal line in (a). The blue and red colors indicate the blueshifted and redshifted parts of the bipolar outflow, respectively (see Section \ref{subsec:mainoutflow}). The two blue+red lines with PA = 135 and 77$\arcdeg$ represent the outflows newly-identified in this study (``Outflow A'' and ``Outflow B''), which are discussed in Section \ref{subsec:newoutflow}. The two white dashed circles indicate faint H$\alpha$ features related to the main outflow and ``Outflow B.'' The small bent line, shown in both (a) and (b), denotes an arc-like H$\alpha$ feature located on the axis of the main outflow.\label{fig:basic}}
\end{figure*}

IGRINS is a cross-dispersed NIR spectrograph equipped with a silicon immersion grating. With a spectral resolution of $\lambda$/$\Delta\lambda$ $\sim$45,000 (corresponding to a velocity resolution of 7.5 km s$^{-1}$), it covers the H-band (1.49--1.80 $\micron$) and K-band (1.96--2.46 $\micron$) wavelength ranges simultaneously \citep{2010SPIE.7735E..1MY,2014SPIE.9147E..1DP,2018SPIE10702E..0QM}. We carried out the IGRINS observations of the MWC 1080 region on 2017 November 26--28 and December 28 using the 4.3 m Lowell Discovery Telescope in Arizona. The IGRINS slit size corresponds to a field of view of 0.$\arcsec$63 $\times$ 9.$\arcsec$3 on the sky when it is mounted on the 4.3 m telescope. We observed a total of ten slit positions on the first observation date (November 26) with a seeing FWHM of $\sim$0.$\arcsec$6 in the K band, which are indicated by blue rectangles in Figure \ref{fig:basic}(a). Slit positions S1--S6 were set to be in the northwest (NW)-southeast (SE) direction (PA = 130$\arcdeg$), which is perpendicular to the axis of the main outflow cavity observed in the $^{13}$CO map. On the other hand, slit positions S7--S10 were aligned parallel to the cavity axis (PA = 40$\arcdeg$). In the figure, the triangles indicate the NIR-identified young, low-mass stars presented by \citet{2008ApJ...673..315W}. We also denote the locations of seven young stars (YSO 1, YSO 2, YSO 3, YSO 4, MWC 1080E, and MWC 1080A+B), which are relevant in this paper, in red dashed circles. Although we additionally observed twelve slit positions in the SE molecular region for the other three observation dates (November 27--28 and December 28), we used this data only for velocity channel maps in the Appendix because of bad weather conditions.

We took the off-source exposures between every on-source exposure to subtract backgrounds such as the telluric emission, bias signal, dark current, etc. The off-source position was chosen to be $\sim$2$\arcmin$ north from each on-source position. The total on-source exposure time for each slit position was 10 minutes (two 5-minute exposures) for S1--S8 and 20 minutes (four 5-minute exposures) for S9 and S10. Auto-guiding was performed with the K-band slit-viewing camera, which covers a $\sim$1.$\arcmin$3 $\times$ 2$\arcmin$ field, including the slit position, during the observation. We selected a nearby bright star (YSO 3 in Figure \ref{fig:basic}(a)) as a guiding star. The slit-viewing image taken during the observation was also used to check the planned slit position. We measured dark and flat-field calibration frames at the beginning of each night and observed an A0V-type star HD 223386 as a telluric standard star. To obtain the stellar spectrum of the central, bright star MWC 1080A, we also observed it four times with a 15-second exposure on the first observation date.

Basic data reduction was carried out using the IGRINS Pipeline Package\footnote{\url{https://github.com/igrins/plp}} \citep[PLP,][]{2017zndo....845059L}, which can deal with the removal of bad pixels and cosmic rays, flat fielding, sky subtraction, spectral extraction, and wavelength calibration. The PLP provides signal and variance ($\sigma^{2}$) values for the processed spectra. We also utilized the python software ``Plotspec''\footnote{\url{https://github.com/kfkaplan/plotspec}} designed for analyzing two-dimensional IGRINS spectra of extended sources. It is capable of performing the telluric correction, relative flux calibration, continuum subtraction, etc. \citep{2017ApJ...838..152K}. Using the software, we constructed two data cubes for every detected emission line in the sky field including all slit positions. One is a continuum-subtracted data cube, which was used to make velocity-integrated maps and PVDs for each emission line. The other is a data cube that contains continuum signals, which was used to measure spectral line profiles for the individual targets. The data cubes have three axes of right ascension (RA), declination (DEC), and radial velocity ($V$). Here, $V$ was adjusted to represent the relative radial velocity with respect to the systemic local standard of rest (LSR) velocity of the MWC 1080 system. The adopted systemic LSR velocity is $-$30 km s$^{-1}$, measured from the $^{12}$CO observation by \citet{1989A&AS...80..149W}. We note that the above data reduction method could not completely remove the telluric OH lines because of the time-variability of the line fluxes. Therefore, we manually confirmed that the residual OH lines did not severely contaminate any spectroscopic results presented in this paper. We assumed that the residual fluxes measured at the OH line wavelength, given in \citet{2000A&A...354.1134R}, are uniform along the spatial direction of the slit. Some strong residuals were masked out when we measured the line profiles.

\subsection{IPHAS H$\alpha$ Data} \label{subsec:iphas}

We obtained the continuum-subtracted H$\alpha$ image for the MWC 1080 region from the calibrated IPHAS data \citep{2014MNRAS.444.3230B}. We used the same method as described in \citet{2018ApJS..238...28K}, except that we retained the original pixel size ($\sim$0.$\arcsec$33) of the IPHAS data. For the narrow-band H$\alpha$ and broad-band $r$ filters, we extracted a total of three IPHAS field data covering the MWC 1080 region with an angular size of 3$\arcmin$ $\times$ 3$\arcmin$. We confirmed that the data satisfy the IPHAS data quality criteria \citep{2014MNRAS.444.3230B}. In addition, the pixels with a ``confidence'' value $<$70 in the ``confidence'' map for each IPHAS field data \citep{2008MNRAS.388...89G} were masked out. Using the photometric zero points and zero magnitudes for the IPHAS H$\alpha$ and $r$ filters, the pixel values were transformed into calibrated flux units of W m$^{-2}$ \AA{$^{-1}$} arcsec$^{-2}$. To match the sky background levels among the individual field images, we subtracted the median background from each field image. All field images in each band were then combined into a single mosaic image using the Montage software.

Because the wavelength range of the broad-band $r$ filter entirely includes that of the narrow-band H$\alpha$ filter \citep{2005MNRAS.362..753D}, we multiplied the H$\alpha$ filter image minus the $r$ filter image by a correction factor of $(\Sigma T_{H\alpha} \times \Sigma T_r) / (\Sigma T_r - \Sigma T_{H\alpha})$, where $\Sigma T_{H\alpha}$ and $\Sigma T_r$ are the integrals of the H$\alpha$- and $r$-filter transmission curves over the wavelength, respectively. This factor prevents the continuum-subtracted H$\alpha$ image from over-subtracting the background continuum, which can arise in a simple-subtraction procedure. It also converts the pixel unit of the continuum-subtracted H$\alpha$ image into the in-band flux unit of W m$^{-2}$ arcsec$^{-2}$. We note that this IPHAS H$\alpha$ flux can include some contributions from the {[}\ion{N}{2}{]} $\lambda\lambda$6548, 6584 \AA{} lines, which are contained within the IPHAS H$\alpha$ filter. Figure \ref{fig:basic}(b) shows the resulting continuum-subtracted H$\alpha$ mosaic image of the MWC 1080 region. The region close to MWC 1080A is saturated in the IPHAS H$\alpha$- and $r$-filter images, and thus it was masked out. Also, there remained imaging artifacts near other bright stars after the continuum subtraction, which is probably due to the difference of the point-spread function (PSF) between the H$\alpha$ and $r$ bands. Although the PSF sizes measured in the H$\alpha$- and $r$-filter images were similar (FWHM $\sim$1.$\arcsec$2 and $\sim$1.$\arcsec$3, respectively), the PSF matching was not accomplished well. Therefore, we masked out one more bright star (YSO 4), showing strong artifacts in the central region of the figure. We note, however, that our analysis on the extended sources was hardly affected by the PSF difference.

In Figure \ref{fig:basic}(b), the H$\alpha$ emission is found to extend out to at least $\sim$1$\arcmin$ from MWC 1080A, as indicated by a solid circle. This radius corresponds to $\sim$0.4 pc at the parallax distance of MWC 1080 \citep[1.373 $\pm$ 0.174 kpc,][]{2018yCat.1345....0G}. We found that an arc-like H$\alpha$ feature lies on the approximate axis of the main outflow cavity, as shown by a red bent line in the figure. As can be seen in Figure \ref{fig:basic}(a), this H$\alpha$ feature is well coincident with the $^{13}$CO boundary forming the NE edge of the main outflow cavity. The diagonal red line passing through MWC 1080A with PA = 40$\arcdeg$ passes the apex of the arc-like H$\alpha$ feature almost exactly. We also identified a few faint H$\alpha$ features far away in the SW direction from MWC 1080A. In Figure \ref{fig:basic}(b), they are distributed between (RA, DEC) $\sim$ (23$^{\mathrm{h}}$17$^{\mathrm{m}}$24$^{\mathrm{s}}$, +60$\arcdeg$49$\arcmin$50$\arcsec$) and (RA, DEC) $\sim$ (23$^{\mathrm{h}}$17$^{\mathrm{m}}$18$^{\mathrm{s}}$, +60$\arcdeg$50$\arcmin$35$\arcsec$), two of which are indicated by the white dashed circles. In the figure, the blue+red line representing the main outflow is the same as the diagonal red line in Figure \ref{fig:basic}(a); here, the blue and red colors indicate the blueshifted and redshifted parts of the bipolar outflow, respectively, which will be discussed in Section \ref{subsec:mainoutflow}. We also overlaid two additional blue+red lines with PA = 135 and 77$\arcdeg$ that represent two outflows newly-identified in this study, ``Outflow A'' and ``Outflow B,'' respectively. We will discuss these outflows in Section \ref{subsec:newoutflow}.

\section{Results} \label{sec:result}

\subsection{Hydrogen Brackett Lines} \label{subsec:br}

\begin{figure*}
\centering
\includegraphics[scale=0.4]{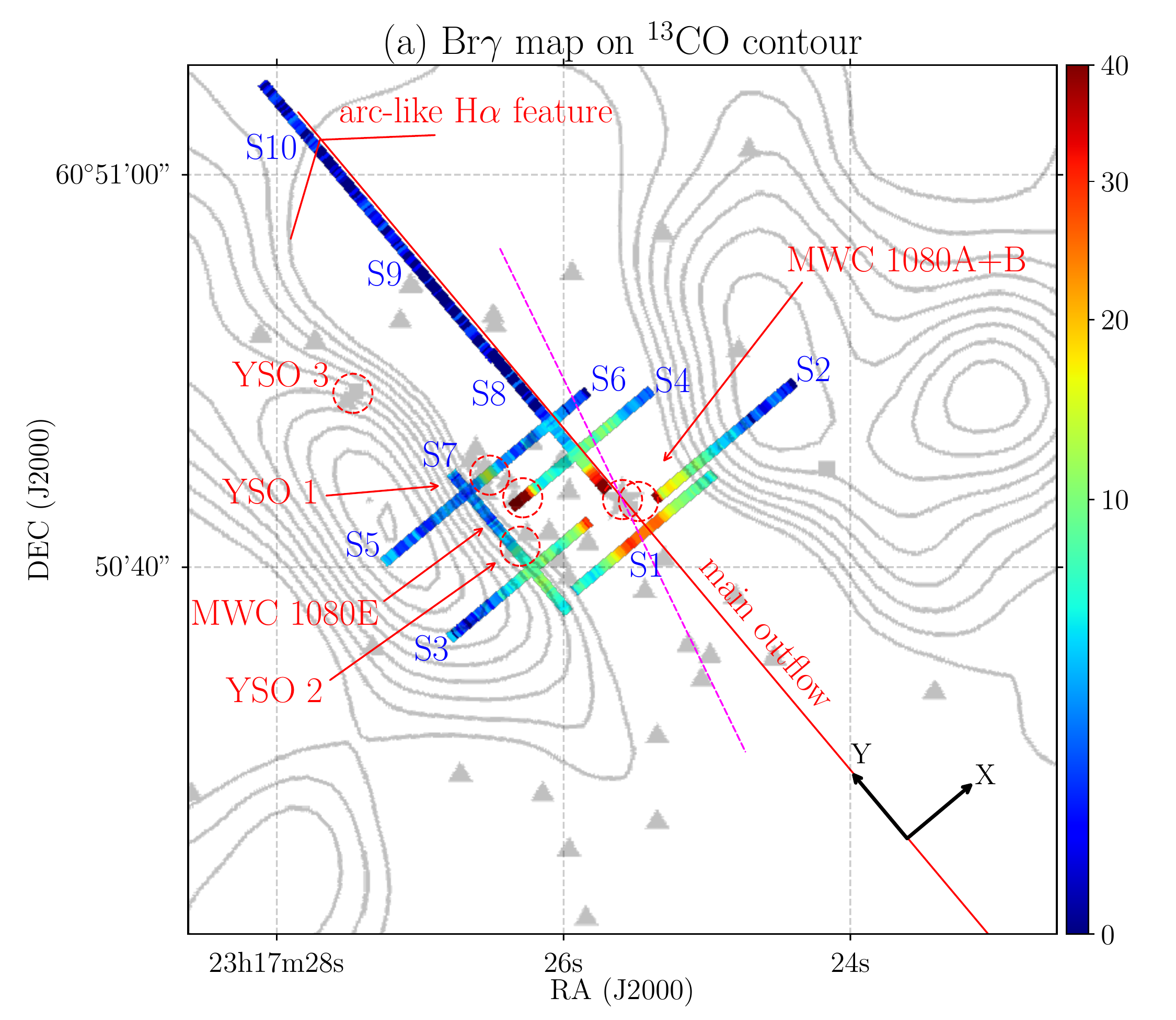}
\includegraphics[scale=0.39]{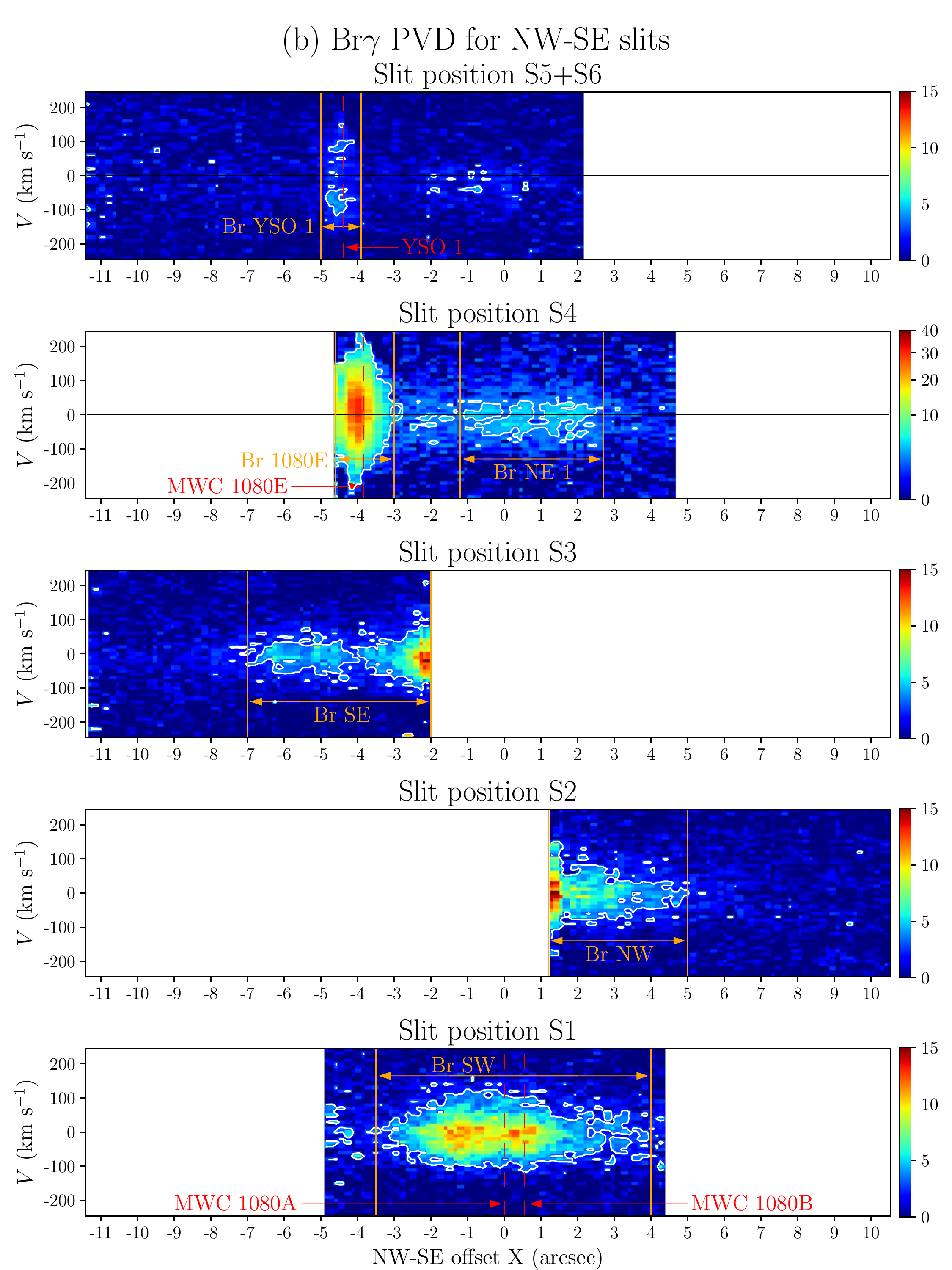}\includegraphics[scale=0.39]{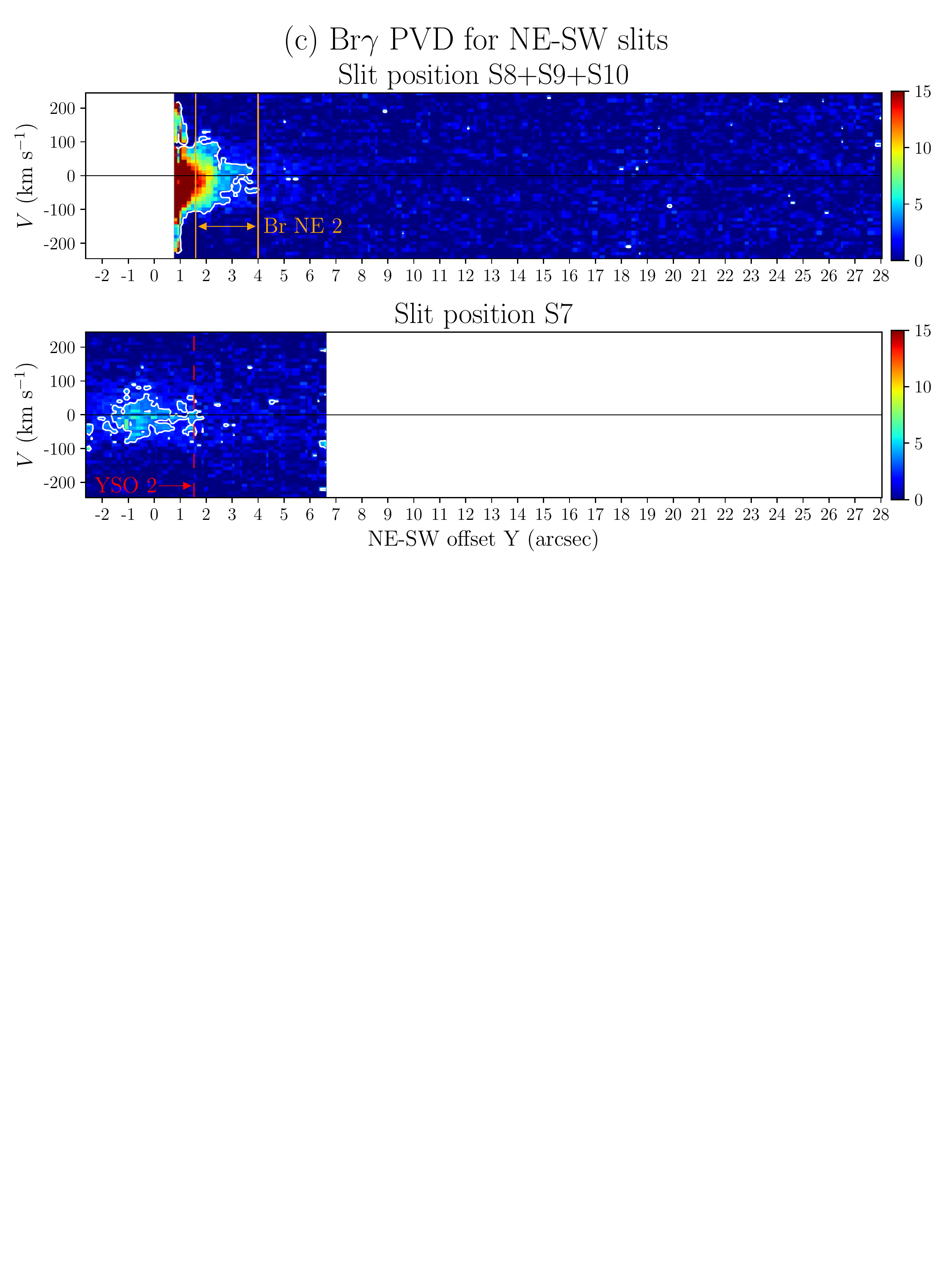}
\caption{(a) Continuum-subtracted Br$\gamma$ map integrated over $-$200 km s$^{-1}$ $\leq$ $V$ $\leq$ +200 km s$^{-1}$, and PVDs for (b) the NW-SE slits S1--S6 and (c) NE-SW slits S7--S10. The color scales are all in units of the detection significance (signal/$\sigma$). In panel (a), the overlaid $^{13}$CO contours, triangles, red dashed circles (young stars), red diagonal line (axis of the main outflow cavity), and red bent line (arc-like H$\alpha$ feature) are the same as in Figure \ref{fig:basic}(a). The magenta diagonal dashed line is another possible axis of the main outflow cavity (see Section \ref{subsec:brs}). In panels (b) and (c), the origin of the NW-SE offset X and NE-SW offset Y is the position of MWC 1080A; their positive directions are indicated in (a). The contours on the PVDs represent a significance level of 3$\sigma$. Seven Br$\gamma$ sources identified in this study are also denoted, together with the vertical solid lines that represent the spatial extent, in the PVDs. The vertical dashed lines indicate the X and Y positions of the young stars denoted in (a).\label{fig:brgmap}}
\end{figure*}

\begin{figure*}
\centering
\includegraphics[scale=0.4]{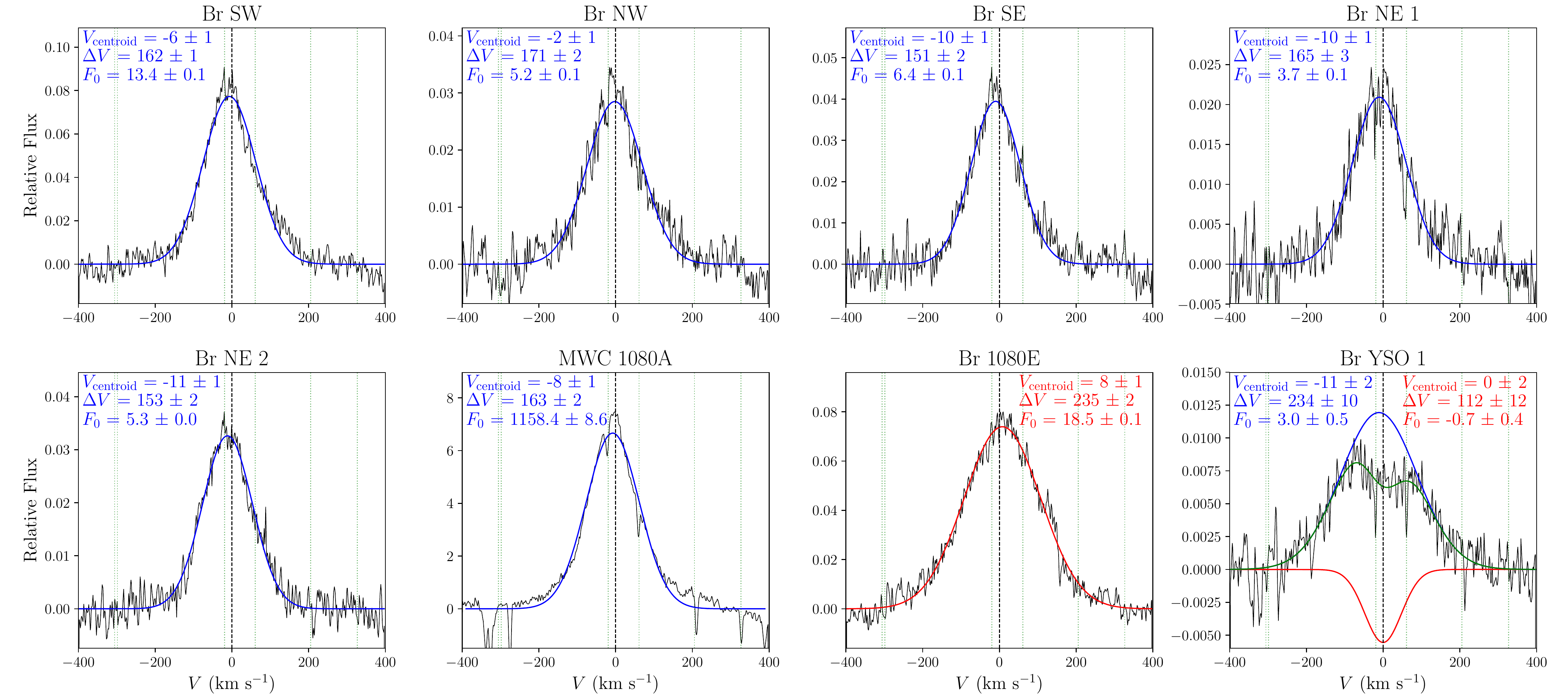}
\caption{Br$\gamma$ line profiles for the diffuse sources, integrated over the source region that is bounded by the vertical solid lines in Figures \ref{fig:brgmap}(b) and (c). The line profile for the star MWC 1080A is also shown for comparison. Each line profile was fitted with a Gaussian function plus a linear continuum (For Br YSO 1, we used two Gaussian functions), and the resulting linear continuum was subtracted from the line profile. The best-fit Gaussian function is shown in blue (negative centroid velocity or blueshift) or red (positive centroid velocity or redshift). The best-fit result obtained from two Gaussians for Br YSO 1 is also overplotted with the green curve. The centroid velocity $V_{\mathrm{centroid}}$, FWHM velocity width $\Delta V$, and line flux $F_{0}$ obtained from the fitting are shown in each panel and also listed in Table \ref{table:brglp}. The green dotted lines indicate the positions of the telluric OH lines \citep{2000A&A...354.1134R}. There are no telluric residuals that severely contaminate the Br$\gamma$ line profiles. The instrumental profiles ($\Delta V$ $\simeq$ 6.5 km s$^{-1}$ estimated from telluric OH lines) have not been removed from the line profiles, which gives only a difference of $<$1 km s$^{-1}$ for $\Delta V$ $>$ 20 km s$^{-1}$.\label{fig:brglp}}
\end{figure*}

We detected six hydrogen Brackett lines (Br$\gamma$, 10, 11, 12, 13, and 14) from the diffuse sources around MWC 1080 using IGRINS. Figure \ref{fig:brgmap}(a) shows the continuum-subtracted Br$\gamma$ map integrated over $-$200 km s$^{-1}$ $\leq$ $V$ $\leq$ +200 km s$^{-1}$. The X and Y axes in the figure denote the spatial direction in the PDVs for slits S1--S6 and S7--S10, respectively. We note that the slit directions (and thus the X and Y directions) are perpendicular or parallel to the axis of the main outflow cavity. The origin of the X and Y offsets was set to be the position of MWC 1080A. Following the definition, the Br$\gamma$ PVDs are shown for the slit positions S1--S10 in Figures \ref{fig:brgmap}(b) and (c). A total of seven Br$\gamma$ sources (5 extended sources and 2 compact sources) are identified in the PVDs. In the figure, the source identification names are also shown together with the vertical boundary lines, indicating the spatial extents of the sources. Five of the Br$\gamma$ sources are widely extended around MWC 1080A. They are referred to as Br SW, Br NW, Br SE, Br NE 1, and Br NE 2 according to their positions relative to MWC 1080A. The extended feature found in S7 is identical to Br SE in S3, and thus, its identification name was not denoted in S7 for clarity. Two of the Br$\gamma$ sources are relatively compact and confined within small areas close to the young stars MWC 1080E and YSO 1; they are referred to as Br 1080E and Br YSO 1, respectively.

To measure the Br$\gamma$ line profiles for the Br$\gamma$ sources, we integrated the cube data over the spatial axis that is bounded by the vertical solid lines in Figures \ref{fig:brgmap}(b) and (c). The resulting Br$\gamma$ line profiles are shown in Figure \ref{fig:brglp}. We fitted the individual line profiles with a Gaussian function plus a linear continuum and measured the centroid velocity $V_{\mathrm{centroid}}$, FWHM velocity width $\Delta V$, and line flux $F_{0}$. Regarding Br YSO 1, two Gaussian functions were applied to resolve the emission and absorption components. The best-fit results are presented in each panel of Figure \ref{fig:brglp} and also listed in the first row of Table \ref{table:brglp}. In Figure \ref{fig:brglp}, we also plot the Br$\gamma$ line profile obtained from the observation of the stellar source MWC 1080A. Because the Br$\gamma$ profile for MWC 1080A has a broad redshifted wing extending up to $\sim$300 km s$^{-1}$, the best-fit Gaussian function shows a slight disagreement with the actual line profile. However, the best-fit parameters $V_{\mathrm{centroid}}$ = $-$8 km s$^{-1}$ and $\Delta V$ = 163 km s$^{-1}$ appear to provide moderate estimates for the centroid velocity and velocity width of the Br$\gamma$ line profile. All of the Br$\gamma$ sources show broad ($\Delta V$ $>$ 150 km s$^{-1}$) and generally symmetric line profiles except Br YSO 1.

For the Br$\gamma$ sources, the other Brackett lines (Br 10 to Br 14) were also analyzed by the same method as applied to Br$\gamma$. Some strong telluric OH residuals found near the Br 10, Br 12, Br 13, and Br 14 lines were masked out before the line profiles were measured. The results for the lines with the fluxes measured at a significance level of $\ge$3$\sigma$ are shown in Table \ref{table:brglp}. For Br YSO 1, no Br 10--14 lines were detected with $\ge$3$\sigma$ significance. The best-fit parameters $V_{\mathrm{centroid}}$ and $\Delta V$ for the Br 10--14 lines were found to be slightly different from those for Br$\gamma$. However, the Br 10--14 line profiles are also broad ($\Delta V$ $>$ 100 km s$^{-1}$) and nearly symmetric, except for MWC 1080A, as found in Br$\gamma$. For the star MWC 1080A, the Br 10--14 line profiles were found to be broad, but show redshifted wing features, as in Br$\gamma$; hence, the estimated $V_{\mathrm{centroid}}$, $\Delta V$, and line flux $F$ of Br 10--14 for MWC 1080A are less reliable.

\subsubsection{Direct and Scattered Br$\gamma$ Sources} \label{subsec:brs}

\begin{figure*}
\centering
\includegraphics[scale=0.4]{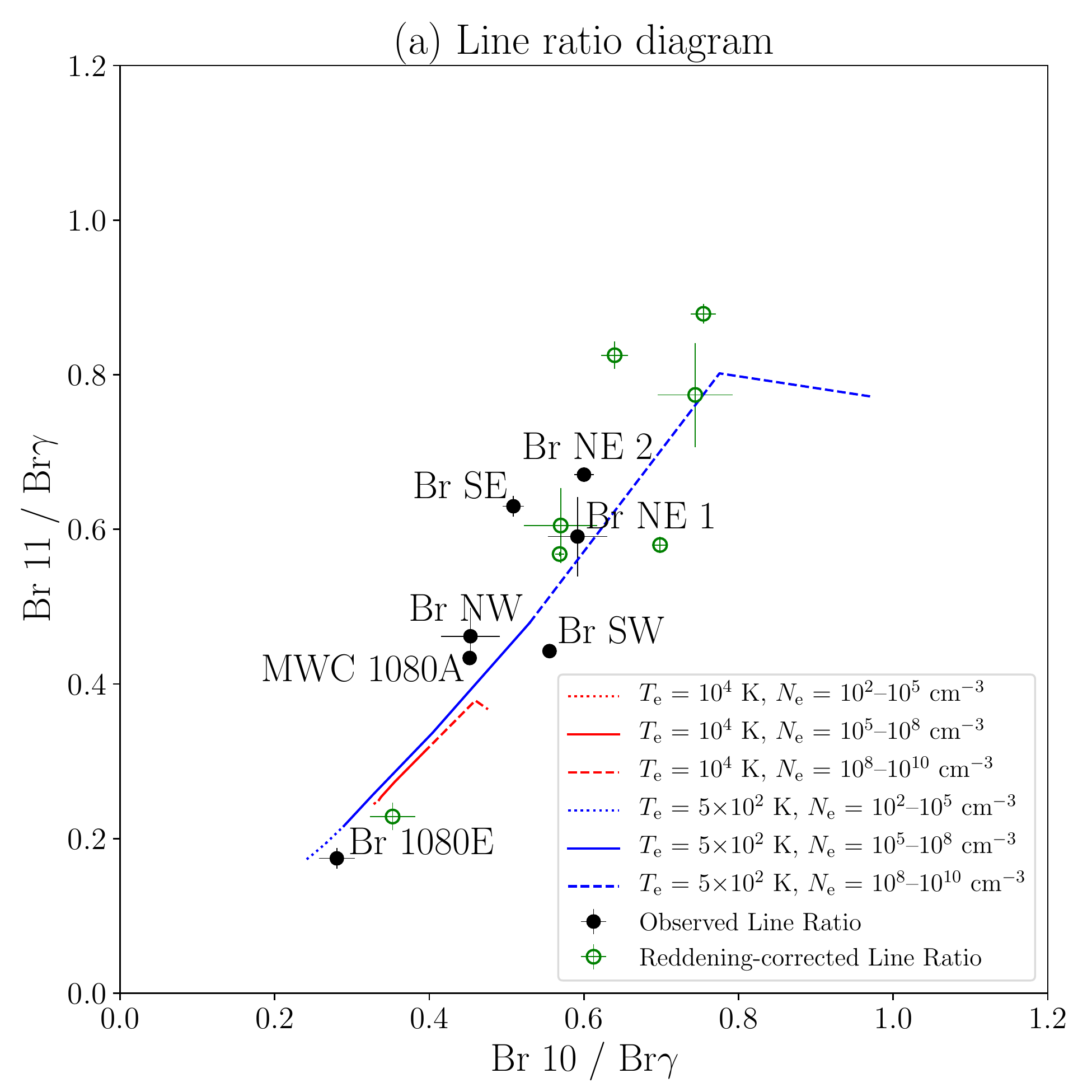}\includegraphics[scale=0.4]{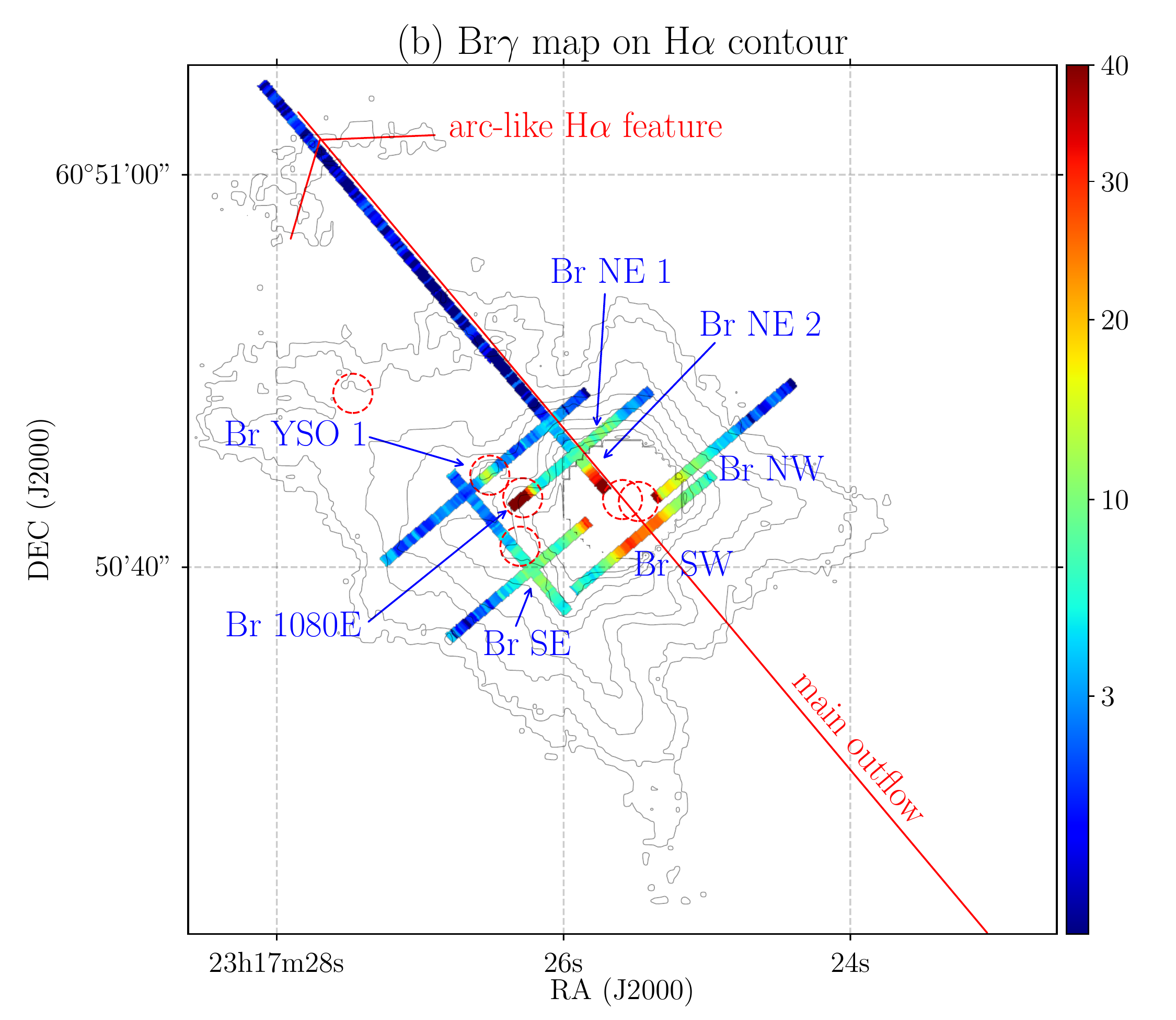}
\caption{(a) Br 10 / Br$\gamma$ vs. Br 11 / Br$\gamma$ line ratio diagram for the Br$\gamma$ sources and MWC 1080A, and (b) the Br$\gamma$ map overlaid on the H$\alpha$ contours. In panel (a), the observed line ratios (with 1$\sigma$ error bars) were calculated from the normalized line fluxes $F/F_{0}$ in Table \ref{table:brglp}. The green open circles indicate the line ratios reddening-corrected assuming $A_V$ = 5 mag. The dotted, solid, and dashed lines represent the theoretical ratios of the hydrogen recombination lines \citep[Case B approximation,][]{1995MNRAS.272...41S}. In panel (b), the Br$\gamma$ map is the same as in Figure \ref{fig:brgmap}(a), and the seven Br$\gamma$ sources identified in the PVDs (Figures \ref{fig:brgmap}(b) and (c)) are indicated by their assigned names. The H$\alpha$ contours were obtained from Figure \ref{fig:basic}(b). The contour levels are 8, 12, 30, 50, 100, 150, 200, and 250 $\times$ 10$^{-19}$ W m$^{-2}$ arcsec$^{-2}$. Red dashed circles (young stars), a red diagonal line (axis of the main outflow cavity), and a red bent line (arc-like H$\alpha$ feature) are the same as in Figure \ref{fig:basic}(a).\label{fig:brlrd}}
\end{figure*}

As described in Section \ref{subsec:br}, seven Br$\gamma$ sources were identified from the PVDs (Figure \ref{fig:brgmap}). We now classify them according to their origins in two groups: direct Br$\gamma$ sources and scattered ones. Five of them (Br SW, Br NW, Br SE, Br NE 1, and Br NE 2) are spatially extended sources, and two of them (Br 1080E and Br YSO 1) are point-like, compact sources. We note that the compact sources Br 1080E and Br YSO 1 were found to be spatially coincident with the stars MWC 1080E and YSO 1, respectively, and thus, their identification names were given following the corresponding stars. Figure \ref{fig:brglp} shows the Br$\gamma$ line profiles, which are all broad ($\Delta V$ $>$ 150 km s$^{-1}$) and generally symmetric about the line center except Br YSO 1. The best-fit results of the Gaussian-line fitting for the line profiles show that the extended Br$\gamma$ sources (Br SW, Br NW, Br SE, Br NE 1, and Br NE 2) have negative centroid velocities of $V_{\mathrm{centroid}}$ = $-$11 to $-$2 km s$^{-1}$ and relatively narrow velocity widths of $\Delta V$ = 151 to 171 km s$^{-1}$. However, the compact source Br 1080E has a positive $V_{\mathrm{centroid}}$ = +8 km s$^{-1}$ and broader $\Delta V$ = 235 km s$^{-1}$. The difference of the velocity widths between Br 1080E and the extended sources is also shown clearly in the Br$\gamma$ PVDs (see Figure \ref{fig:brgmap}). In Figure \ref{fig:brglp}, the Br$\gamma$ line profile of the other compact source Br YSO 1 was fitted with two Gaussians that represent the emission and absorption components, respectively. The best-fit parameters for the emission component were obtained to be $V_{\mathrm{centroid}}$ = $-$11 km s$^{-1}$ and $\Delta V$ = 234 km s$^{-1}$. This result indicates that the Br$\gamma$ emission line profile of Br YSO 1 is also distinct from those of the extended Br$\gamma$ sources: a similar $V_{\mathrm{centroid}}$, but a much broader $\Delta V$. In Figure \ref{fig:brglp}, the Br$\gamma$ line profile for the star MWC 1080A shows $V_{\mathrm{centroid}}$ = $-$8 km s$^{-1}$ and $\Delta V$ = 163 km s$^{-1}$, which is similar to those obtained from the extended Br$\gamma$ sources. The estimated best-fit parameters are slightly inconsistent (differences of $<$9 km s$^{-1}$ for $V_{\mathrm{centroid}}$ and $<$20 km s$^{-1}$ for $\Delta V$) due to the broad wing component; however, the shapes of the line profiles agree well with each other. The similarity in the Br$\gamma$ line profiles between the central star MWC 1080A and the extended sources implies that the extended sources (Br SW, Br NW, Br SE, Br NE 1, and Br NE 2) would be dusty reflection nebulae that scatter the stellar Br$\gamma$ emission from MWC 1080A. On the other hand, the compact sources Br 1080E and Br YSO 1 are most likely to be direct emissions from the young stars MWC 1080E and YSO 1, respectively.

In Figure \ref{fig:brlrd}(a), we compare the line ratios Br 10/Br$\gamma$ and Br 11/Br$\gamma$ for six Br$\gamma$ sources and the star MWC 1080A, from which both the Br 10 and Br 11 lines were significantly detected at a level of $>$10$\sigma$. We used the normalized line fluxes $F/F_{0}$ for Br 10 and Br 11 presented in Table \ref{table:brglp}. In the figure, we also show the ratios reddening-corrected by assuming $A_V$ = 5 mag and utilizing the extinction curve of \citet{1989ApJ...345..245C} with $R_V$ = 3.1. Figure \ref{fig:brlrd}(a) also shows the theoretical line ratios calculated assuming the Case B recombination \citep{1995MNRAS.272...41S} for a combination of two temperatures of $T_{\mathrm{e}}$ = 5$\times$10$^{2}$ K (blue) and 10$^{4}$ K (red) and three density ranges of $N_{\mathrm{e}}$ = 10$^{2}$--10$^{5}$ cm$^{-3}$ (dotted), 10$^{5}$--10$^{8}$ cm$^{-3}$ (solid), and 10$^{8}$--10$^{10}$ cm$^{-3}$ (dashed). We also calculated the theoretical line ratios for a wide range of electron temperature $T_{\mathrm{e}}$ = 5$\times$10$^{2}$--10$^{5}$ K and electron density $N_{\mathrm{e}}$ = 10$^{2}$--10$^{14}$ cm$^{-3}$, but they are not shown in the figure.

Figure \ref{fig:brlrd}(a) shows that all the line ratios measured for the extended Br$\gamma$ sources (Br SW, Br NW, Br SE, Br NE 1, and Br NE 2) are higher than those for the star MWC 1080A. We note that the line ratios for MWC 1080A could be affected by the broad wing components, as mentioned in Section \ref{subsec:br}. However, we also found that the line ratios for the extended sources are higher than those theoretically-predicted for most of $T_{\mathrm{e}}$ and $N_{\mathrm{e}}$; the line ratios are found to be comparable only with the predictions for the lowest temperature $T_{\mathrm{e}}$ = 5$\times$10$^{2}$ K and the highest density $N_{\mathrm{e}}$ = 10$^{8}$--10$^{10}$ cm$^{-3}$. Moreover, even this extreme case was not able to reproduce the observed line ratios if the dust extinction toward the sources is higher than $A_V$ = 5 mag (see the green open circles in the figures). The excess of the line ratios (Br 10/Br$\gamma$ and Br 11/Br$\gamma)$ strongly supports our conclusion, inferred from the line profiles, that the extended Br$\gamma$ sources are dusty nebulae illuminated by MWC 1080A. The lines Br 10 and Br 11 have shorter wavelengths than Br$\gamma$ and thus are more easily scattered by dust. The scattered light in the Br 10 and Br 11 lines then gives higher line ratios than those theoretically-predicted for the intrinsic line ratios of the direct light from MWC 1080A. In contrast, the line ratios for the compact source Br 1080E were found to be much lower than those for the extended sources and MWC 1080A. They are also lower than those predicted for most combinations of $T_{\mathrm{e}}$ and $N_{\mathrm{e}}$ except for the extreme case of $T_{\mathrm{e}}$ = 5$\times$10$^{2}$ K and $N_{\mathrm{e}}$ = 10$^{2}$--10$^{5}$ cm$^{-3}$. This result supports that the compact source Br 1080E, which is spatially coincident with MWC 1080E, is most likely a direct emission from MWC 1080E. Because no scattered light will contribute to the point-like source, Br 10 and Br 11 are attenuated more strongly than Br$\gamma$. This makes the observed line ratios of Br 10/Br$\gamma$ and Br 11/Br$\gamma$ for Br 1080E much lower than the intrinsic line ratios. Hence, we can estimate the extinction $A_V$ toward MWC 1080E by assuming Case B and, for instance, $T_{\mathrm{e}}$ = 10$^{4}$ K and $N_{\mathrm{e}}$ = 10$^{6}$ cm$^{-3}$. The extinction $A_V$ toward MWC 1080E is then derived as 4.0 $\pm$ 1.8 mag from Br 10/Br$\gamma$ and 6.9 $\pm$ 1.4 mag from Br 11/Br$\gamma$. These values are comparable with $A_V$ $\simeq$ 5 mag obtained toward MWC 1080A \citep{1979ApJS...41..743C,1992ApJ...397..613H,1997MNRAS.291..797O}, although the extinctions toward MWC 1080A and MWC 1080E can be somewhat different.

In Figure \ref{fig:brlrd}(b), the Br$\gamma$ map of Figure \ref{fig:brgmap}(a) is overlaid on the H$\alpha$ contours for comparison. The H$\alpha$ contours were obtained from Figure \ref{fig:basic}(b) but smoothed using a Gaussian function with $\sigma$ = 5.0 pixel. The Br$\gamma$ morphology is, in general, consistent with the H$\alpha$ image. However, the Br$\gamma$ emission at a significance level of $>$3$\sigma$ was observed only in the regions with the H$\alpha$ intensities higher than $\sim$10$^{-17}$ W m$^{-2}$ arcsec$^{-2}$ (the fifth lowest contour level). This intensity value is actually the upper limit of the sole H$\alpha$ intensity, considering the {[}\ion{N}{2}{]} contribution to the IPHAS H$\alpha$ filter. The Br$\gamma$ counterpart for the arc-like H$\alpha$ feature was not detected. The lack of the counterpart is likely due to the Br$\gamma$ detection limit of our IGRINS observation. The Brackett line ratios shown in Figure \ref{fig:brlrd}(a) suggest that much of the extended H$\alpha$ emission around MWC 1080A is likely stellar emission from MWC 1080A scattered by dust, although the emissions in the saturated region close to MWC 1080A might originate from an ionized stellar or disk wind. On the other hand, we note that two compact Br$\gamma$ sources (Br 1080E and Br YSO 1) are spatially coincident with two local H$\alpha$ peaks, as can be seen in Figure \ref{fig:brlrd}(b). Therefore, the two H$\alpha$ peaks close to MWC 1080E and YSO 1 also likely originate from the individual young stars as the compact Br$\gamma$ sources do.

The IGRINS Br$\gamma$ map partially supplements the brightest region close to MWC 1080A that was saturated in the IPHAS images and thus masked out. The morphology of the Br$\gamma$-reflection nebulae is able to provide clues on the spatial distribution of dusty material and outflow around MWC 1080A. We note that Br SW shows two peaks separated by $\sim$1.$\arcsec$8 in slit position S1 of Figure \ref{fig:brgmap}(b). The X positions of the close binary MWC 1080A and MWC 1080B are overlaid by two red, vertical dashed lines in the PVD of S1. One of the Br$\gamma$ peaks is coincident with the position of MWC 1080B (X $\simeq$ +0.$\arcsec$5), but the other one is offset by about $-$1.$\arcsec$3 from MWC 1080A. We also note that another extended source Br NE 1 shows two peaks in its PVD of slit position S4, although they are much fainter than those of Br SW. The separation $(\sim$2$\arcsec$) between the two peaks in Br NE 1 is similar to that for Br SW ($\sim$1.$\arcsec$8). The locations of the four peaks with respect to MWC 1080A and the similarity between the peak separations suggest a connection of the peaks in Br SW and Br NE 1 with the bipolar, main outflow from MWC 1080A. The midpoint between the two peaks in Br SW lies at X $\simeq$ $-$0.$\arcsec$4, and the midpoint for Br NE 1 lies at X $\simeq$ +1$\arcsec$. Figure \ref{fig:brgmap}(a) shows the magenta dashed line connecting MWC 1080A and the two midpoints, which has a PA of $\sim$ 25$\arcdeg$. This is somewhat different from the PA of 40$\arcdeg$ for the red solid line that passes through MWC 1080A and the apex of the arc-like H$\alpha$ feature. However, both the diagonal lines seem to, at least approximately, follow the axis of the main outflow cavity inferred from the $^{13}$CO map. In other words, the four peaks found in Br SW and Br NE 1 are likely to trace the main outflow driven by MWC 1080A.

\subsection{Molecular Hydrogen Lines} \label{subsec:h2}

\begin{figure*}
\centering
\includegraphics[scale=0.4]{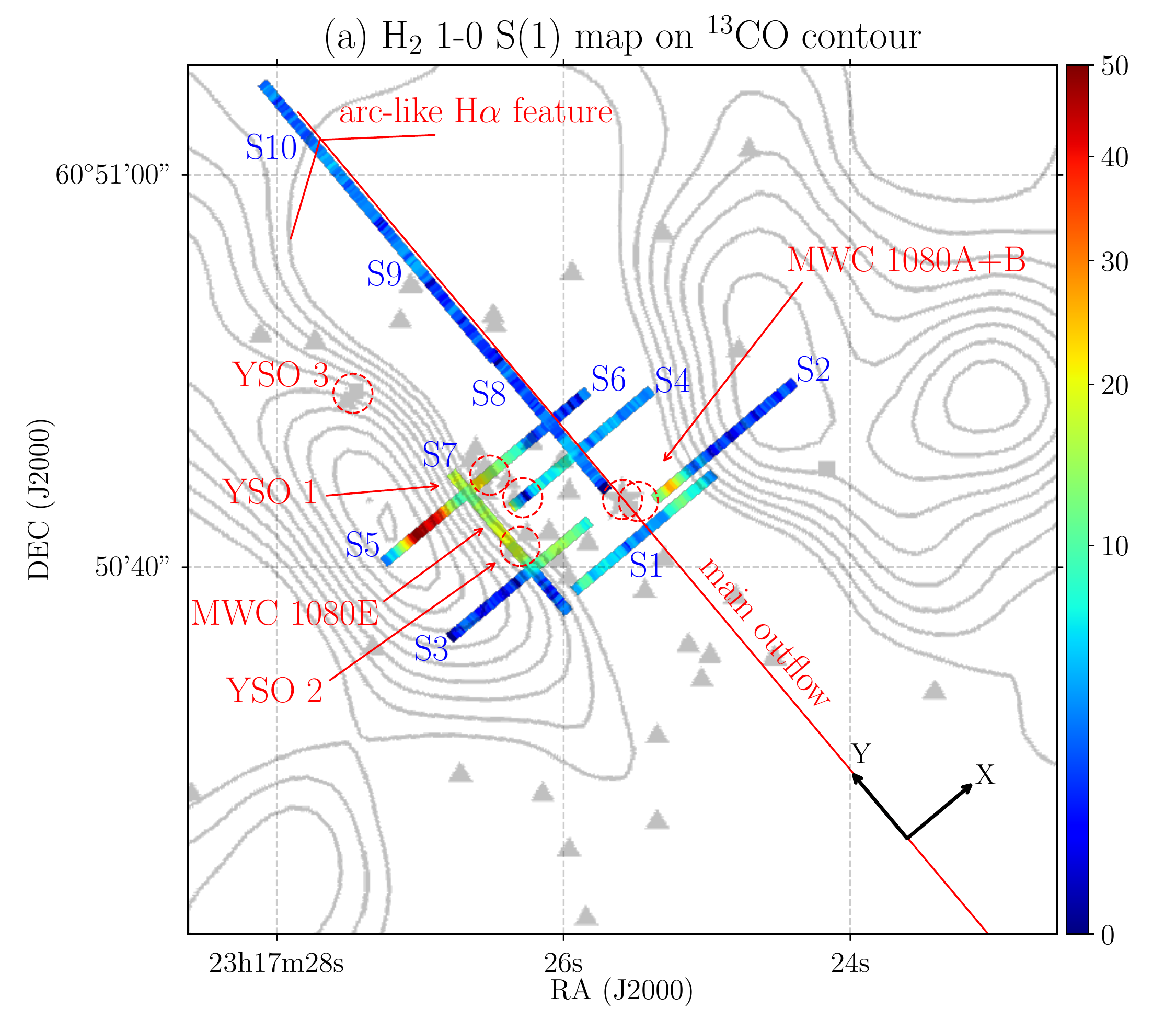}
\includegraphics[scale=0.39]{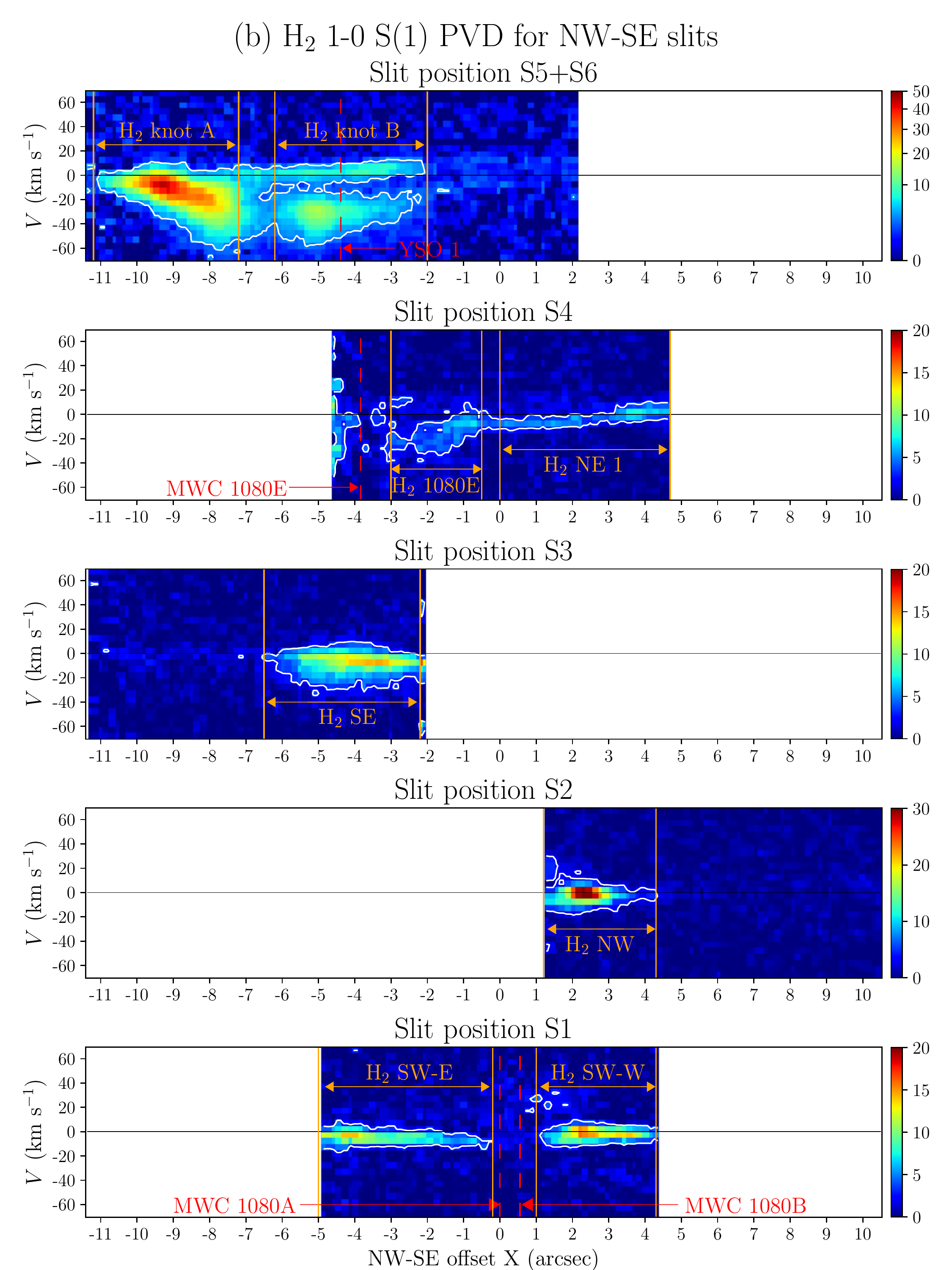}\includegraphics[scale=0.39]{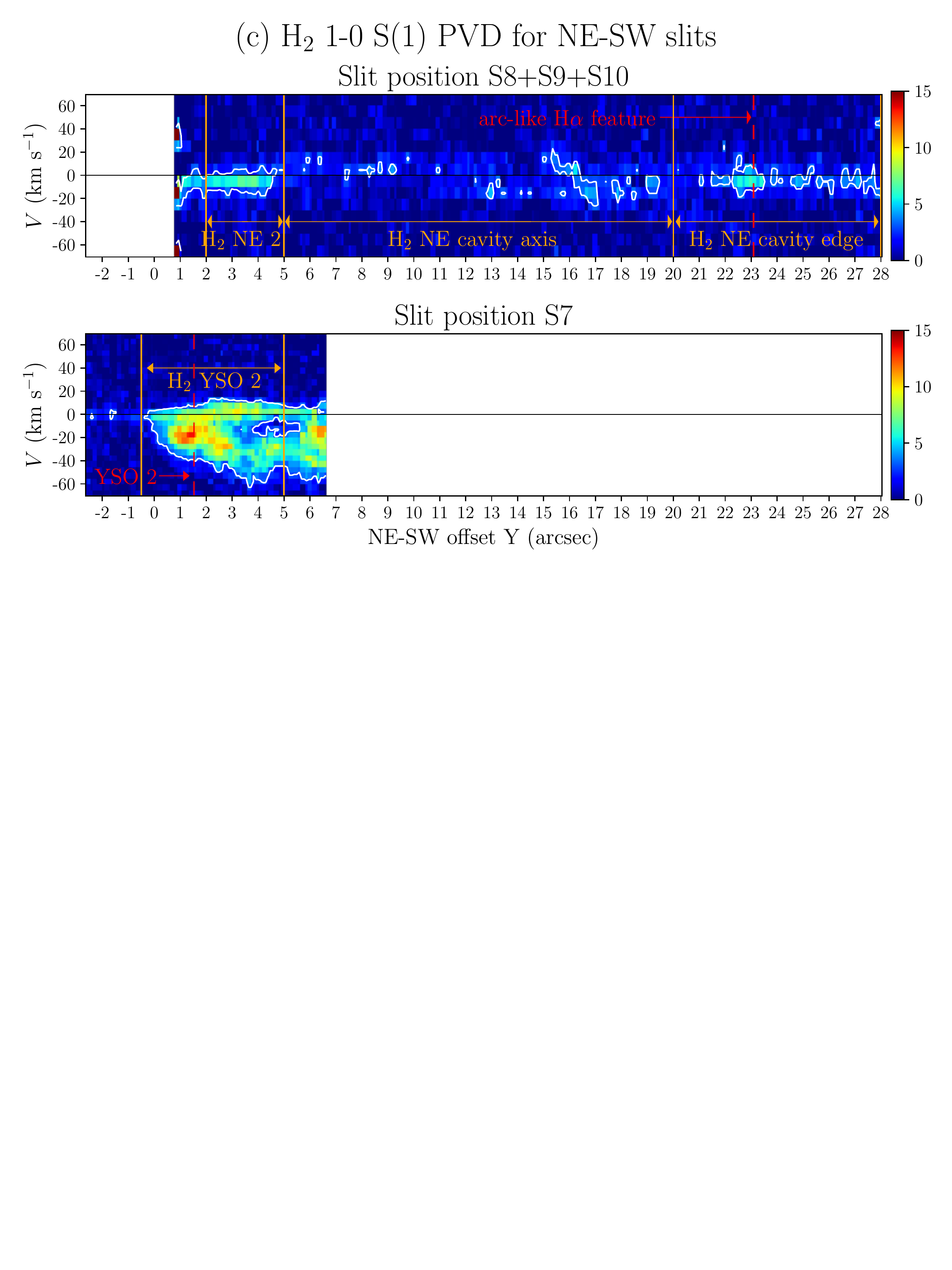}
\caption{Continuum-subtracted H$_{2}$ 1-0 S(1) map integrated over $-$60 km s$^{-1}$ $\leq$ $V$ $\leq$ +60 km s$^{-1}$, and PVDs for (b) the NW-SE slits S1--S6 and (c) NE-SW slits S7--S10. The color scales are all in units of the detection significance (signal/$\sigma$). In panel (a), the overlaid $^{13}$CO contours, triangles, red dashed circles (young stars), red diagonal line (axis of the main outflow cavity), and red bent line (arc-like H$\alpha$ feature) are the same as in Figure \ref{fig:basic}(a). In panels (b) and (c), the origin of the NW-SE offset X and NE-SW offset Y is the position of MWC 1080A; their positive directions are indicated in (a). The contours on the PVDs represent a significance level of 3$\sigma$. The twelve H$_{2}$ 1-0 S(1) sources identified in this study are also denoted, together with the vertical solid lines that represent the spatial extent, in the PVDs. The vertical dashed lines indicate the X or Y positions of the young stars and the arc-like H$\alpha$ feature denoted in (a).\label{fig:h2map}}
\end{figure*}

\begin{figure*}
\centering
\includegraphics[scale=0.4]{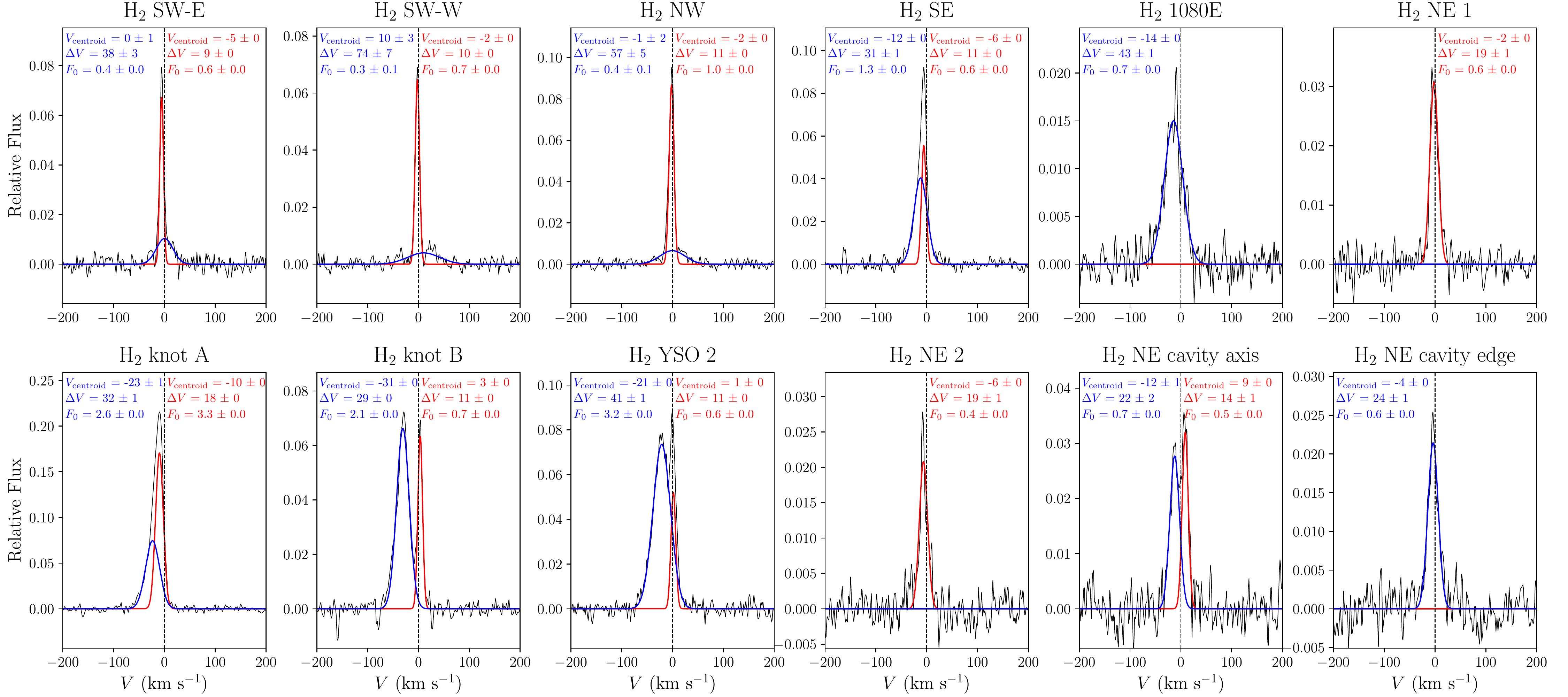}
\caption{H$_{2}$ 1-0 S(1) line profiles of the diffuse sources, integrated over the source region that is bounded by the vertical solid lines in Figures \ref{fig:h2map}(b) and (c). Each line profile was fitted with two Gaussians plus a linear continuum, and the resulting linear continuum was subtracted from the line profile. The best-fit Gaussian functions are overplotted with the blue (broad component; $\Delta V$ $\ge$ 20 km s$^{-1}$) and red (narrow component; $\Delta V$ $<$ 20 km s$^{-1}$) curves. The centroid velocity $V_{\mathrm{centroid}}$, FWHM velocity width $\Delta V$, and line flux $F_{0}$ obtained from the fitting are shown in each panel and also listed in Tables \ref{table:h2lp1} and \ref{table:h2lp2}. There are no strong telluric OH lines near the H$_{2}$ 1-0 S(1) line \citep{2000A&A...354.1134R}. The instrumental profiles ($\Delta V$ $\simeq$ 6.5 km s$^{-1}$ estimated from telluric OH lines) have not been removed from the line profiles, which gives only a difference of $<$1 km s$^{-1}$ for $\Delta V$ $>$ 20 km s$^{-1}$ and $\sim$2.8 km s$^{-1}$ for $\Delta V$ = 9 km s$^{-1}$.\label{fig:h2lp}}
\end{figure*}

We detected seven NIR H$_{2}$ lines from the diffuse sources around MWC 1080 with IGRINS. We first carried out a kinematic analysis for the brightest line, H$_{2}$ 1-0 S(1). Figure \ref{fig:h2map}(a) shows a continuum-subtracted H$_{2}$ 1-0 S(1) map integrated over $-$60 km s$^{-1}$ $\leq$ $V$ $\leq$ +60 km s$^{-1}$. The H$_{2}$ 1-0 S(1) PVDs for slit positions S1--S10 are presented in Figures \ref{fig:h2map}(b) and (c). In the PVDs, we found a total of twelve H$_{2}$ 1-0 S(1) sources. The figures show their identification names and spatial boundaries. H$_{2}$ SW-E, H$_{2}$ SW-W, H$_{2}$ NW, H$_{2}$ NE 1, and H$_{2}$ NE 2 have narrow velocity widths centered at the systemic LSR velocity ($V$ = 0 km s$^{-1}$) and are located within the main outflow cavity near MWC 1080A. On the other hand, H$_{2}$ SE, H$_{2}$ 1080E, H$_{2}$ knot A, H$_{2}$ knot B, and H$_{2}$ YSO 2 show much broader velocity widths extending blueward and lie near/in the SE molecular region. Along the NE axis of the main outflow cavity (slit position S8+S9+S10), two faint sources were detected in a velocity range of $-$20 km s$^{-1}$ $\lesssim$ $V$ $\lesssim$ +20 km s$^{-1}$, out of which the brightest one is located near the NE edge of the cavity. They are referred to as H$_{2}$ NE cavity axis and H$_{2}$ NE cavity edge.

To measure the H$_{2}$ 1-0 S(1) line profiles of the H$_{2}$ sources, we integrated the cube data over the source area that is bounded by the vertical solid lines in Figures \ref{fig:h2map}(b) and (c). The resulting H$_{2}$ line profiles are shown in Figure \ref{fig:h2lp}. In the figure, most of the H$_{2}$ sources appear to have at least two velocity components. We, therefore, fitted the individual line profiles with two Gaussian functions plus a linear function and measured the centroid velocity $V_{\mathrm{centroid}}$, FWHM velocity width $\Delta V$, and line flux $F_{0}$ for each velocity component. The best-fit profiles are shown in each panel of Figure \ref{fig:h2lp}. Only the velocity components detected with a significance level of $\ge$3$\sigma$ are presented in the figure. For the sake of convenience in the identifications, we classified the detected velocity components into two categories: narrow component with $\Delta V$ $<$ 20 km s$^{-1}$ (red curves) and broad one with $\Delta V$ $\ge$ 20 km s$^{-1}$ (blue curves). The second row of Table \ref{table:h2lp1} shows the best-fit parameters for the narrow components and Table \ref{table:h2lp2} for the broad components. The broad components for H$_{2}$ SW-E, H$_{2}$ SW-W, and H$_{2}$ NW were found to be relatively negligible, as shown in Figure \ref{fig:h2lp}, and thus are not presented in Table \ref{table:h2lp2}.

The other six H$_{2}$ lines, which are fainter than H$_{2}$ 1-0 S(1), are also shown in Tables \ref{table:h2lp1} and \ref{table:h2lp2}. We also extracted their line profiles and fitted them individually using the same method as for H$_{2}$ 1-0 S(1). Some strong telluric OH residuals found near the H$_{2}$ lines were masked out before the line profiles were measured. We found that the lines are also well represented by two velocity components, as in the H$_{2}$ 1-0 S(1) line profile. We also classified the velocity components like the H$_{2}$ 1-0 S(1) line components. The best-fit results for the velocity components detected with a significance level of $\ge$3$\sigma$ are presented in Tables \ref{table:h2lp1} and \ref{table:h2lp2}. The narrow component for these weak H$_{2}$ lines was detected in most of the sources except for two sources, H$_{2}$ NE 2 and H$_{2}$ NE cavity axis. On the other hand, the broad component was not found in several sources, as shown in Table \ref{table:h2lp2}. In particular, the broad component of H$_{2}$ 2-1 S(1) was detected only in H$_{2}$ 1080E.

\subsubsection{Fluorescent and Thermal H$_{2}$ Sources} \label{subsec:h2s}

\begin{figure*}
\centering
\includegraphics[scale=0.4]{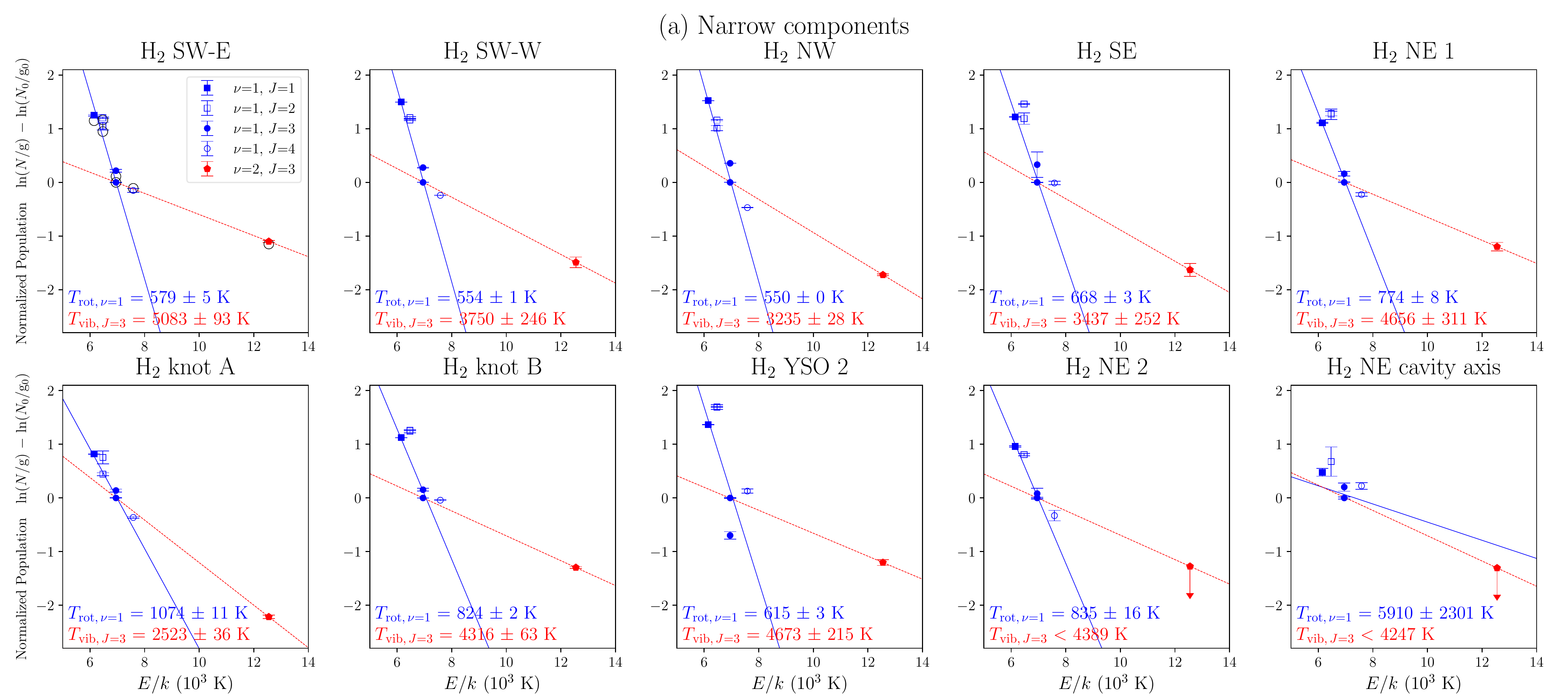}
\includegraphics[scale=0.4]{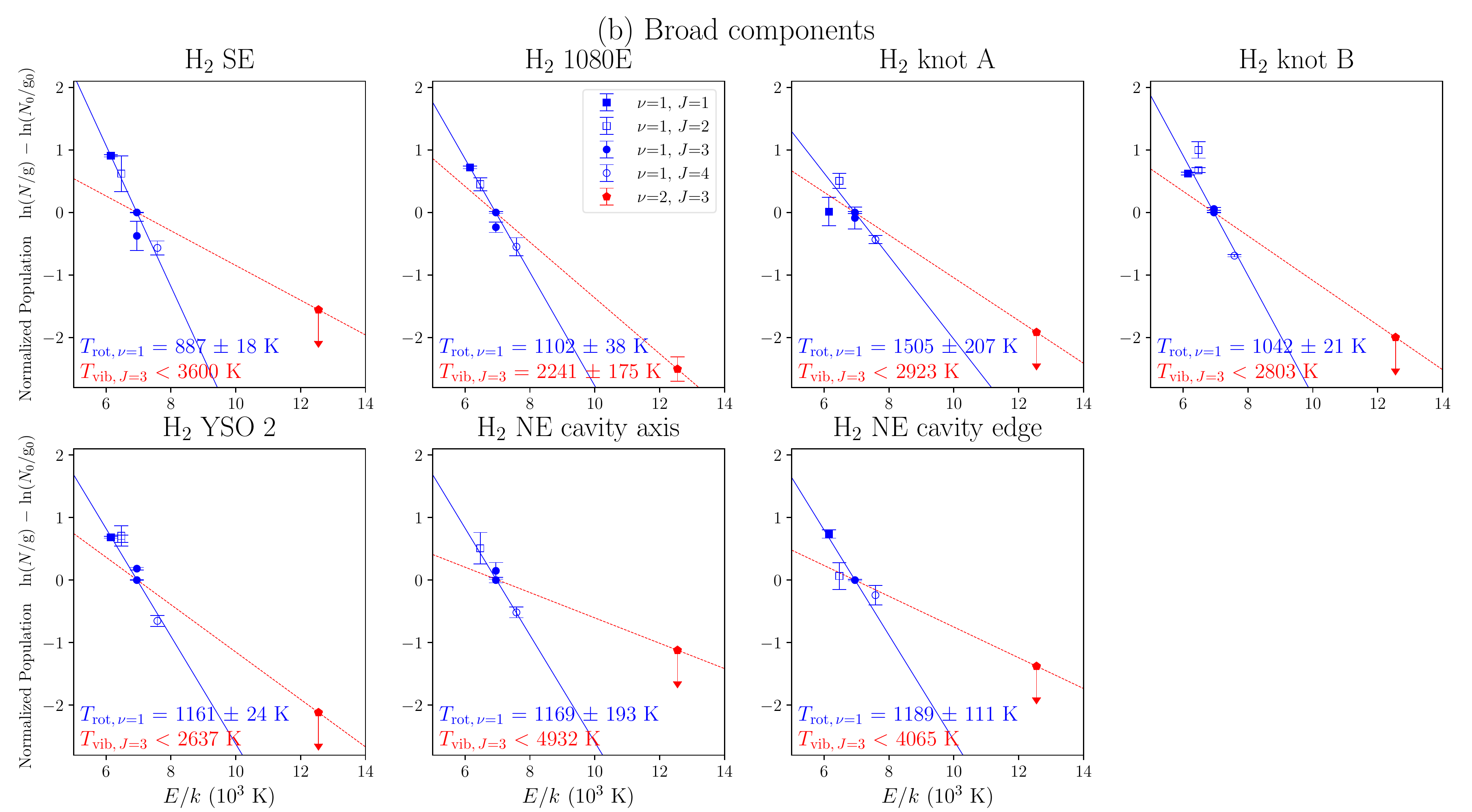}
\caption{Level population diagrams of (a) the narrow and (b) broad components for H$_{2}$ sources. The normalized level populations (with 1$\sigma$ error bars) were calculated from the normalized line fluxes $F/F_{0}$ in Tables \ref{table:h2lp1} and \ref{table:h2lp2}. For eight sources without the H$_{2}$ 2-1 S(1) detections (red points), the upper limits of the H$_{2}$ 2-1 S(1) fluxes were used. Rotation temperatures $T_{\mathrm{rot},\,v=1}$ were derived from the best-fit slopes (the blue lines) for $v$ = 1, $J$ = 1--4 levels, and vibration temperatures $T_{\mathrm{vib},\,J=3}$ were from the best-fit slopes (the red lines) for $v$ = 1--2, $J$ = 3 levels, for which the results are given in each panel. In the first panel, the open circles indicate the results obtained from the reddening correction with $A_V$ = 5 mag, which caused only minor changes.\label{fig:h2lpd}}
\end{figure*}

In the H$_{2}$ 1-0 S(1) PVDs of Section \ref{subsec:h2}, twelve H$_{2}$ sources were identified (see Figures \ref{fig:h2map}(b) and (c)). We now classify them according to their excitation mechanisms in two groups: fluorescent H$_{2}$ sources and thermal ones. As shown in Figure \ref{fig:h2lp}, many of them have both narrow and broad components. In particular, two sources (H$_{2}$ knot B and H$_{2}$ YSO 2) show two clearly-separated components not only in the spectra but also in the H$_{2}$ 1-0 S(1) PVDs. As listed in Table \ref{table:h2lp1}, the narrow components of H$_{2}$ 1-0 S(1) have velocity widths of $\Delta V$ = 9--19 km s$^{-1}$ and centroid velocities of $V_{\mathrm{centroid}}$ = $-$10 to +9 km s$^{-1}$, which are not significantly different from the systemic LSR velocity. On the other hand, the broad components ($\Delta V > 20$ km s$^{-1}$) of H$_{2}$ 1-0 S(1) appear to have substantially blueshifted line-centers (see Table \ref{table:h2lp2}). The broad components for H$_{2}$ SE, H$_{2}$ 1080E, H$_{2}$ knot A, H$_{2}$ knot B, and H$_{2}$ YSO 2 have $V_{\mathrm{centroid}}$ = $-$31 to $-$12 km s$^{-1}$ and a much broader $\Delta V$ of 29 to 43 km s$^{-1}$. The broad components for H$_{2}$ NE cavity axis and H$_{2}$ NE cavity edge, detected along the NE cavity region, have $(V_{\mathrm{centroid}},\,\Delta V)$ = $(-$12, 22) and ($-$4, 24) km s$^{-1}$, respectively. The difference in the line profiles suggests that the narrow and broad components are excited by different mechanisms. The narrow components with $\left|V_{\mathrm{centroid}}\right|\lesssim 10$ km s$^{-1}$, indicating no substantial motion relative to the systemic LSR motion, would be due to the fluorescent H$_{2}$ lines from the photodissociation regions (PDRs). On the other hand, the broad components with blueshifted line centers $(V_{\mathrm{centroid}}<0)$ are likely caused by thermal H$_{2}$ emission from shock-heated regions.

To estimate the relative contributions of the fluorescent and thermal excitations in the H$_{2}$ lines, we examined the H$_{2}$ line ratios observed for the individual sources. We detected only seven H$_{2}$ lines belonging to the $v$ = 1--0 or 2--1 transition and thus utilized a simple method by \citet{1989ApJ...336..207T}, which is suitable to analyze a handful number of lines. No H$_{2}$ lines from the $v$ = 3--2 transition were detected. Using the H$_{2}$ line fluxes normalized to the H$_{2}$ 1-0 S(1) line flux given in Tables \ref{table:h2lp1} and \ref{table:h2lp2}, we constructed the level population diagrams of both narrow and broad components for each H$_{2}$ source and presented the results in Figure \ref{fig:h2lpd}. To derive the normalized level population ln($N$/g) $-$ ln($N_{0}$/g$_{0}$), we used the transition probabilities from \citet{1998ApJS..115..293W} and Equations (4)--(6) of \citet{2017ApJ...838..152K}. For eight sources without the H$_{2}$ 2-1 S(1) detection, we used the upper limits of the H$_{2}$ 2-1 S(1) flux, which were calculated from the RMS noises in the spectra. The wavelength differences between the H$_{2}$ lines are so small (at most $\sim$0.3 $\micron$) that there is no significant variation in extinction between the lines. Therefore, no reddening correction was made; the reddening-correction did not significantly alter the present result. We show a sample of reddening-corrected results assuming $A_V$ = 5 mag in the first panel of Figure \ref{fig:h2lpd}(a). As shown in the figure, the reddening-corrected line ratios denoted by open circles show only minor changes.

In the excitation diagram of Figure \ref{fig:h2lpd}, the slope of the best-fit straight line gives the excitation temperature (i.e., rotational or vibrational temperature). Using uncertainty-weighted least-squares fitting, the rotation temperature $T_{\mathrm{rot},\,v=1}$ was derived from the relative populations between $J$ = 1--4 levels in the $v=1$ state (the blue lines), and the vibration temperature $T_{\mathrm{vib},\,J=3}$ was from the relative populations between the $(v, J)$ = (1, 3) and $(v, J)$ = (2, 3) states (the red lines). The derived temperatures are denoted in each panel. For the narrow components in Figure \ref{fig:h2lpd}(a), the rotational temperature for all of the H$_{2}$ sources except for two sources (H$_{2}$ knot A and H$_{2}$ NE cavity axis) was found to be in a range of $T_{\mathrm{rot},\,v=1}$ = 550--835 K. The vibrational temperature for the narrow components was derived as $T_{\mathrm{vib},\,J=3}$ = 3235--5083 K, except for the source H$_{2}$ NE 2 without the H$_{2}$ 2-1 S(1) detection. On the other hand, the narrow component for H$_{2}$ knot A was found to have a relatively higher $T_{\mathrm{rot},\, v=1}$ (1074 K) and a lower $T_{\mathrm{vib},\,J=3}$ (2523 K) than the other narrow components. An abnormally high $T_{\mathrm{rot},\,v=1}$ (5910 $\pm$ 2301 K), albeit with a large uncertainty, was found for the narrow component of H$_{2}$ NE cavity axis. For the broad components in Figure \ref{fig:h2lpd}(b), all H$_{2}$ sources appear to have $T_{\mathrm{rot},\,v=1}$ = 887--1505 K, which is, in general, higher than $T_{\mathrm{rot},\,v=1}$ for the narrow components. We also note that the only one source (H$_{2}$ 1080E) was found to have a non-zero flux in the broad component of the H$_{2}$ 2-1 S(1) line. Hence, the vibrational temperature of the broad component is available only from H$_{2}$ 1080E; the resulting vibrational temperature of the broad component ($T_{\mathrm{vib},\,J=3}$ = 2241 K) was found to be lower than those of the narrow components observed in other sources (note that this source has no narrow component detected).

\begin{figure*}
\centering
\includegraphics[scale=0.4]{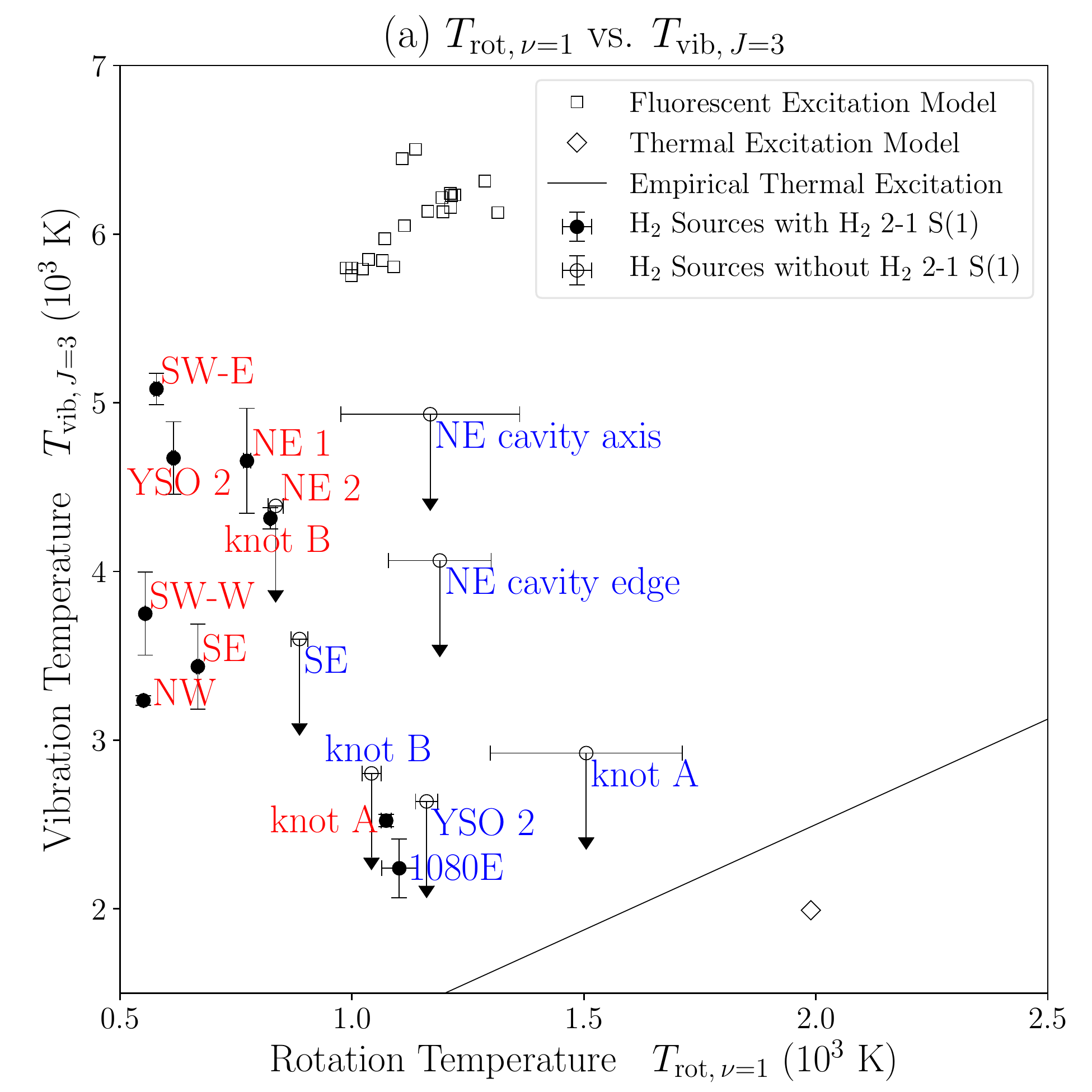}\includegraphics[scale=0.4]{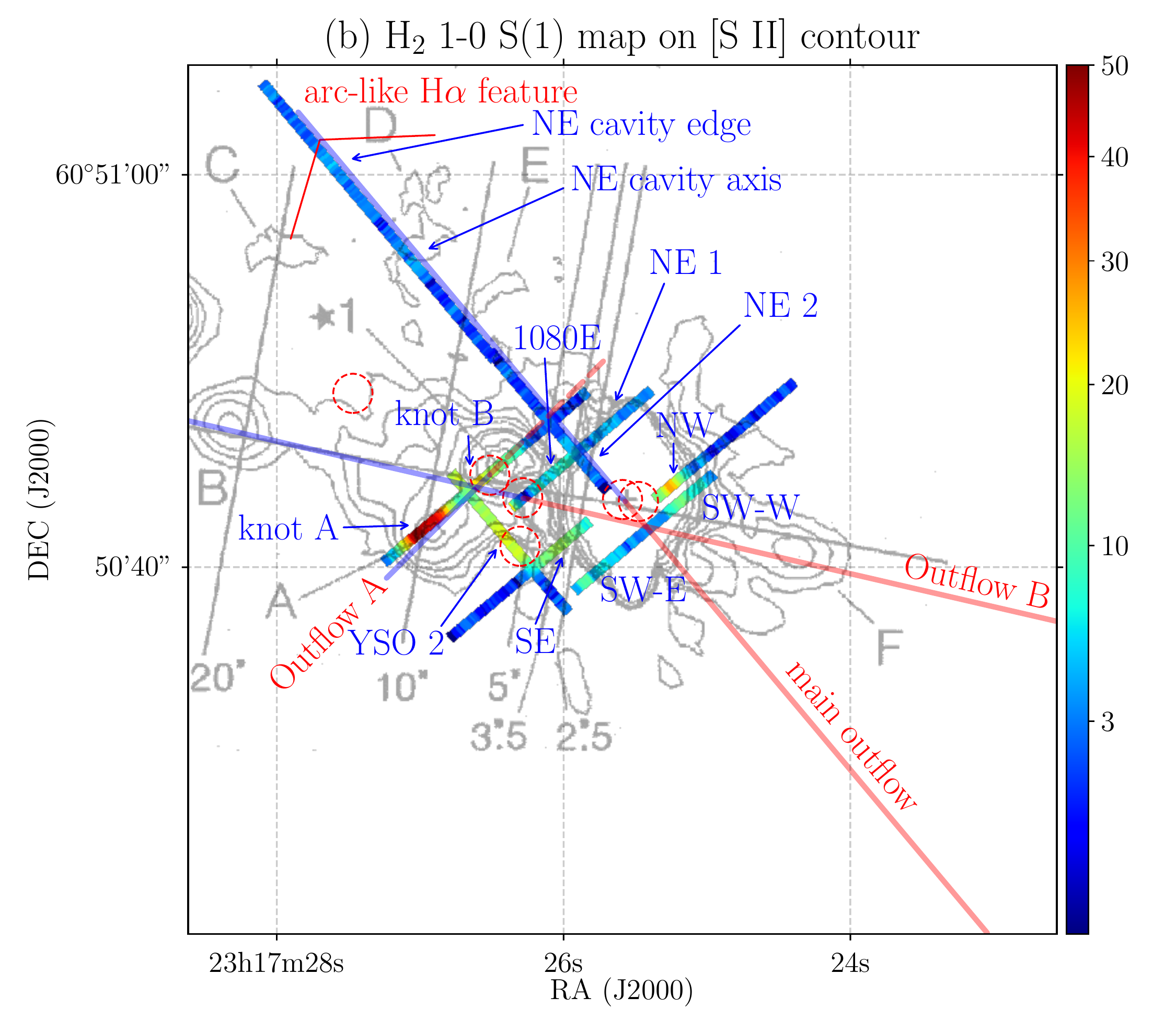}
\caption{(a) Rotation temperature $T_{\mathrm{rot},\,v=1}$ vs. vibration temperature $T_{\mathrm{vib},\,J=3}$ for H$_{2}$ sources, and (b) H$_{2}$ 1-0 S(1) map overlaid on {[}\ion{S}{2}{]} (continuum-subtracted) contours. In panel (a), red source names are for the narrow components and blue ones for the broad components. For sources without the H$_{2}$ 2-1 S(1) detection, the upper limits of the vibration temperatures are applied. The square (fluorescent excitation) and diamonds (thermal excitation) denote the values derived from the line ratios presented in the theoretical models of \citet{1987ApJ...322..412B}. The solid line comes from Figure 1 of \citet{1989ApJ...336..207T} and indicates an empirical relationship ($T_{\mathrm{rot},\,v=1}$ = 0.8 $\times$ $T_{\mathrm{vib},\,J=3}$) for purely the thermal excitation case. In panel (b), the H$_{2}$ 1-0 S(1) map is the same as in Figure \ref{fig:h2map}(a), and the twelve H$_{2}$ 1-0 S(1) sources identified in the PVDs (Figures \ref{fig:h2map}(b) and (c)) are indicated by their assigned names (``H$_{2}$'' was omitted). The {[}\ion{S}{2}{]} contours come from Figure 8(d) of \citet{1992A&A...262..229P}. Red dashed circles (young stars), blue+red lines (axes of the main outflow, ``Outflow A,'' and ``Outflow B''), and a red bent line (arc-like H$\alpha$ feature) are the same as in Figure \ref{fig:basic}.\label{fig:h2ttd}}
\end{figure*}

\citet{1989ApJ...336..207T} discussed that the rotation and vibration temperatures are similar around 2000 K in purely thermal excitation cases. As the contribution of the fluorescent excitation increases, the rotation temperature gets lower and the vibration temperature becomes higher. Their Figure 2 shows several population diagrams for the thermal, fluorescent, and mixed excitation cases, which can be compared with our population diagrams. In Figure \ref{fig:h2lpd}(a), most of the narrow components appear to belong to the mixed or nearly fluorescent cases. Exceptionally, the narrow component of H$_{2}$ knot A seems to be much closer to the thermal case than the other narrow components, although it is still in the mixed case. Similarly, the broad component for H$_{2}$ 1080E in Figure \ref{fig:h2lpd}(b) is also likely to be the mixed case dominated by the thermal excitation. Because the other broad components have no H$_{2}$ 2-1 S(1) detected, their $T_{\mathrm{vib},\,J=3}$ were calculated as the upper limits. The broad components for H$_{2}$ knot A, H$_{2}$ knot B, and H$_{2}$ YSO 2 have the upper limits of $T_{\mathrm{vib},\,J=3}$ that are as low as for the case close to thermal excitation.

In Figure \ref{fig:h2ttd}(a), we compare $T_{\mathrm{rot},\,v=1}$ and $T_{\mathrm{vib},\,J=3}$. We excluded the narrow component of H$_{2}$ NE cavity axis because it has a too high $T_{\mathrm{rot},\,v=1}$ with a large uncertainty (5910 $\pm$ 2301 K). The points with red labels are for the narrow components and those with blue ones are for the broad components. \citet{1987ApJ...322..412B} presented various theoretical models for fluorescent and thermal excitations of the H$_{2}$ lines. Using the H$_{2}$ line intensities in their Table 2, we calculated $T_{\mathrm{rot},\,v=1}$ and $T_{\mathrm{vib},\,J=3}$ predicted from their models. We used only the H$_{2}$ lines that were detected in our observation. In Figure \ref{fig:h2ttd}(a), the open squares located near the upper left corner indicate the results from Models 1--32 of \citet{1987ApJ...322..412B}, which are for the fluorescent excitation cases. On the other hand, the open diamond near the lower right corner denotes the result from their Model S2 for the thermal excitation case. In addition, adopting from Figure 1 of \citet{1989ApJ...336..207T}, an empirical relationship between the rotation and vibration temperatures for purely the thermal excitation case is overlaid by a solid line. Figure \ref{fig:h2ttd}(a) confirms the results of the previous paragraph more clearly. Most of the narrow components, except that of H$_{2}$ knot A, are closer to the fluorescent excitation than to the thermal excitation. On the other hand, the broad components of H$_{2}$ 1080E, H$_{2}$ knot A, H$_{2}$ knot B, and H$_{2}$ YSO 2 appear to be closer to the thermal excitation. The broad components of H$_{2}$ SE, H$_{2}$ NE cavity axis, and H$_{2}$ NE cavity edge have too high upper-limits of $T_{\mathrm{vib},\,J=3}$ to determine a dominant excitation mechanism. However, their $T_{\mathrm{rot},\,v=1}$ (887, 1169, and 1189 K) are higher than $T_{\mathrm{rot},\,v=1}$ (550--835 K) for the narrow components close to the fluorescent excitation, which suggests a possibility for the dominance of the thermal excitations. We also note that the narrow component of H$_{2}$ knot A, unlike the other narrow components, lies close to the thermal excitation despite its narrow velocity width.

\begin{figure*}
\centering
\includegraphics[scale=0.65]{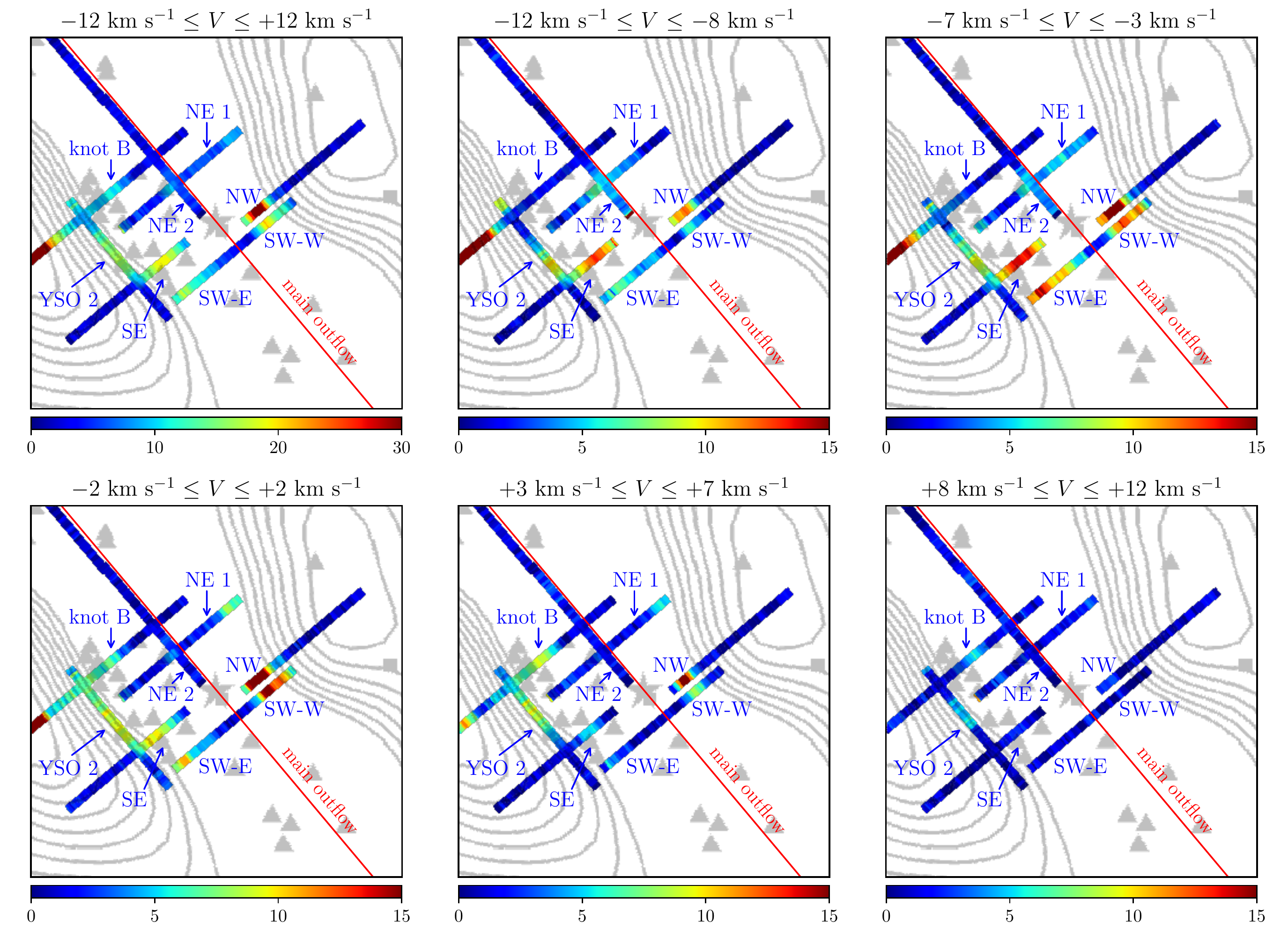}
\caption{H$_{2}$ 1-0 S(1) velocity channel maps ($-$12 km s$^{-1}$ $\leq$ $V$ $\leq$ +12 km s$^{-1}$) around MWC 1080A. The color scales are all in units of the detection significance (signal/$\sigma$). The integrated velocity range is indicated at the top of each channel map. The eight H$_{2}$ sources with fluorescent H$_{2}$ lines are indicated by their assigned names. The overlaid $^{13}$CO contours, triangles (young stars), and red diagonal line (axis of the main outflow cavity) are the same as in Figure \ref{fig:basic}(a).\label{fig:h2ch}}
\end{figure*}

In Figure \ref{fig:h2ttd}(b), the H$_{2}$ 1-0 S(1) map of Figure \ref{fig:h2map}(a) is overlaid on the continuum-subtracted {[}\ion{S}{2}{]} contours for comparison. The {[}\ion{S}{2}{]} contours together with six slit positions for the {[}\ion{S}{2}{]} spectroscopy (solid lines) were adopted from Figure 8(d) of \citet{1992A&A...262..229P}. Six {[}\ion{S}{2}{]} HH objects (``Knot A'' to ``Knot F'') identified by \citet{1992A&A...262..229P} are also shown in the figure. The correlation between the H$_{2}$ 1-0 S(1) and {[}\ion{S}{2}{]} morphologies seems to be rather weak. However, we found that the brightest H$_{2}$ 1-0 S(1) emission (H$_{2}$ knot A) was detected near the {[}\ion{S}{2}{]} peak of ``Knot A.'' This knot is the brightest one among the six {[}\ion{S}{2}{]} HH objects. As already discussed, the H$_{2}$ line ratios in Figure \ref{fig:h2ttd}(a) indicate that both the narrow and broad components of H$_{2}$ knot A are likely thermal H$_{2}$ emission from shock-heated gas. This suggests that ``Knot A'' is probably an outflow feature forming shocked regions. We also note that the broad components detected along the NE axis of the main outflow cavity (H$_{2}$ NE cavity axis and H$_{2}$ NE cavity edge) might be related to ``Knot D'' and ``Knot E'' although the IGRINS slit positions are slightly displaced from the {[}\ion{S}{2}{]} HH objects. In particular, H$_{2}$ NE cavity edge is almost coincident with the arc-like H$\alpha$ feature (indicated by a red bent line in the figure); this is also shown in Figure \ref{fig:h2map}(c). Figure \ref{fig:h2ttd}(a) shows that the broad components of H$_{2}$ NE cavity axis and H$_{2}$ NE cavity edge might be a thermal H$_{2}$ emission. If this is the case, at least some portions of ``Knot D'', ``Knot E'', and the arc-like H$\alpha$ feature could be associated with shock-heated regions.

The IGRINS kinematic results for the narrow components of H$_{2}$ can provide information on the geometry of the main outflow cavity near MWC 1080A. The H$_{2}$ line ratios imply that the narrow components for the eight H$_{2}$ sources around MWC 1080A (H$_{2}$ SW-E, H$_{2}$ SW-W, H$_{2}$ NW, H$_{2}$ SE, H$_{2}$ NE 1, H$_{2}$ knot B, H$_{2}$ YSO 2, and H$_{2}$ NE 2) are the H$_{2}$ lines induced predominantly by the fluorescent excitation, as shown in Figure \ref{fig:h2ttd}(a). Therefore, we suggest that the narrow components of the H$_{2}$ lines originate from the PDRs formed on the internal surfaces of the main outflow cavity. In Figure \ref{fig:h2ch}, we plot the H$_{2}$ 1-0 S(1) velocity channel maps at an interval of 5 km s$^{-1}$ over the velocity range of $-12\le V\le 12$ km s$^{-1}$ where the narrow components are dominant. The first channel map was obtained over the whole velocity range of $-$12 to +12 km s$^{-1}$. In this map, H$_{2}$ 1-0 S(1) appears much brighter along the NW and SE boundaries of the cavity than near the central cavity axis. This property is shown more clearly in the channel map for $-$2 km s$^{-1}$ $\leq$ $V$ $\leq$ +2 km s$^{-1}$ near the systemic LSR velocity. These bright regions are likely to be limb-brightened PDRs on the main outflow cavity walls. Large column densities along the lines of sight to the boundaries make the boundaries brighter than the other sightlines that are close to the cavity axis. We also note that the brightest peak (H$_{2}$ NW) is almost coincident with the apex of the NW $^{13}$CO boundary, which is the cavity boundary closest to MWC 1080A. The sources H$_{2}$ SW-E, H$_{2}$ SE, H$_{2}$ NE 1, and H$_{2}$ NE 2 that are close to the central cavity axis become slightly brighter in the two, negative (blueshift) velocity channel maps for $-$12 km s$^{-1}$ $\leq$ $V$ $\leq$ $-$3 km s$^{-1}$. In particular, the brightest locations of H$_{2}$ SW-E, H$_{2}$ SE, and H$_{2}$ NE 1 appear to move closer to the cavity axis in the higher-blueshift velocity maps. These variations appear more clearly in the H$_{2}$ 1-0 S(1) PVDs (Figure \ref{fig:h2map}(b)), where H$_{2}$ SW-E and H$_{2}$ NE 1 show notable radial velocity gradients along the X direction (higher blueshift towards X = 0$\arcsec$). From these results, we conclude that the narrow components of H$_{2}$ SW-E, H$_{2}$ SE, H$_{2}$ NE 1, and H$_{2}$ NE 2 trace the PDRs on the front surface of the main outflow cavity wall, which is expanding outward with a velocity of $\sim$10--15 km s$^{-1}$. In the positive (redshift) velocity channel maps for +3 km s$^{-1}$ $\leq$ $V$ $\leq$ +12 km s$^{-1}$, H$_{2}$ knot B appears slightly brighter than in the velocity channel map for $-$2 km s$^{-1}$ $\leq$ $V$ $\leq$ +2 km s$^{-1}$. As can be seen in Figure \ref{fig:h2map}(b), the narrow component of H$_{2}$ knot B also shows a radial velocity gradient along the X direction in the H$_{2}$ 1-0 S(1) PVD (higher redshift towards X = 0$\arcsec$). This tendency suggests that the region traced by H$_{2}$ knot B may be a PDR formed on the part of the rear surface of the cavity wall. It is also noticeable that H$_{2}$ knot B, tracing the rear part of the cavity wall, is much fainter than those tracing the front part.

\subsection{{[}\ion{Fe}{2}{]} $\lambda$1.644 $\micron$ Line} \label{subsec:fe}

\begin{figure*}
\centering
\includegraphics[scale=0.4]{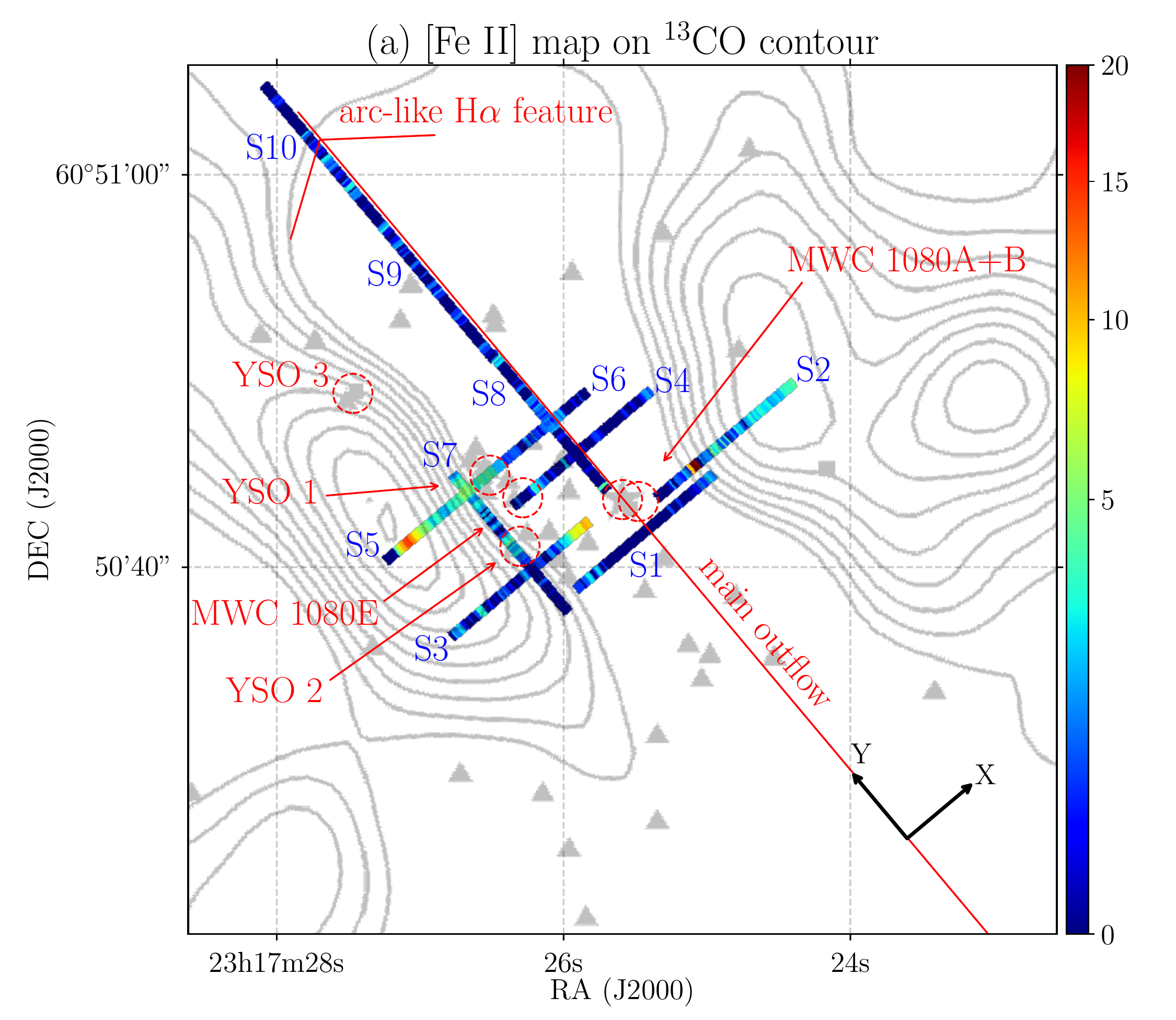}
\includegraphics[scale=0.39]{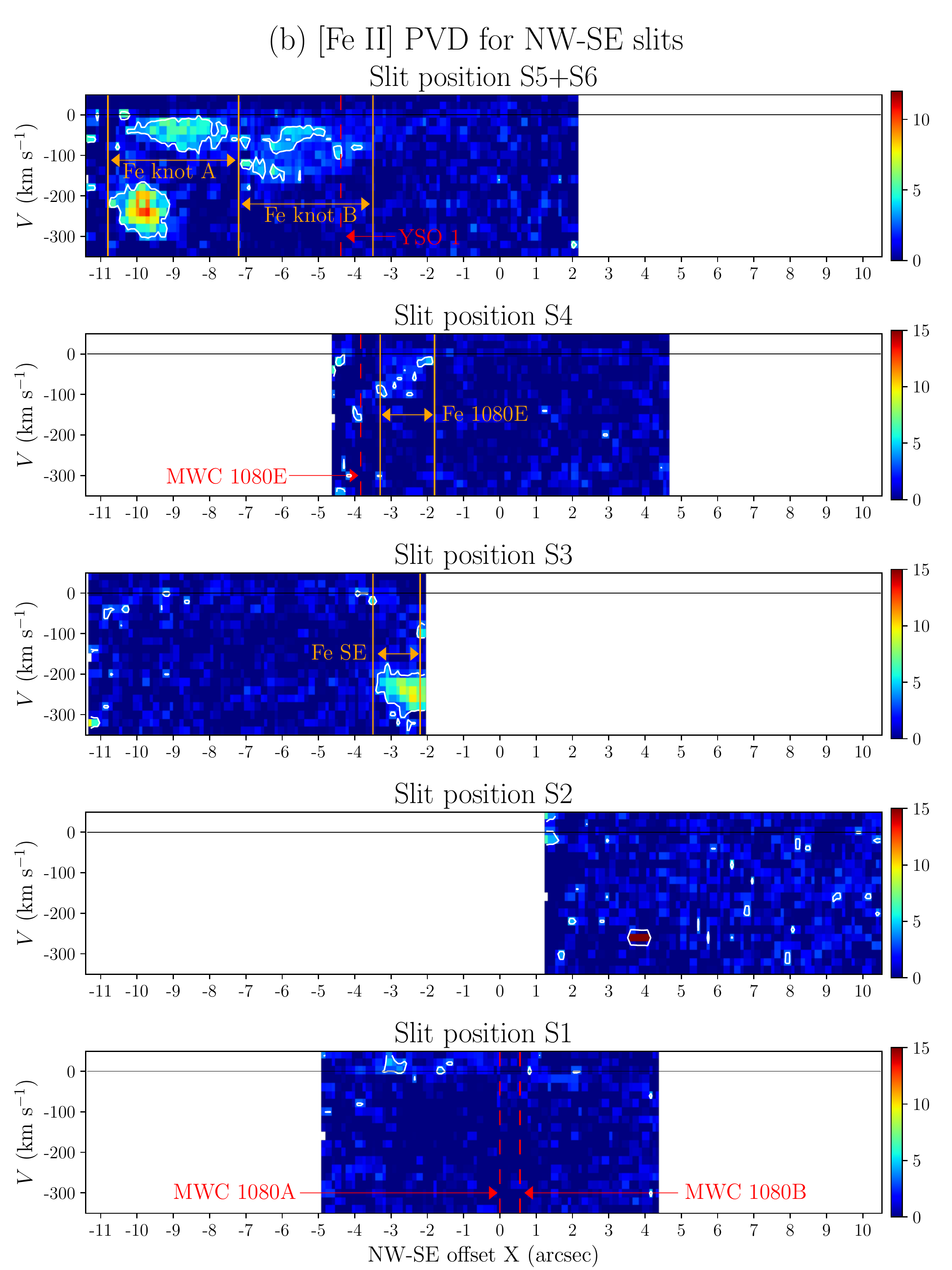}\includegraphics[scale=0.39]{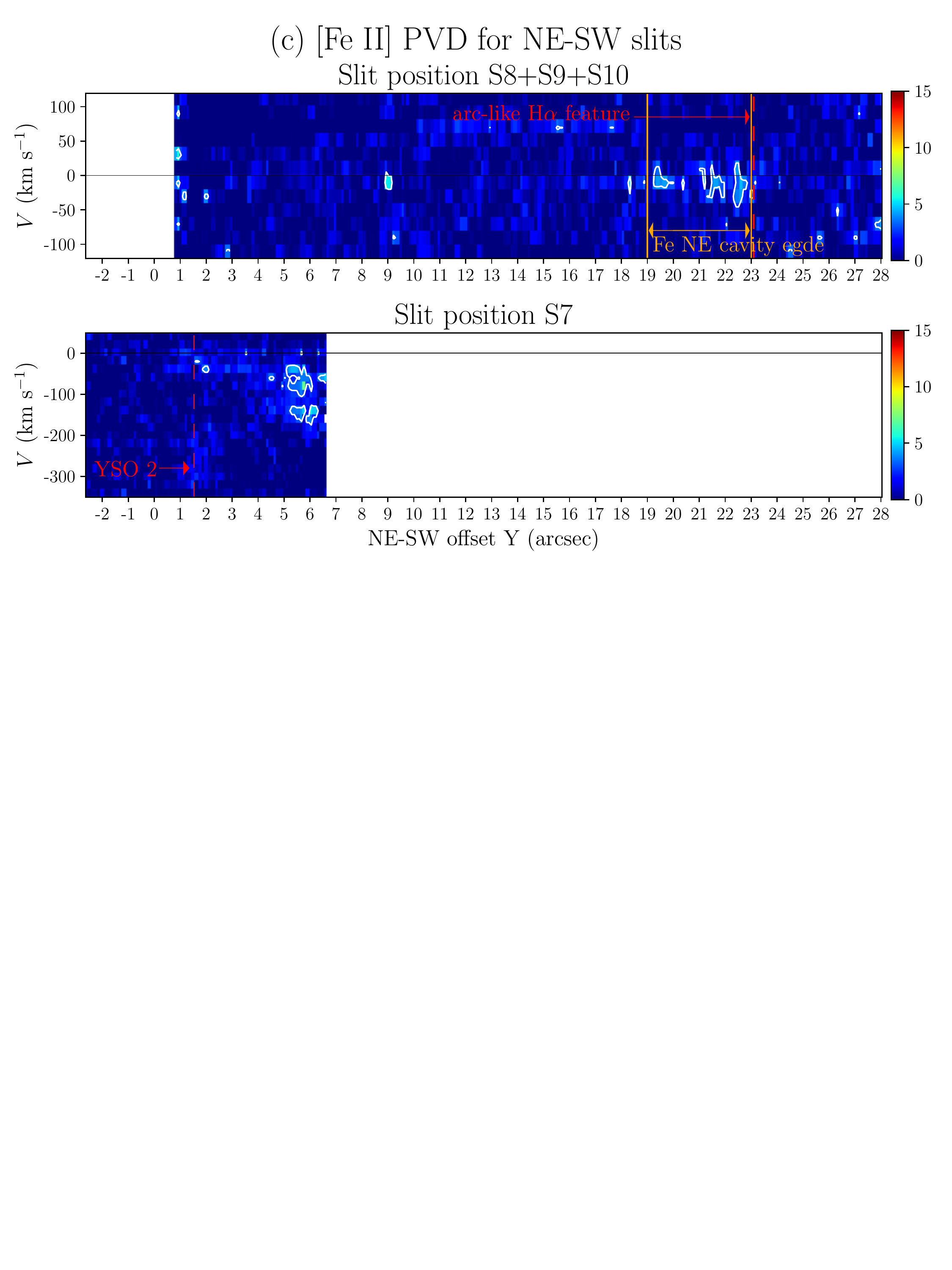}
\caption{Continuum-subtracted {[}\ion{Fe}{2}{]} map integrated over $-$300 km s$^{-1}$ $\leq$ $V$ $\leq$ +50 km s$^{-1}$, and PVDs for (b) the NW-SE slits S1--S6 and (c) NE-SW slits S7--S10. The color scales are all in units of the detection significance (signal/$\sigma$). In panel (a), the overlaid $^{13}$CO contours, triangles, red dashed circles (young stars), red diagonal line (axis of the main outflow cavity), and red bent line (arc-like H$\alpha$ feature) are the same as in Figure \ref{fig:basic}(a). In panels (b) and (c), the origin of the NW-SE offset X and NE-SW offset Y is the position of MWC 1080A; their positive directions are indicated in (a). The contours on the PVDs represent a significance level of 3$\sigma$. The five {[}\ion{Fe}{2}{]} sources identified in this study are also denoted, together with the vertical solid lines that represent the spatial extent, in the PVDs. The vertical dashed lines indicate the X or Y positions of the young stars and the arc-like H$\alpha$ feature denoted in (a).\label{fig:femap}}
\end{figure*}

\begin{figure*}
\centering
\includegraphics[scale=0.4]{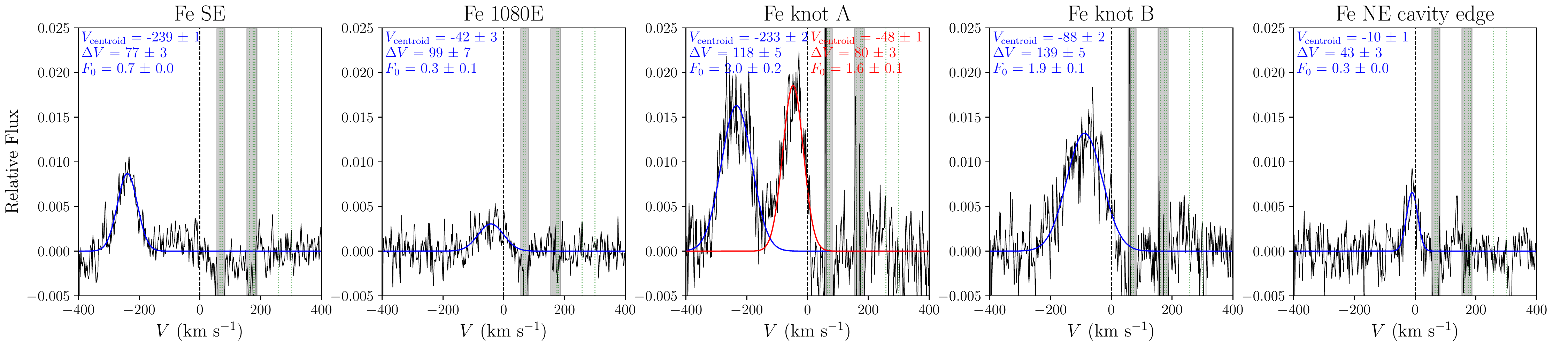}
\caption{{[}\ion{Fe}{2}{]} line profiles for the diffuse sources, integrated over the source region that is bounded by the vertical solid lines in Figures \ref{fig:femap}(b) and (c). Each line profile was fitted with a Gaussian function plus a linear continuum (For Fe knot A, we used two Gaussian functions), and the resulting linear continuum was subtracted from the line profile. The best-fit Gaussian functions are overplotted with the blue or red curves as well. The centroid velocity $V_{\mathrm{centroid}}$, FWHM velocity width $\Delta V$, and line flux $F_{0}$ obtained from the fitting are shown in each panel and also listed in Table \ref{table:felp}. The green dotted lines indicate the positions of the telluric OH lines \citep{2000A&A...354.1134R}. There are several telluric residuals in the positive $V$ domain, but the blueshifted part of the {[}\ion{Fe}{2}{]} lines was not severely affected by them. The two shaded areas denote the $V$ domains containing strong telluric residuals, which were masked out when the line profiles were measured. The instrumental profiles ($\Delta V$ $\simeq$ 6.5 km s$^{-1}$ estimated from telluric OH lines) have not been removed from the line profiles, which gives only a difference of $<$1 km s$^{-1}$ for $\Delta V$ $>$ 20 km s$^{-1}$.\label{fig:felp}}
\end{figure*}

We also detected the {[}\ion{Fe}{2}{]} $\lambda$1.644 $\micron$ forbidden line from the diffuse sources around MWC 1080 with IGRINS. This is the only Fe line detected in our observations. Figure \ref{fig:femap}(a) shows a continuum-subtracted {[}\ion{Fe}{2}{]} map integrated over $-$300 km s$^{-1}$ $\leq$ $V$ $\leq$ +50 km s$^{-1}$. The {[}\ion{Fe}{2}{]} PVDs for slit positions S1--S10 are also presented in Figures \ref{fig:femap}(b) and (c). In the PVDs, we identified a total of five {[}\ion{Fe}{2}{]} sources. Their identification names and spatial boundaries are also shown in the figure. The features seen in S7 are a part of Fe knot B observed in S5+S6. The sharp peak found at X $\simeq$ 3.$\arcsec$8 in S2 is likely due to a cosmic ray event that was not removed by the basic data reduction. In S5+S6, the {[}\ion{Fe}{2}{]} line profile of Fe knot A appears to consist of two separate velocity components. One is the high-blueshift component that is found in a velocity range of $-$300 to $-$200 km s$^{-1}$, and the other is the low-blueshift component in a velocity range of $-$100 to 0 km s$^{-1}$. The low-blueshift velocity component extends more toward YSO 1 than the high-blueshift one. In the same silt position, Fe knot B was also detected in a velocity range of $-200\le V \le 0$ km s$^{-1}$, at a location close to YSO 1. In S4, Fe 1080E was detected with $V$ $\simeq$ $-$100 to 0 km s$^{-1}$, although it is somewhat faint. Fe SE found in S3 has a high-blueshift component in $-300\le V \le -200$ km s$^{-1}$, which is similar to the high-blueshift component of Fe knot A. In S8+S9+S10, weak, sparse {[}\ion{Fe}{2}{]} features were detected near the NE edge of the main outflow cavity; those are altogether referred to as Fe NE cavity edge. This {[}\ion{Fe}{2}{]} source shows a low-blueshift component in $-50\le V\le 0$ km s$^{-1}$.

Figure \ref{fig:felp} shows the {[}\ion{Fe}{2}{]} line profiles for each source region. To measure the centroid velocity $V_{\mathrm{centroid}}$, FWHM velocity width $\Delta V$, and line flux $F_{0}$ for the five {[}\ion{Fe}{2}{]} sources, we fitted them with the same method as for Br$\gamma$ and H$_{2}$ 1-0 S(1); the best-fit profiles are shown in Figure \ref{fig:felp}. Two velocity intervals, containing strong telluric OH residuals, were masked out before the line profiles were measured. Only the velocity components detected with a detection significance $F_{0}/\sigma_{F_{0}}$ $>$3 are presented in the figure and also listed in Table \ref{table:felp}. All of the {[}\ion{Fe}{2}{]} sources appear to have a negative (blueshift) $V_{\mathrm{centroid}}$ ($<$$-$10 km s$^{-1}$) and broad $\Delta V$ ($>$43 km s$^{-1}$). In particular, Fe knot A was fitted to have two velocity components of $V_{\mathrm{centroid}}$ = $-$233 and $-$48 km s$^{-1}$.

\subsubsection{Shock-induced {[}\ion{Fe}{2}{]} Sources} \label{subsec:fes}

\begin{figure*}
\centering
\includegraphics[scale=0.4]{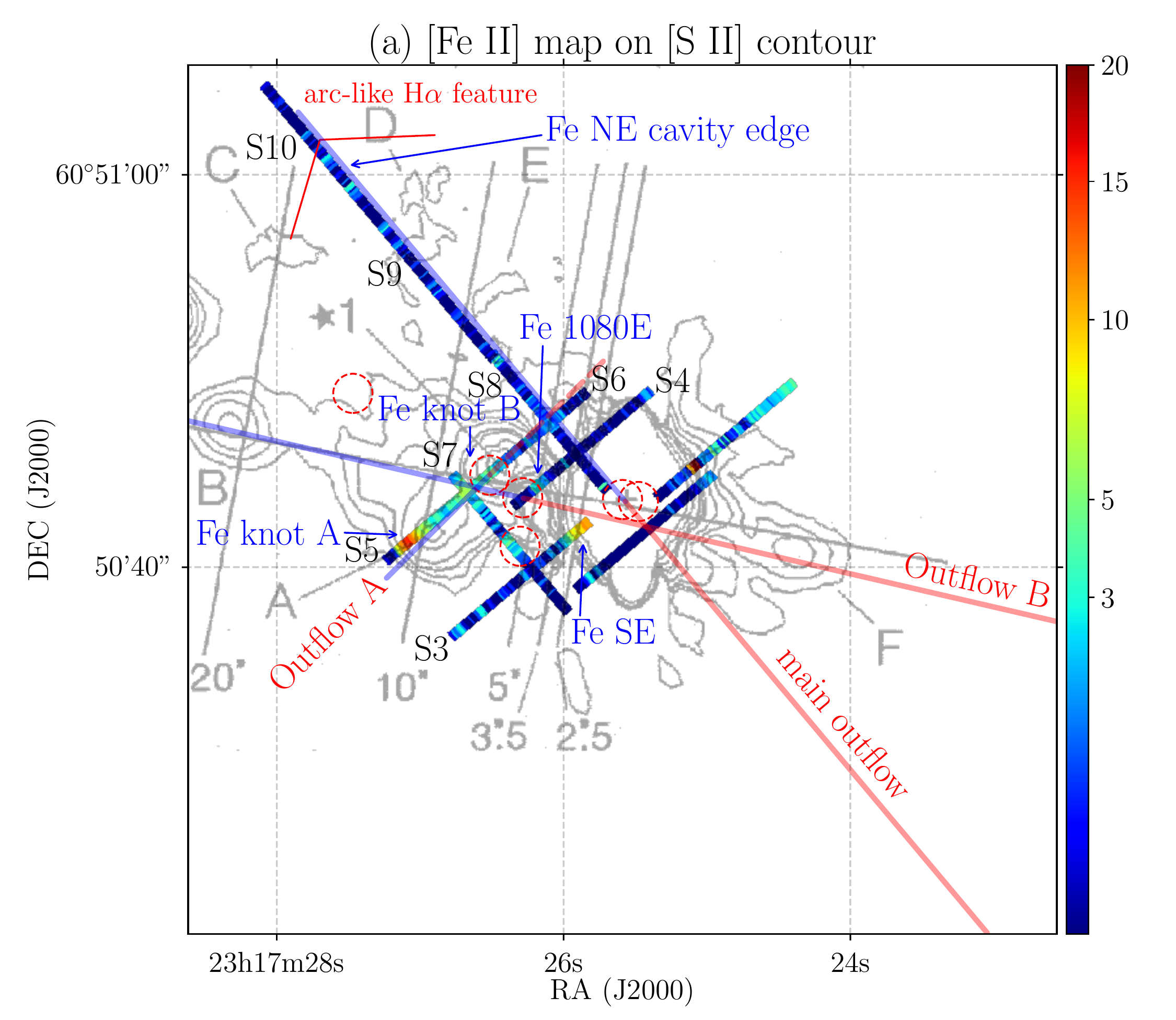}
\includegraphics[scale=0.39]{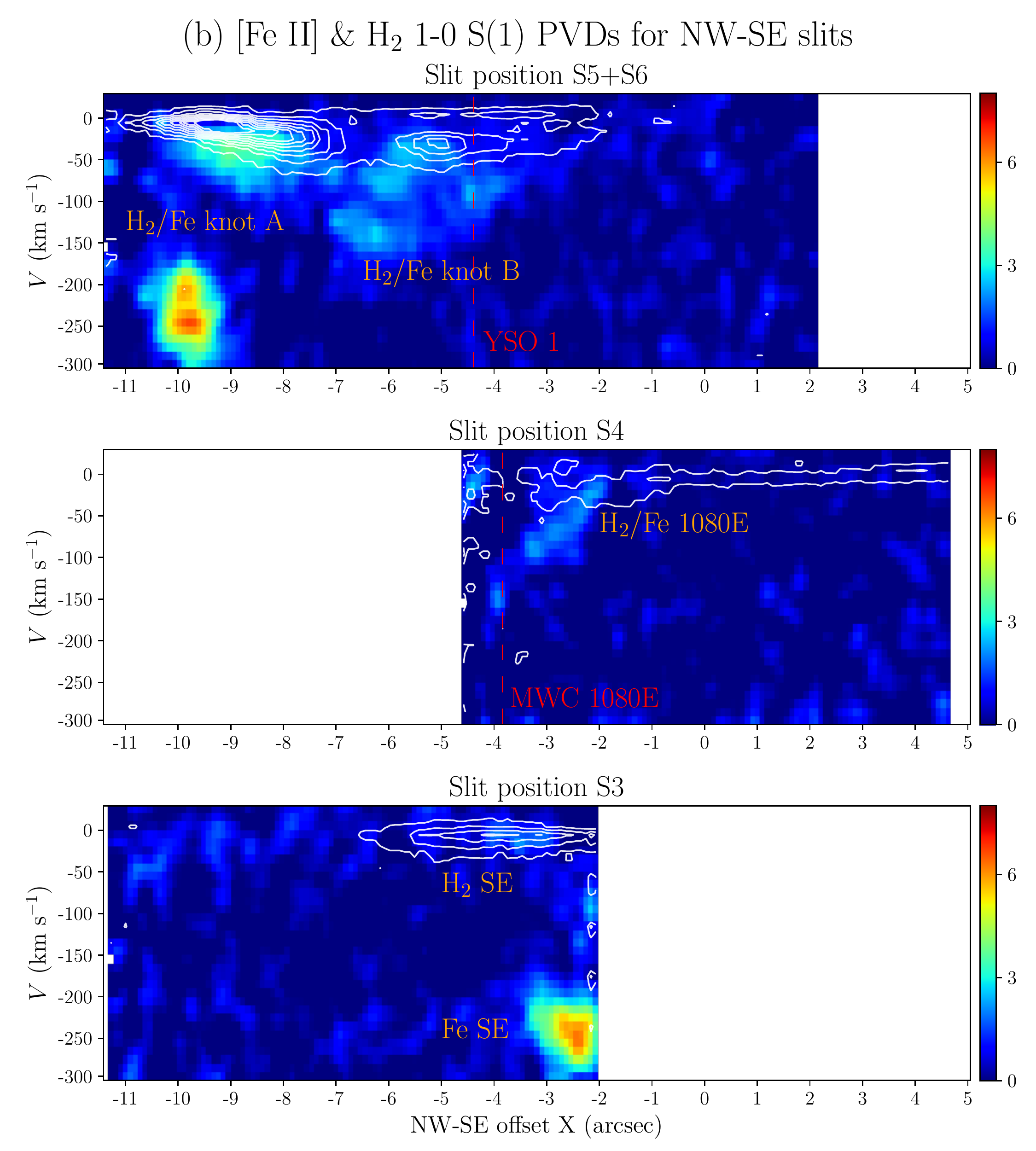}\includegraphics[scale=0.39]{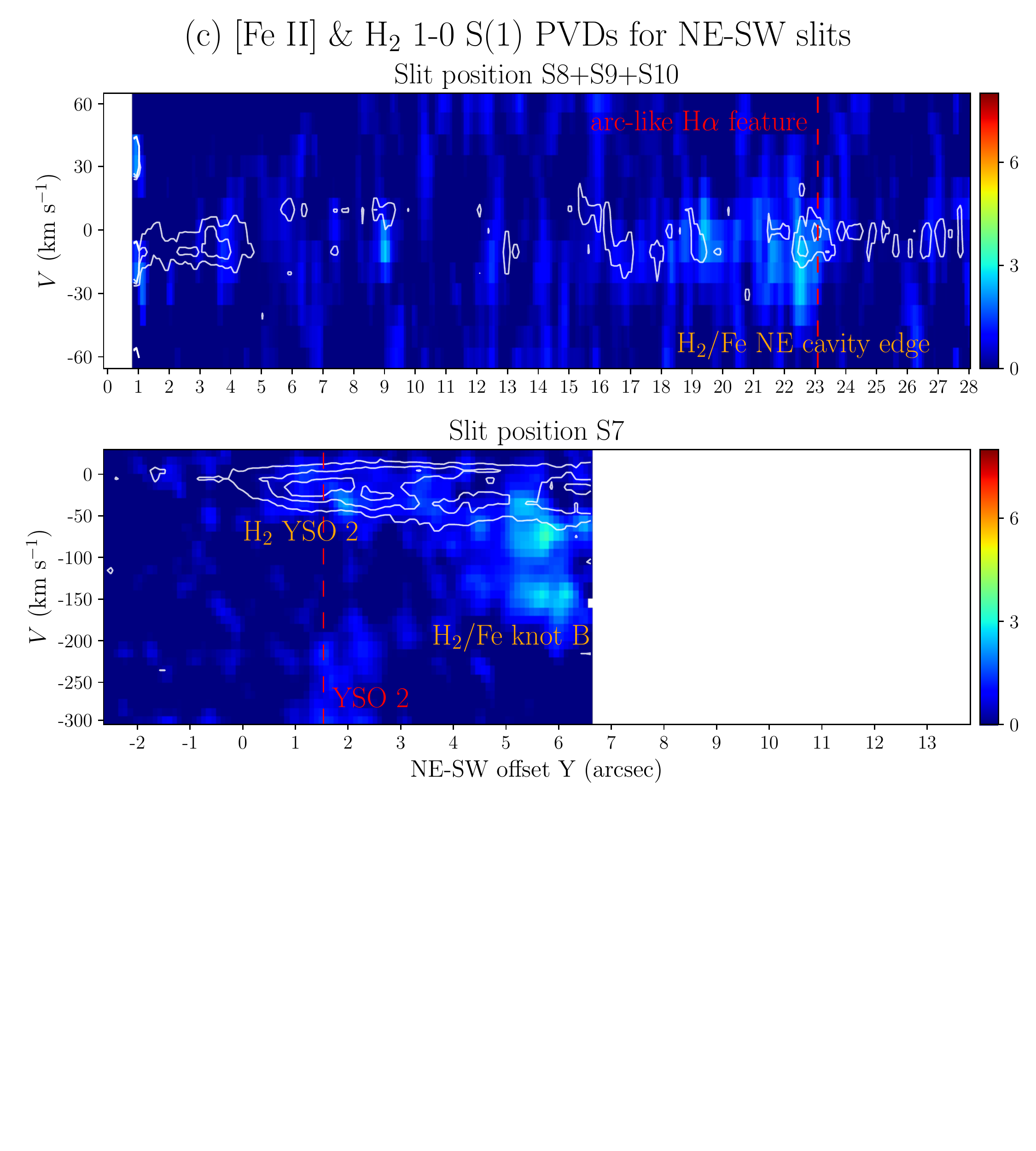}
\caption{(a) {[}\ion{Fe}{2}{]} map overlaid on {[}\ion{S}{2}{]} (continuum-subtracted) contours, and {[}\ion{Fe}{2}{]} (color image) and H$_{2}$ 1-0 S(1) (contours) PVDs for (b) the NW-SE slits S3--S6 and (c) NE-SW slits S7--S10. In panel (a), the {[}\ion{Fe}{2}{]} map is the same as in Figure \ref{fig:femap}(a), and the five {[}\ion{Fe}{2}{]} sources identified in the PVDs (Figures \ref{fig:femap}(b) and (c)) are indicated by their assigned names. The {[}\ion{S}{2}{]} contours come from Figure 8(d) of \citet{1992A&A...262..229P}. Red dashed circles (young stars), blue+red lines (axes of the main outflow, ``Outflow A,'' and ``Outflow B''), and a red bent line (arc-like H$\alpha$ feature) are the same as in Figure \ref{fig:basic}. In panels (b) and (c), the H$_{2}$ 1-0 S(1) contours come from Figures \ref{fig:h2map}(b) and (c), and range from 3$\sigma$ to 43$\sigma$ in a 5$\sigma$ equal interval (except S8+S9+S10, showing 3$\sigma$, 5$\sigma$, and 7$\sigma$ levels). The {[}\ion{Fe}{2}{]} PVDs are from Figures \ref{fig:femap}(b) and (c) but were smoothed using a Gaussian function with $\sigma$ = 1.0 pixel. The H$_{2}$ and {[}\ion{Fe}{2}{]} sources are indicated by their names. The vertical dashed lines indicate the X or Y positions of the young stars and the arc-like H$\alpha$ feature denoted in (a).\label{fig:feh2pvd}}
\end{figure*}

In the {[}\ion{Fe}{2}{]} PVDs of Section \ref{subsec:fe}, five {[}\ion{Fe}{2}{]} sources were identified (see Figure \ref{fig:femap}). As shown in Figure \ref{fig:felp} and Table \ref{table:felp}, all {[}\ion{Fe}{2}{]} lines, except one source (Fe NE cavity edge), are found to be blueshifted by $V_{\mathrm{centroid}}<-42$ km s$^{-1}$ and have a broad velocity width $\Delta V>77$ km s$^{-1}$; Fe NE cavity edge has the lowest blueshift ($V_{\mathrm{centroid}} = -10$ km s$^{-1}$) and the narrowest width ($\Delta V= 43$ km s$^{-1}$). These line profiles indicate that all the {[}\ion{Fe}{2}{]} sources likely originate from the shock-heated regions.

In Figure \ref{fig:feh2pvd}(a), the {[}\ion{Fe}{2}{]} map of Figure \ref{fig:femap}(a) is overlaid on the {[}\ion{S}{2}{]} (continuum-subtracted) contours for comparison. The {[}\ion{S}{2}{]} contours are the same as in Figure \ref{fig:h2ttd}(b). The brightest {[}\ion{Fe}{2}{]} emission (Fe knot A) was detected near the {[}\ion{S}{2}{]} peak of ``Knot A,'' just like the source H$_{2}$ knot A was. The detection of the shock-induced {[}\ion{Fe}{2}{]} and H$_{2}$ emission at the location of ``Knot A'' strongly supports that ``Knot A'' is an outflow forming shock-heated region. The {[}\ion{Fe}{2}{]} $\lambda$1.644 $\micron$ emission was also detected throughout the region between the {[}\ion{S}{2}{]} peak and YSO 1, which overlaps with ``Knot A.'' The source Fe NE cavity edge, together with H$_{2}$ NE cavity edge, was detected near the arc-like H$\alpha$ feature (indicated by a red bent line in the figure). Although Fe NE cavity edge has a somewhat small blueshift $V_{\mathrm{centroid}}=-$10 km s$^{-1}$ and a relatively narrow velocity width $\Delta V = 43$ km s$^{-1}$, it also could be shock-induced; this will be discussed in Section \ref{subsec:mainoutflow}.

To compare the spatial and kinematic distributions between the {[}\ion{Fe}{2}{]} and H$_{2}$ sources in detail, we overplotted the H$_{2}$ 1-0 S(1) PVDs (contours) on the {[}\ion{Fe}{2}{]} PVDs (color image) for slit positions S3--S10 in Figures \ref{fig:feh2pvd}(b) and (c). In the figures, the associations between the {[}\ion{Fe}{2}{]} and H$_{2}$ sources are found at several positions. In $-$10.$\arcsec$0 $\la$ X $\la$ $-$7.$\arcsec$5 of slit position S5+S6, the low-blueshift component of Fe knot A ($V_{\mathrm{centroid}}$ = $-$48 km s$^{-1}$) seems to be associated with both the narrow and broad components of H$_{2}$ knot A. In the velocity space, the peak position of H$_{2}$ knot A appears to align along with the lowest blueshift boundary ($V$ $\simeq$ $-$25 to $-$10 km s$^{-1}$) of the low-blueshift component of Fe knot A, which strongly supports the physical association between them. On the other hand, the high-blueshift component of Fe knot A ($V_{\mathrm{centroid}}$ = $-$233 km s$^{-1}$) has no counterpart in the H$_{2}$ emission. In $-$6.$\arcsec$0 $\la$ X $\la$ $-$4.$\arcsec$5, the lowest blueshift boundary ($V$ $\simeq$ $-$30 km s$^{-1}$) of Fe knot B shows an association with the broad component of H$_{2}$ knot B, although the H$_{2}$ source extends much more towards the NW direction (up to X $\simeq$ $-$2.$\arcsec$0) than Fe knot B. In slit position S4, Fe 1080E appears to overlap with H$_{2}$ 1080E although the correlation between them is not clear. In slit position S3, Fe SE and H$_{2}$ SE are observed at similar locations, but they show neither clear spatial nor kinematic relationships. In slit position S8+S9+S10, Fe NE cavity edge and H$_{2}$ NE cavity edge are spatially coincident and are located just inside the arc-like H$\alpha$ feature, as indicated by a vertical dashed line in the figure; this suggests that Fe NE cavity edge and H$_{2}$ NE cavity edge could be associated with the arc-like H$\alpha$ feature. In addition, in the velocity space, the H$_{2}$ peak seems to lie near the lowest blueshift boundary ($V$ $\simeq$ $-$5 km s$^{-1}$) of Fe NE cavity edge; the property is the same as that found for Fe knot A/H$_{2}$ knot A and Fe knot B/H$_{2}$ knot B in S5+S6. We also note that the {[}\ion{Fe}{2}{]} and H$_{2}$ features seen in 5.$\arcsec$0 $\la$ Y $\la$ 6.$\arcsec$5 of slit position S7 are identical to Fe knot B and H$_{2}$ knot B, respectively, in slit position S5+S6. In 0.$\arcsec$0 $\la$ Y $\la$ 5.$\arcsec$0 of slit position S7, the broad component of H$_{2}$ YSO 2 was detected with no corresponding {[}\ion{Fe}{2}{]} source. A portion of it seems to be connected to the SW part of H$_{2}$ knot B. However, its peak location (at $V$ $\simeq$ $-$20 km s$^{-1}$) is much closer to YSO 2, suggesting its association with YSO 2, rather than with H$_{2}$ knot B.

\section{Discussion} \label{sec:discussion}

\subsection{Stellar Outflows} \label{subsec:outflow}

\subsubsection{Main Outflow from MWC 1080A} \label{subsec:mainoutflow}

It has been observed in the outflows of young stars that the H$_{2}$ emission is found in relatively low-velocity ranges while the {[}\ion{Fe}{2}{]} emission is found in higher velocity ranges \citep[e.g.,][]{2002ApJ...570..724P,2003A&A...397..693D,2006ApJ...641..357T,2016ApJ...817..148O}. It has been interpreted that the {[}\ion{Fe}{2}{]} and H$_{2}$ emissions, respectively, originate from high-velocity collimated outflows and low-velocity entrained ambient molecular gas. This suggests that H$_{2}$/Fe NE cavity edge shown in Figure \ref{fig:feh2pvd}(c) is caused by shock-excitation due to a stellar outflow, although the H$_{2}$ and {[}\ion{Fe}{2}{]} emissions were detected very weakly. More specifically, we suggest that the NE main outflow from MWC 1080A would be the excitation source of H$_{2}$/Fe NE cavity edge, based on the spatial coincidence between H$_{2}$/Fe NE cavity edge, the arc-like H$\alpha$ feature, and the NE $^{13}$CO boundary (see Figures \ref{fig:basic} and \ref{fig:feh2pvd}(c)).

The centroid velocity $V_{\mathrm{centroid}}$ of the {[}\ion{Fe}{2}{]} line in H$_{2}$/Fe NE cavity edge was observed to be relatively low ($-$10 km s$^{-1}$ in Figure \ref{fig:felp}) compared to those observed from the outflows of other young stars. This difference can be explained if the inclination angle of the NE main outflow from MWC 1080A is close to 0$\arcdeg$ with respect to the sky plane. \citet{2008ApJ...673..315W} showed that the overall velocity distributions of $^{13}$CO and CS are mostly blueshifted by $-$5 to 0 km s$^{-1}$, and most of them are located around the NE cavity region, as can be seen in Figure \ref{fig:basic}(a). Therefore, their velocity distributions agree well with our kinematic estimation for the NE main outflow of MWC 1080A. Additionally, \citet{2009A&A...497..117A} obtained an inclination angle of $\sim$80$\arcdeg$ for the circumstellar disk and large toroidal envelope around MWC 1080A by SED fitting. The nearly edge-on orientation of the circumstellar disk is compatible with our low inclination angle inferred for the main outflow of MWC 1080A. \citet{2003A&A...397..693D} reported a YSO, HH 379 IRS, with the $V_{\mathrm{centroid}}$ near the systemic LSR velocity for both H$_{2}$ and {[}\ion{Fe}{2}{]}, and presented the same explanation as for MWC 1080A. They also argued that {[}\ion{Fe}{2}{]} still traces the high-velocity collimated outflow based on the result that $\Delta V$ for {[}\ion{Fe}{2}{]} is about twice as broad as $\Delta V$ for H$_{2}$. Our velocity widths derived from H$_{2}$/Fe NE cavity edge ($\Delta V$ = 24 km s$^{-1}$ for H$_{2}$ 1-0 S(1) and $\Delta V$ = 43 km s$^{-1}$ for {[}\ion{Fe}{2}{]}) show the same trend as that for HH 379 IRS.

The broad component of H$_{2}$ NE cavity axis may also be caused by the NE main outflow, although no {[}\ion{Fe}{2}{]} counterpart was detected and the H$_{2}$ line ratios could not obviously confirm the thermal origin of the lines. The kinematic results of the broad component ($V_{\mathrm{centroid}}$ = $-$12 km s$^{-1}$ and $\Delta V$ = 22 km s$^{-1}$ for H$_{2}$ 1-0 S(1)) support the possibility that it might be related to shock-heated regions by the NE main outflow. As we suggested in Section \ref{subsec:h2s}, the broad component can be associated with two {[}\ion{S}{2}{]} HH objects, ``Knot D'' and ``Knot E'', located near the NE axis of the main outflow cavity.

In fact, the narrow component of H$_{2}$ was also detected along the NE axis of the main outflow cavity. In Figure \ref{fig:h2lp}, H$_{2}$ NE cavity axis has both the narrow and broad components. Although the H$_{2}$ line ratios obtained in this study could not reveal the dominant excitation mechanism for the narrow component, its kinematic results of the redshifted $V_{\mathrm{centroid}}$ (+9 km s$^{-1}$) and narrow $\Delta V$ (14 km s$^{-1}$) imply that the narrow component unlikely belongs to the thermal H$_{2}$ lines related to the NE main outflow. Rather, it may come from PDRs formed on the rear surfaces of the NE cavity wall. The exciting photons emitted from MWC 1080A can reach the distant surfaces effectively through the low-density medium inside the NE cavity.

In Section \ref{subsec:iphas}, we reported a few faint H$\alpha$ features located far away in the SW direction from MWC 1080A. We suggest that the H$\alpha$ features around (RA, DEC) $\sim$ (23$^{\mathrm{h}}$17$^{\mathrm{m}}$21.6$^{\mathrm{s}}$, +60$\arcdeg$50$\arcmin$08$\arcsec$) in Figure \ref{fig:basic}(b) (denoted by a dashed circle) could be related to the SW counter outflow corresponding to the NE main outflow. As indicated by a blue+red line with PA = 40$\arcdeg$ in the figure, the SW faint H$\alpha$ features lie near the SW extension of the diagonal line passing through MWC 1080A and the apex of the NE arc-like H$\alpha$ feature. They are away from MWC 1080A by about twice the distance between MWC 1080A and the NE arc-like H$\alpha$ feature. As shown in Figure \ref{fig:basic}(a), the $^{13}$CO boundary is opened toward the SW direction from MWC 1080A, which means that the mediums toward the SW direction have a relatively low density. This can make the SW main outflow move much farther from MWC 1080A and be seen much more faintly. Because the NE main outflow was found to be a blueshifted outflow, the SW main outflow is expected to be a redshifted outflow, although it also likely has a low inclination angle with respect to the sky plane. The blueshifted and redshifted parts of the main outflow are denoted by the blue and red colors on the line with PA = 40$\arcdeg$ in Figure \ref{fig:basic}(b).

\subsubsection{New Outflows A and B} \label{subsec:newoutflow}

\citet{1987ApJ...316..323H} presented bow shock models of HH outflows for a wide range of shock velocities (20--400 km s$^{-1}$) and various inclination angles. One of their results is that double-peaked line profiles (high velocity close to shock velocity and low velocity close to 0 km s$^{-1}$) observed from the HH outflows are predicted from bow shocks with high ($>$150 km s$^{-1}$) shock velocities and large ($>$45$\arcdeg$) inclination angles with respect to the sky plane. This can explain the high- and low-blueshift components of Fe knot A shown in Figure \ref{fig:feh2pvd}(b). Adopting the model, the high-blueshift {[}\ion{Fe}{2}{]} component ($V_{\mathrm{centroid}}$ = $-$233 km s$^{-1}$) would originate mostly from the apex of a bow shock, whereas the low-blueshift {[}\ion{Fe}{2}{]} component ($V_{\mathrm{centroid}}$ = $-$48 km s$^{-1}$) from a portion of the bow wings. We note that the seeing (FWHM $\sim$0.$\arcsec$6) during the observations was comparable to the IGRINS slit width (0.$\arcsec$63), meaning that they are both narrower than the whole size of the bow shock. Therefore, a substantial portion of the emission from the bow wings could be missed, leading to a weakening of the low-blueshift {[}\ion{Fe}{2}{]} component compared to the high-blueshift one. In fact, the high-blueshift {[}\ion{Fe}{2}{]} component has two velocity peaks of $V$ $\simeq$ $-$250 and $-$200 km s$^{-1}$, as can be seen in Figure \ref{fig:feh2pvd}(b). This can be due to a slight displacement of slit position S5 from the exact location of the bow apex. The bow shock model of \citet{1987ApJ...316..323H} also shows that the shock velocity is similar to the full width at zero intensity (FWZI) of the line profiles observed from an HH outflow independent of its inclination angles. From the PVD in Figure \ref{fig:feh2pvd}(b) and the line profile fitting in Figure \ref{fig:felp}, the FWZI of Fe knot A is estimated to be about 300--400 km s$^{-1}$. Therefore, a shock velocity of 300--400 km s$^{-1}$ and an inclination angle of $>$45$\arcdeg$ are expected for the outflow related to Fe knot A. We refer to this outflow as ``Outflow A.''

In Figure \ref{fig:feh2pvd}(b), H$_{2}$/Fe knot A shows a velocity structure in which the H$_{2}$ peak is coincident with the lowest blueshift boundary of the low-blueshift {[}\ion{Fe}{2}{]} component. This is the velocity structure that is expected from the ambient H$_{2}$ gas entrained by the {[}\ion{Fe}{2}{]} outflow shock \citep[e.g.,][]{2001MNRAS.326..524D,2002ApJ...570..724P,2003A&A...397..693D,2016ApJ...817..148O}. Therefore, H$_{2}$ knot A (both the narrow and broad components), associated with the low-blueshift component of Fe knot A, is likely molecular gas entrained by the shocks propagating in the bow wing regions of ``Outflow A.'' The estimated shock velocity of ``Outflow A'' (about 300--400 km s$^{-1}$) highly exceeds the H$_{2}$ dissociation speed (generally 35--50 km s$^{-1}$), which would prevent the H$_{2}$ emission associated with the high-blueshift component of Fe knot A. On the other hand, in the bow wing regions, the normal component of the shock velocity with respect to the bow shock front can be below the H$_{2}$ dissociation speed.

We associate ``Outflow A'' with the brightest {[}\ion{S}{2}{]} HH object, ``Knot A.'' As can be seen in Figures \ref{fig:h2ttd}(b) and \ref{fig:feh2pvd}(a), slit position S5 is aligned well with the SE extension of ``Knot A,'' and the brightest locations of the detected {[}\ion{Fe}{2}{]} and H$_{2}$ emissions are almost coincident with the {[}\ion{S}{2}{]} peak of ``Knot A,'' considering the seeing size of the {[}\ion{S}{2}{]} map \citep[at most 2.$\arcsec$0,][]{1992A&A...262..229P}. \citet{1992A&A...262..229P} also presented the {[}\ion{S}{2}{]} spectroscopic result for a long slit passing through ``Knot A'' (the 10$\arcsec$ line in Figures \ref{fig:h2ttd}(b) and \ref{fig:feh2pvd}(a)). They detected two separate velocity components ($-$215 and $-$85 km s$^{-1}$) at the {[}\ion{S}{2}{]} peak of ``Knot A'' (see their Figure 12(a)). Considering their observed radial velocities with respect to the systemic velocity of the MWC 1080 system \citep[about $-$39 km s$^{-1}$ heliocentric,][]{1992A&A...262..229P}, the {[}\ion{S}{2}{]} components (converted to $-$176 and $-$46 km s$^{-1}$) agree well with the high-blueshift ($V_{\mathrm{centroid}}$ = $-$233 km s$^{-1}$) and low-blueshift ($V_{\mathrm{centroid}}$ = $-$48 km s$^{-1}$) {[}\ion{Fe}{2}{]} components of Fe knot A, respectively. The difference in velocities between the high-blueshift components can be due to a slight mismatch between their slit position and our slit position S5. Based on the above spatial and kinematic coincidence, we conclude that ``Knot A'' represents ``Outflow A,'' a highly-blueshifted outflow (with a shock velocity of 300--400 km s$^{-1}$ and an inclination angle of $>$45$\arcdeg$), and H$_{2}$/Fe knot A is a shock-induced line emission by this outflow. However, it is difficult to identify the driving young star for ``Outflow A'' in this study. Nevertheless, based on the morphology of ``Knot A'' extending towards the SE direction in the {[}\ion{S}{2}{]} map, we propose that MWC 1080E and YSO 1 are the most probable candidates. As presented in Section \ref{subsec:brs}, the detection of the Br$\gamma$ emission from them in the present work supports the existence of their active accretion and winds. In Figures \ref{fig:basic}(b), \ref{fig:h2ttd}(b), and \ref{fig:feh2pvd}(a), we denoted the blueshifted and redshifted parts of ``Outflow A'' in the blue and red lines with PA = 135$\arcdeg$, assuming that YSO 1 is the driving source of ``Outflow A.'' In this study, we could not find any evidence for the redshifted part of ``Outflow A,'' but its plausible position was indicated by a red dashed line.

As for H$_{2}$/Fe knot A, a strong kinematic correlation between the H$_{2}$ and {[}\ion{Fe}{2}{]} emissions from H$_{2}$/Fe knot B shown in Figure \ref{fig:feh2pvd}(b) suggests that they are also likely related to an outflow shock and ambient H$_{2}$ gas entrained by the shock. However, the driving source of H$_{2}$/Fe knot B cannot be YSO 1 because their blueshifted components extend toward both the SE and NW directions from YSO 1, as can be seen in Figure \ref{fig:feh2pvd}(b). One possibility is that they might be related to another {[}\ion{S}{2}{]} HH object, ``Knot B.'' The faint H$\alpha$ features around (RA, DEC) $\sim$ (23$^{\mathrm{h}}$17$^{\mathrm{m}}$18.5$^{\mathrm{s}}$, +60$\arcdeg$50$\arcmin$30$\arcsec$) in Figure \ref{fig:basic}(b), as denoted by a dashed circle, are possibly related to the counter outflow corresponding to ``Knot B.'' This is based on the identical bipolar outflowing direction between them, as indicated by a blue+red line with PA = 77$\arcdeg$ in the figure. We suggest that ``Knot B'' and the faint H$\alpha$ features could be related to another bipolar outflow from a young star around MWC 1080A. We refer to this bipolar outflow as ``Outflow B.'' Like the main outflow from MWC 1080A, ``Outflow B'' also has a similar morphological characteristic: bright and close in the NE direction, and faint and distant in the SW direction.

As can be seen in Figures \ref{fig:h2ttd}(b) and \ref{fig:feh2pvd}(a), H$_{2}$/Fe knot B is located near the line of PA = 77$\arcdeg$ representing the axis of ``Outflow B.'' \citet{1992A&A...262..229P} detected a $-$141 km s$^{-1}$ (with respect to the systemic velocity) {[}\ion{S}{2}{]} component at $\sim$3$\arcsec$ north from the {[}\ion{S}{2}{]} peak of ``Knot A'' in their slit position of 10$\arcsec$ (see their Figure 12(a)). Its position lies almost on the line of PA = 77$\arcdeg$ (the intersection of the 10$\arcsec$ line and the red dashed line for ``Outflow B'' in Figure \ref{fig:feh2pvd}(a)). The {[}\ion{S}{2}{]} component has a similar velocity to $V$ $\simeq$ $-$150 km s$^{-1}$ for the {[}\ion{Fe}{2}{]} component seen at $-$7.$\arcsec$0 $\la$ X $\la$ $-$6.$\arcsec$0 in slit position S5+S6 of Figure \ref{fig:feh2pvd}(b). Moreover, \citet{1992A&A...262..229P} detected a $-$156 km s$^{-1}$ {[}\ion{S}{2}{]} component for ``Knot B'' itself in their slit position of 20$\arcsec$ (see their Figure 12(b)). The fact that ``Knot B,'' Fe knot B, and an intermediate component between them all have similar blueshifts supports their associations. Therefore, we propose that H$_{2}$/Fe knot B might be related to ``Knot B'' corresponding to the blueshifted part of ``Outflow B.'' Considering active young stars located near the axis of ``Outflow B,'' MWC 1080E and MWC 1080B can be possible candidates for the driving young star for ``Outflow B.'' In Figures \ref{fig:basic}(b), \ref{fig:h2ttd}(b), and \ref{fig:feh2pvd}(a), the blueshifted and redshifted parts of ``Outflow B'' are denoted in the blue and red lines with PA = 77$\arcdeg$ assuming that MWC 1080E is the driving source of ``Outflow B.''

Fe SE and Fe 1080E were also detected near the axis of ``Outflow B,'' as can be seen in Figure \ref{fig:feh2pvd}(a). \citet{1992A&A...262..229P} detected a $-$241 km s$^{-1}$ (with respect to the systemic velocity) {[}\ion{S}{2}{]} component in their slit positions passing through Fe SE (see their Figure 11). Its velocity is almost the same as $V_{\mathrm{centroid}}$ for Fe SE ($-$239 km s$^{-1}$). If Fe SE and/or Fe 1080E are also truly related to the blueshifted part of ``Outflow B,'' MWC 1080E cannot be the driving source for ``Outflow B'' because they are located in the opposite direction from MWC 1080E with respect to ``Knot B.'' Fe SE, Fe 1080E, and ``Knot B'' all have negative (blueshift) velocities. In fact, there are many driving source candidates around MWC 1080A, and their outflowing axes can overlap with each other on the projected sky plane. Therefore, it is difficult to find the actual origins of Fe SE and Fe 1080E in this study. The similar velocity ranges ($V$ $\simeq$ $-$300 to $-$200 km s$^{-1}$) between the high-blueshift component of Fe knot A and the source Fe SE, imply another possibility of their associations as well. Further observations are needed to reveal their origins definitely.

In the Appendix, we present the H$_{2}$ 1-0 S(1) and {[}\ion{Fe}{2}{]} velocity channel maps using the data obtained from twelve, additional slit observations. Despite bad weather conditions during these observations, we were able to obtain a few meaningful results. In Figure \ref{fig:add}(a), the bright H$_{2}$ 1-0 S(1) regions appear roughly along the CS molecular boundary. As the radial velocity varies from $-$50 km s$^{-1}$ to $-$1 km s$^{-1}$, the bright H$_{2}$ regions move into the deeper inside of the CS molecular region, which agrees with the H$_{2}$ emission observed within slit position S5. This shows a possibility that ``Outflow A'' can be related to the CS molecular boundary distorted toward the extending direction of ``Knot A'' (see the yellow line with PA = 135$\arcdeg$ in the figure). However, we note that the $^{13}$CO molecular boundary near ``Knot A'' shows no distorted shape, as can be seen in Figure \ref{fig:h2map}(a). It might be because the CS map traces molecular gas with a higher density than the $^{13}$CO map \citep{2008ApJ...673..315W}. In Figure \ref{fig:add}(b), an {[}\ion{Fe}{2}{]} region was detected additionally at $-$180 km s$^{-1}$ $<$ $V$ $<$ $-$121 km s$^{-1}$, which is located near the axis of ``Outflow B'' (see the blue line with PA = 77$\arcdeg$ in the figure). This {[}\ion{Fe}{2}{]} emission may be the counterpart for the $-$141 km s$^{-1}$ {[}\ion{S}{2}{]} component detected by \citet{1992A&A...262..229P} at the same location, and it supports the possible association between H$_{2}$/Fe knot B and ``Knot B.''

\subsection{Origins of Atomic Hydrogen Line Emission}

\subsubsection{Extended Emission}

\begin{figure*}
\centering
\includegraphics[scale=0.39]{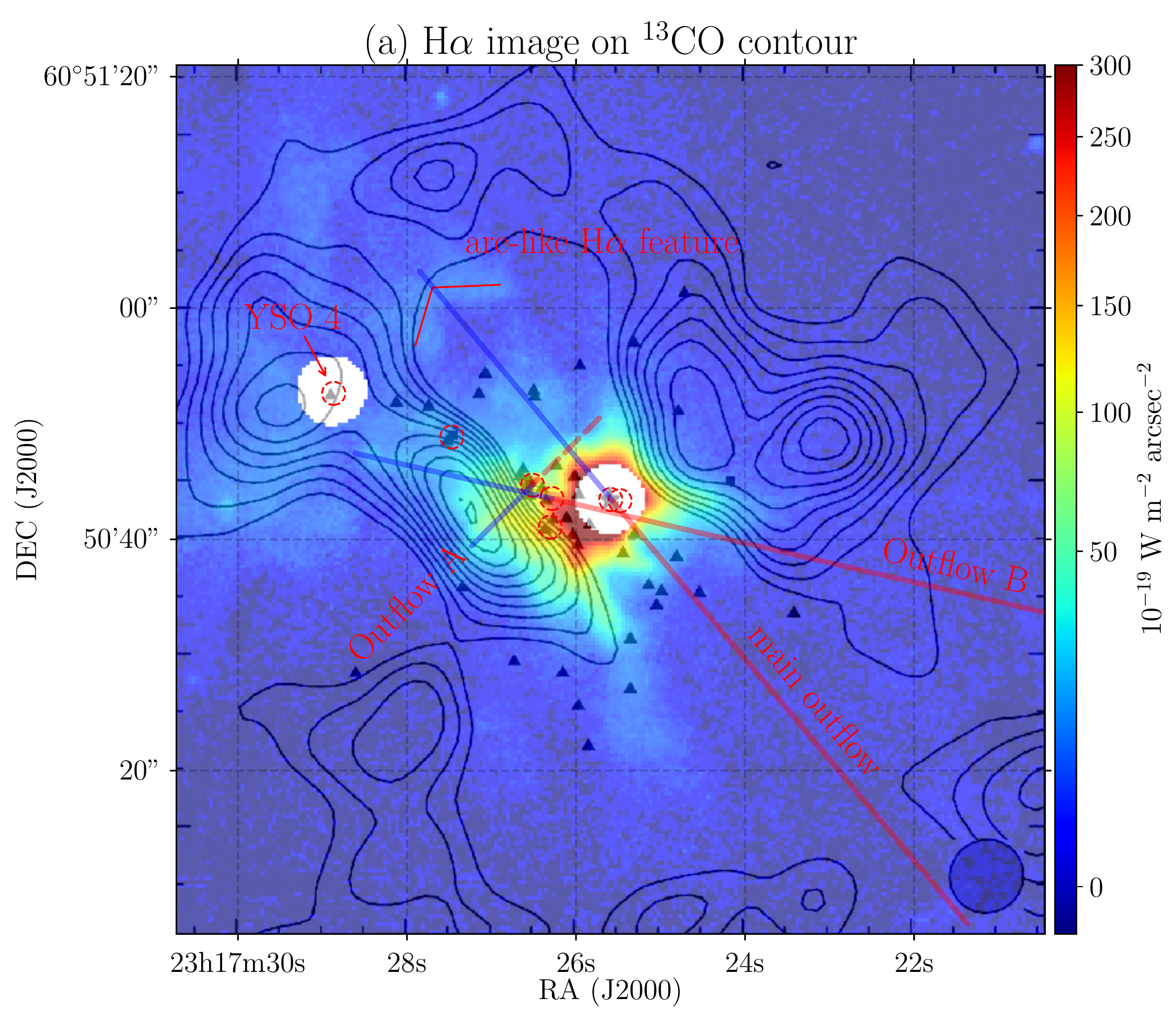}\includegraphics[scale=0.39]{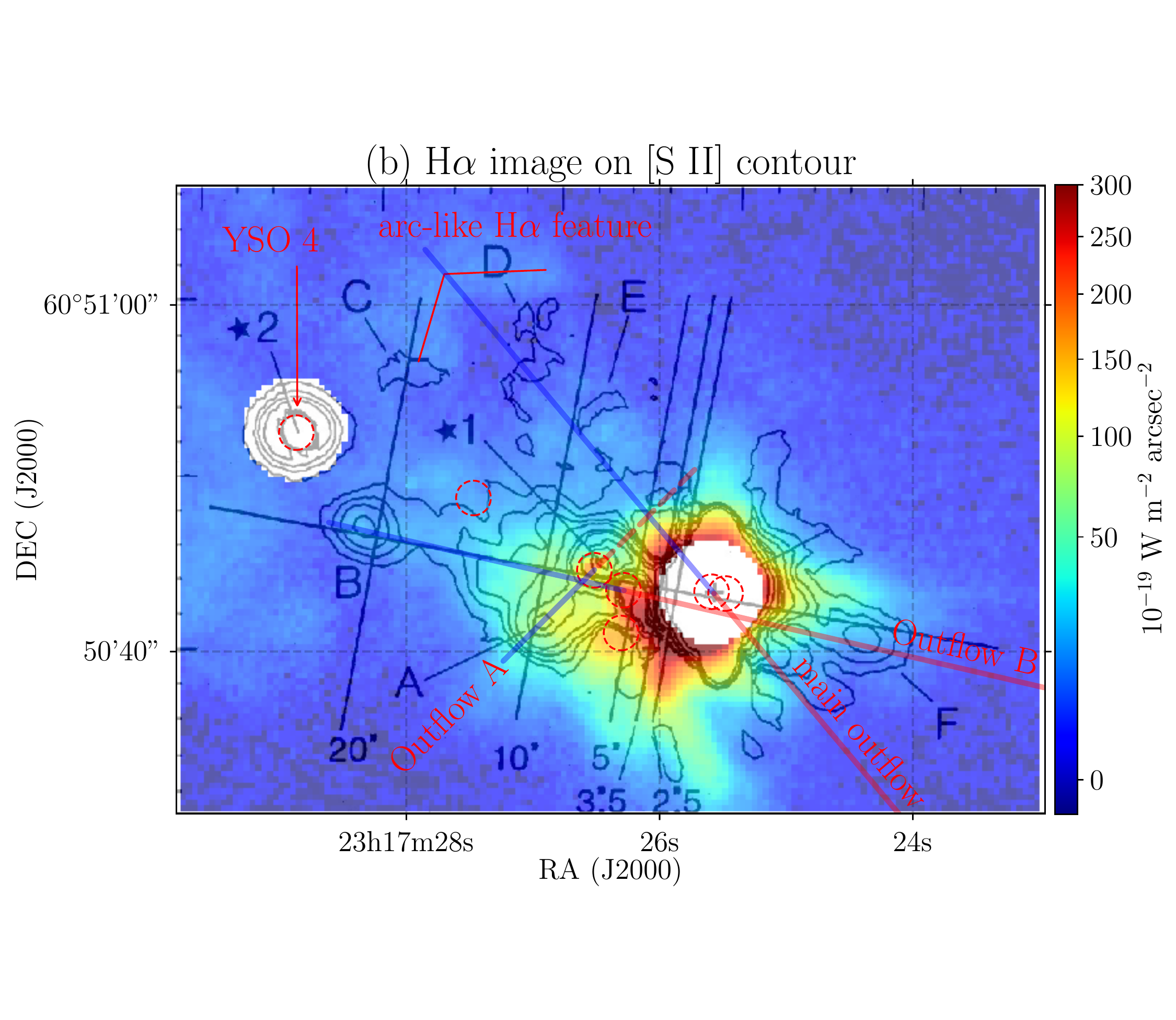}
\includegraphics[scale=0.39]{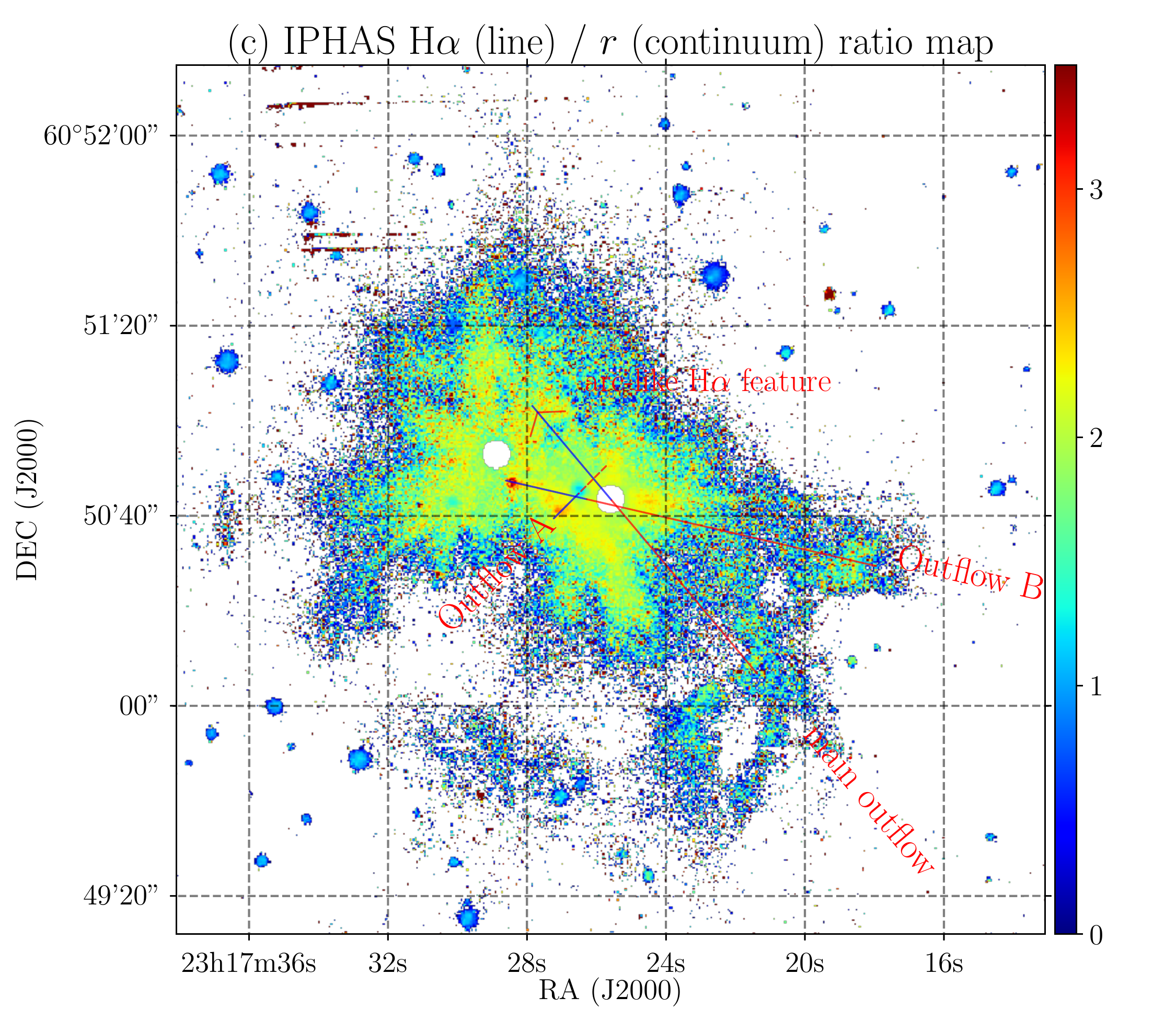}\includegraphics[scale=0.39]{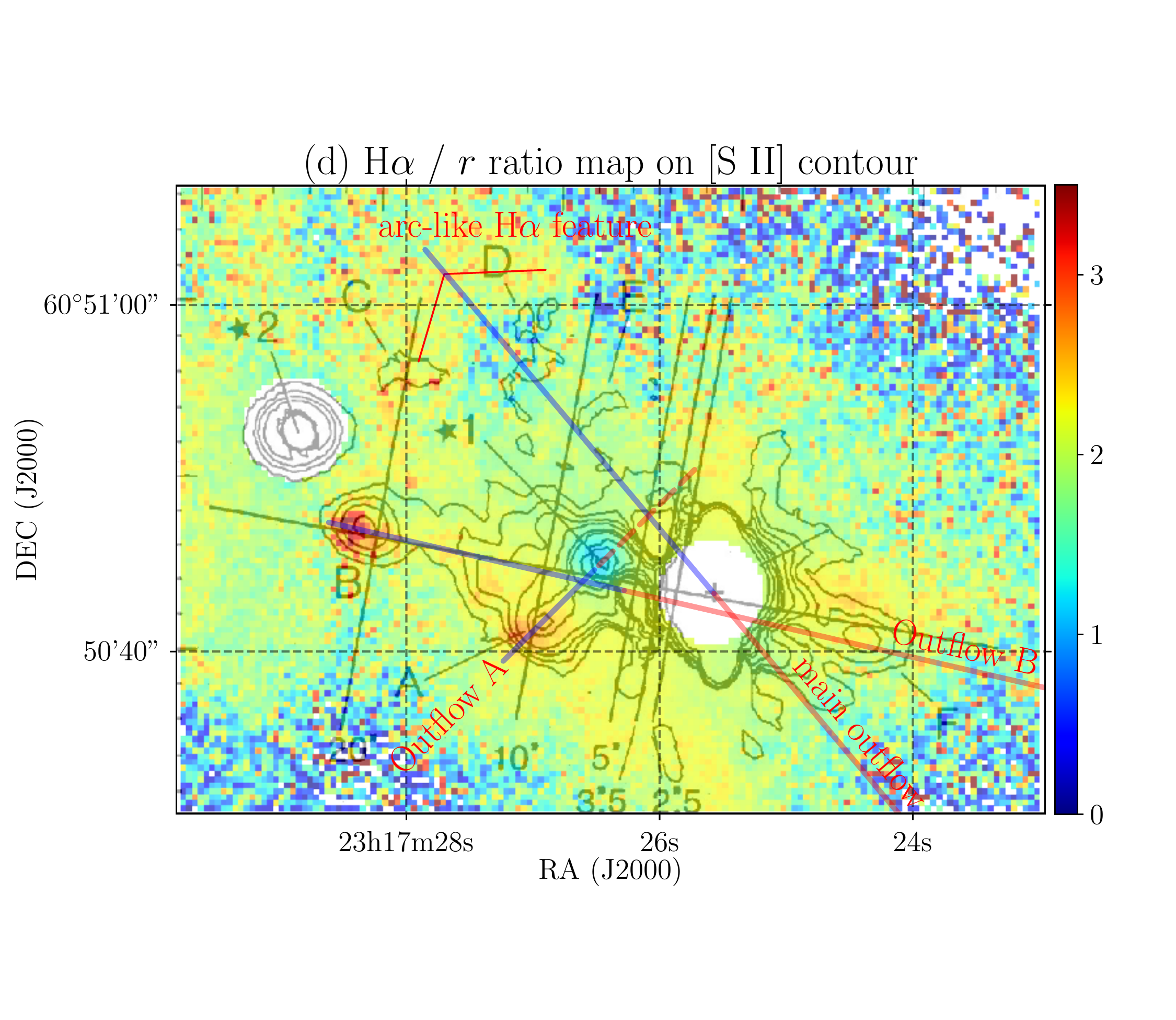}
\caption{IPHAS H$\alpha$ images overlaid on (a) $^{13}$CO and (b) {[}\ion{S}{2}{]} contours. The H$\alpha$ image comes from Figure \ref{fig:basic}(b), but its color scale has been changed slightly. The $^{13}$CO contours were adapted from Figure 2 of \citet{2008ApJ...673..315W} and are the same as in Figure \ref{fig:basic}(a). The {[}\ion{S}{2}{]} contours were adapted from Figure 8(d) of \citet{1992A&A...262..229P} and are the same as in Figures \ref{fig:h2ttd}(b) and \ref{fig:feh2pvd}(a). Triangles, red dashed circles (young stars), blue+red lines (axes of the main outflow, ``Outflow A,'' and ``Outflow B''), and a red bent line (arc-like H$\alpha$ feature) are the same as in Figure \ref{fig:basic}. Panel (c) shows an H$\alpha$-to-continuum ratio map for the whole MWC 1080 region, where the three outflows and the arc-like H$\alpha$ feature are indicated. The ratio map was made from the IPHAS H$\alpha$- and $r$-filter images. A close-up view of the central region of panel (c) is overlaid on the {[}\ion{S}{2}{]} contours in panel (d).\label{fig:hacosii}}
\end{figure*}

From the analysis of the kinematics and line ratio for the hydrogen Brackett emission lines, a large portion of the extended H$\alpha$ emission around MWC 1080A shown in Figure \ref{fig:basic}(b) was confirmed to trace dusty nebulae that scatter the stellar H$\alpha$ emission from MWC 1080A. In Figure \ref{fig:hacosii}(a), the H$\alpha$ image of Figure \ref{fig:basic}(b) is overlaid on the $^{13}$CO contours of Figure \ref{fig:basic}(a). In general, the H$\alpha$ emission around MWC 1080A correlates well with the main outflow cavity apparent in the $^{13}$CO map. The H$\alpha$ emission is bright along the NW and SE boundaries of the cavity near MWC 1080A. This is very similar to the 11.2 $\micron$ morphology shown in Figure 5 of \citet{2014ApJ...796...74L}. Based on the mid-IR (11.2, 11.6, and 18.5 $\micron$) morphology correlated well with the cavity structure, they discussed that the extended mid-IR emission is emitted by warm dust located close to the main outflow cavity walls. Similarly, the H$\alpha$ emission brightening along the cavity boundaries is likely due to the emission scattered by dust on the surfaces of the cavity.

The bright Br$\gamma$ emission detected in the region close to MWC 1080A, masked out because of saturation in the IPHAS H$\alpha$ image, can trace dusty materials existing within the main outflow cavity. In particular, Br SW and Br NE 1 detected near the cavity axis were found to show two separate Br$\gamma$ peaks, for which the association with the bipolar main outflow from MWC 1080A was proposed in Section \ref{subsec:brs}. Br SW is one of the brightest Br$\gamma$ sources detected in our IGRINS observation and found to have a mid-IR counterpart. \citet{2014ApJ...796...74L} identified three prominent mid-IR features in the vicinity of MWC 1080A, one of which (designated as ``south finger'' in their Figure 2) is spatially coincident with Br SW although their 11.2 $\micron$ image showed no two separate peaks.

In Figure \ref{fig:hacosii}(b), the H$\alpha$ image of Figure \ref{fig:basic}(b) is overlaid on the {[}\ion{S}{2}{]} contours of Figures \ref{fig:h2ttd}(b) and \ref{fig:feh2pvd}(a). We found that several H$\alpha$ features show correlations with the {[}\ion{S}{2}{]} HH objects. ``Knot B'' appears to have a definite H$\alpha$ counterpart, and ``Knot A' also overlaps with a bright H$\alpha$ region. The H$\alpha$ counterpart of ``Knot B'' is likely the H$\alpha$ emission induced by ``Outflow B.'' The bright H$\alpha$ region overlapping with ``Knot A'' is also likely related to ``Outflow A'' although the scattered component of the stellar emission seems to be blended. In fact, in Figure \ref{fig:hacosii}(a), the global H$\alpha$ morphology agrees well with the $^{13}$CO boundaries, but the bright H$\alpha$ region near ``Outflow A'' is located inside the SE $^{13}$CO region. For this region, the H$\alpha$ morphology follows the distorted CS molecular boundary shown in Figure \ref{fig:add}(a) much better. In Section \ref{subsec:newoutflow}, we discussed the possibility of the actual association between ``Outflow A'' and the distorted CS boundary based on the H$_{2}$ 1-0 S(1) channel maps in Figure \ref{fig:add}(a). The fact that this bright H$\alpha$ region is not apparent in the 11.2 $\micron$ image of \citet{2014ApJ...796...74L} (see their Figure 5) also supports its origin in ionized hydrogen, rather than dust scattering. The other {[}\ion{S}{2}{]} HH objects (``Knot C'' to ``Knot F'') have no clear H$\alpha$ counterparts. However, ``Knot D'' and ``Knot E'' seem to be coincident with a boundary of a bright H$\alpha$ region near the NE axis of the main outflow cavity. If they are indeed related to the NE main outflow, as mentioned in Sections \ref{subsec:h2s} and \ref{subsec:mainoutflow}, then at least some H$\alpha$ fluxes of the bright H$\alpha$ region could originate from ionized hydrogen possibly associated with the NE main outflow.

As can be seen in Figure \ref{fig:hacosii}(b), the arc-like H$\alpha$ feature has no {[}\ion{S}{2}{]} counterpart, although ``Knot C'' lies close to it. This arc-like feature is seen well in the red continuum filter r$_{\mathrm{N}}$ image of \citet{1992A&A...262..229P}, but disappears in their {[}\ion{S}{2}{]}$-$r$_{\mathrm{N}}$ image (see their Figures 7(b) and (d)). In fact, this feature also appears in the IPHAS broad-band $r$- and $i$-filter images. Furthermore, we found the same feature in the PanSTARRS broad-band $g$-, $r$-, and $i$-filter images.\footnote{\url{https://ps1images.stsci.edu/cgi-bin/ps1cutouts}} The presence of a similar arc-like feature in the continuum over a wide wavelength range indicates that the arc shape found in H$\alpha$ is highly likely to be due to a reflection nebula that scatters stellar H$\alpha$ emission. The weakness of the H$_{2}$ and {[}\ion{Fe}{2}{]} lines (H$_{2}$/Fe NE cavity edge in Figure \ref{fig:feh2pvd}(c)) also supports that shock excitation is not the main cause of this feature. The spatial coincidence of the arc-like H$\alpha$ feature with the NE $^{13}$CO boundary, as shown in Figure \ref{fig:hacosii}(a), suggests that the arc-like feature is produced by the illumination of starlight on a dusty surface of the NE cavity edge formed by the main outflow from MWC 1080A.

In Figures \ref{fig:hacosii}(a) and (b), there also exist several H$\alpha$ features around YSO 4 (the second masked-out star in the figure). However, they show no correlations with the $^{13}$CO and {[}\ion{S}{2}{]} morphologies. Because these features are also seen clearly in the IPHAS and PanSTARRS broad-band continuum images, they are likely due to dusty materials illuminated by stellar emission. The star YSO 4 is one of the bright stars in the optical images of \citet{1992A&A...262..229P}, designated as ``star \#2'' in their image (see Figure \ref{fig:hacosii}(b)). It also belongs to the NIR-identified young stars \citep{2008ApJ...673..315W}, as indicated by a triangle in Figure \ref{fig:hacosii}(a). Therefore, if YSO 4 is another active young star emitting a strong H$\alpha$ line, besides MWC 1080A, the extended H$\alpha$ features around it could be likely induced by YSO 4; of course, most of the other H$\alpha$ reflection features seem to be illuminated by MWC 1080A.

In addition to the H$\alpha$ emission shown in Figures \ref{fig:hacosii}(a) and (b), we identified a few more, faint H$\alpha$ features far away in the SW direction from MWC 1080A, as indicated in Figure \ref{fig:basic}(b). As discussed in Sections \ref{subsec:mainoutflow} and \ref{subsec:newoutflow}, we associated some parts of them with the counter outflows corresponding to the NE main outflow and ``Knot B.'' However, most of these faint features can also be seen in the IPHAS and PanSTARRS broad-band images. Therefore, they also likely trace the stellar H$\alpha$ emission scattered by dust on surfaces of the cavities formed by the main outflow, ``Outflow B,'' and other unknown outflows. \citet{2008A&A...477..193M} reported a large, diffuse lobe structure including most of the above faint H$\alpha$ features in their own H$\alpha$ image (see their Figure 5(a)). They interpreted the large H$\alpha$ lobe as a photoionized boundary of a cavity formed by outflows from central stars. However, most parts of the lobe are also seen in the IPHAS and PanSTARRS broad-band continuum images. This implies that the lobe structure is likely to be caused by the light scattered on the dusty surfaces of the cavity.

In order to clarify the origin of the extended H$\alpha$ emission in the MWC 1080 region, we made an H$\alpha$-to-continuum ratio map using the IPHAS H$\alpha$- and $r$-filter images, as shown in Figures \ref{fig:hacosii}(c) and (d). In Figure \ref{fig:hacosii}(c), most areas are found to have ratios in a range of 1.5--2.5, including the central region around MWC 1080A, the arc-like feature, the bright features around YSO 4, the SW faint features, and the regions near the axis of the main outflow. Only two small regions, near the heads of ``Knot A'' and ``Knot B'', appear to have ratios as high as 2.5--3.5, as can be seen more clearly in Figure \ref{fig:hacosii}(d). In these regions, the H$\alpha$ emission would originate dominantly from the stellar outflows (``Outflow A'' and ``Outflow B''), rather than from dust scattering.

Finally, it is worth noting that the H$_{2}$ fluorescent lines observed along the surfaces of the main outflow cavity wall, as shown in Section \ref{subsec:h2s}, imply the presence of widespread UV photons (11.2 eV $<$ $h\nu$ $<$ 13.6 eV) within the cavity. The Lyman-Werner radiation is likely to co-exist with Lyman continuum photons ($h\nu$ $>$ 13.6 eV), suggesting that diffuse photo-ionized hydrogen gas may also exist inside the main outflow cavity. However, we found no clear signature for photo-ionized components in the analysis of the hydrogen Brackett emission lines, which could be due to the dominant scattering components in this region.

\subsubsection{Compact Emission}

We identified the compact hydrogen line emissions (Br$\gamma$ and H$\alpha$) in the three young stars MWC 1080A, MWC 1080E, and YSO 1. Their Br$\gamma$ line profiles were found to be broad ($\Delta V$ = 163, 235, and 234 km s$^{-1}$, respectively) and generally symmetric about the line center, except for the central absorption feature (or the double-peaked profile) appearing in YSO 1. The stellar hydrogen emission lines observed in young stars have been associated with infalling accretion and/or outflowing winds \citep[e.g.,][]{1996ApJ...456..292N,2001A&A...365...90F,2006MNRAS.370..580K,2008A&A...489.1157K,2016A&A...590A..97T}. \citet{1996ApJ...456..292N} reported broad (FWZI = 400--700 km s$^{-1}$) and generally symmetric Br$\gamma$ line profiles from seven low-mass YSOs. Even though two of those YSOs have P Cygni absorptions (blueshifted absorptions) in their Balmer lines indicating outflowing winds, they suggested that the Br$\gamma$ line emission arises from infalling accretion rather than outflowing winds. \citet{2001A&A...365...90F} observed the Br$\gamma$ line profiles for 37 T Tauri stars and found that 72\% of them have broad ($\Delta V$ = 100--300 km s$^{-1}$) and generally symmetric Br$\gamma$ line profiles with no strong absorption features. They suggested that those line profiles could be partially explained by magnetospheric accretion models. Regarding the modeling of the H$\alpha$ line emission from T Tauri stars, \citet{2006MNRAS.370..580K} proposed hybrid models including both magnetospheric accretion and a disk wind, which could reproduce a wide variety of types for the H$\alpha$ line profiles observed in T Tauri stars. \citet{2006MNRAS.370..580K} found that symmetric H$\alpha$ line profiles can appear at various inclination angles for the circumstellar disk when the magnetospheric component is dominant. Although there are no definitely-known spectral types for the young stars MWC 1080E and YSO 1, they were classified as low-mass young stars by \citet{2008ApJ...673..315W} from a NIR $K'$-band image. If they are truly low-mass young stars, their Br$\gamma$ emissions can be explained mainly by magnetospheric accretion. The central absorption feature (or the double-peaked profile) in the Br$\gamma$ profile of YSO 1 may be caused by the disk wind component, as shown in the model of \citet{2006MNRAS.370..580K} (see their Figures A1--A3).

If MWC 1080E and YSO 1 are the driving sources of the newly-identified outflows in this study (``Outflow A'' and ``Outflow B''), as discussed in Section \ref{subsec:newoutflow}, they would be massive young stars, considering the large distance to the MWC 1080 system \citep[1.373 $\pm$ 0.174 kpc,][]{2018yCat.1345....0G}. They could belong to Herbig stars as MWC 1080A. Due to the weak magnetic fields detected in Herbig stars, the magnetosphere is expected to be less developed in Herbig stars than in low-mass young stars \citep{2014A&A...562A.104T,2016MNRAS.456..156G,2016A&A...590A..97T}. By modeling the Br$\gamma$ line emission from the magnetosphere, inner gaseous accretion disk, and disk wind region of Herbig AeBe stars, \citet{2016A&A...590A..97T} showed that the disk wind component is the most dominant contributor to the Br$\gamma$ line. Their disk wind model shows that symmetric Br$\gamma$ line profiles appear in disks viewed close to pole-on, whereas double-peaked profiles become prominent in disks viewed closer to edge-on (see their Figure 1). They also suggested that the component from the inner gaseous accretion disk can transform double-peaked profiles into single-peaked (symmetric) ones when the gap between two peaks is not large (see their Figure 8). The high-blueshift components detected in the two outflows ($-$233 km s$^{-1}$ {[}\ion{Fe}{2}{]} in ``Outflow A'' and $-$156 km s$^{-1}$ {[}\ion{S}{2}{]} in ``Outflow B,'' see Section \ref{subsec:newoutflow}) indicate that the corresponding circumstellar disks are not viewed close to edge-on. Therefore, if MWC 1080E and YSO 1 are truly Herbig stars driving ``Outflow A'' and ``Outflow B,'' the symmetric Br$\gamma$ line profile of MWC 1080E and the weakly double-peaked profile of YSO 1 can be explained by dominant disk winds. For MWC 1080E, the contribution of an inner gaseous accretion disk might enhance the symmetry of its profile.

On the other hand, MWC 1080A is expected to harbor a nearly edge-on circumstellar disk, considering the low inclination angle of the main outflow, as discussed in Section \ref{subsec:mainoutflow}. To explain the symmetric and single-peaked Br$\gamma$ line profile observed in the Herbig star VV Ser, which has a nearly edge-on circumstellar disk, \citet{2016MNRAS.456..156G} added a bipolar outflow component into a model with magnetospheric accretion and a disk wind. They showed that the bipolar outflow could reproduce the single-peaked Br$\gamma$ line profile in an edge-on disk system. Therefore, the similar single-peaked Br$\gamma$ line profile observed in MWC 1080A is also likely to be dominated by a stellar wind producing a bipolar outflow. \citet{1992PASJ...44...77Y} observed prominent P Cygni absorptions and broad wings extending up to $\sim$1000 km s$^{-1}$ in the hydrogen Balmer lines of MWC 1080A and concluded that this Herbig star has a strong stellar wind with a velocity of up to 1000 km s$^{-1}$. Although the Br$\gamma$ line of MWC 1080A observed in this study shows no P Cygni absorption, the redshifted wing extending up to $\sim$300 km s$^{-1}$, as seen in Figure \ref{fig:brglp}, could support the association with the stellar wind expected from the Balmer lines. In addition, note that the close (0.$\arcsec$8) binary MWC 1080B of MWC 1080A was observed in the same IGRINS slit position. However, we could not resolve the Br$\gamma$ line of MWC 1080B from that of MWC 1080A. Judging from the more than an order of magnitude higher total luminosity of MWC 1080A than that of MWC 1080B \citep{1997A&A...318..472L}, MWC 1080A could mostly contribute to the observed Br$\gamma$ emission.

\section{Summary and conclusions} \label{sec:conclusions}

We observed the MWC 1080 region by IGRINS with a high spectral resolution of $\lambda$/$\Delta\lambda$ $\sim$45,000 and a broad wavelength coverage of the NIR H (1.49--1.80 $\micron$) and K (1.96--2.46 $\micron$) bands. This study is the first high-resolution NIR spectroscopic observation for the MWC 1080 region. From a total of ten slit positions, we detected six hydrogen Brackett lines including Br$\gamma$ $\lambda$2.166 $\micron$, seven molecular hydrogen lines including H$_{2}$ 1-0 S(1) $\lambda$2.122 $\micron$, and an {[}\ion{Fe}{2}{]} $\lambda$1.644 $\micron$ forbidden line. We identified several diffuse sources in the Br$\gamma$, H$_{2}$ 1-0 S(1), and {[}\ion{Fe}{2}{]} PVDs and analyzed the line profiles and line ratios for the individual sources. From the obtained kinematic and line ratio results, and also the morphological comparisons with the H$\alpha$, {[}\ion{S}{2}{]}, $^{13}$CO, and CS maps, we found the following.

1. Most of the diffuse Br$\gamma$ sources around MWC 1080A were found to be dusty nebulae that scatter the Br$\gamma$ emission from MWC 1080A. Therefore, the extended H$\alpha$ emission around MWC 1080A would also be predominantly due to scattering by dust located on the surface of the main outflow cavity. The Br$\gamma$ morphology close to MWC 1080A shows two separate peaks near the axis of the main outflow (PA $\simeq$ 40$\arcdeg$), which may be due to dusty materials related to the bipolar outflow from MWC 1080A. On the other hand, two compact Br$\gamma$ sources close to the young stars MWC 1080E and YSO 1 were found to be direct Br$\gamma$ emissions from the corresponding stars; this suggests that MWC 1080E and YSO 1 are highly likely young stars with active accretion and/or winds.

2. The H$_{2}$ lines with $V_{\mathrm{centroid}}=$ $-$6 to +3 km s$^{-1}$ near the systemic LSR velocity and narrow $\Delta V$ of 9--19 km s$^{-1}$ were detected in the central region around MWC 1080A. Based on the line ratio analysis, we identified them as fluorescent H$_{2}$ lines. From the H$_{2}$ 1-0 S(1) velocity channel maps, we revealed that the fluorescent H$_{2}$ emission traces PDRs formed on the front, limb, and rear surfaces of the main outflow cavity, and the cavity wall with a cylindrical shape is expanding outward with a velocity of $\sim$10--15 km s$^{-1}$. The observed H$_{2}$ fluorescent lines seem to indirectly suggest the presence of diffuse photo-ionized hydrogen gas inside the main outflow cavity.

3. The H$_{2}$ lines with slightly-blueshifted $V_{\mathrm{centroid}}$ of $-$12 to $-$4 km s$^{-1}$ and moderate $\Delta V$ of 22--24 km s$^{-1}$ were detected along the NE axis of the main outflow cavity. The line ratio analysis shows that they are possibly thermal H$_{2}$ lines. In particular, just inside an arc-like H$\alpha$ feature located at the NE cavity edge, the H$_{2}$ line emission does peak together with the shock-induced {[}\ion{Fe}{2}{]} line emission with $V_{\mathrm{centroid}}$ = $-$10 km s$^{-1}$ and $\Delta V$ = 43 km s$^{-1}$. This suggests that the H$_{2}$ and {[}\ion{Fe}{2}{]} emissions could be induced by shock heating in the NE main outflow from MWC 1080A, although most parts of the arc-like H$\alpha$ feature were revealed to be reflection nebulae scattering stellar H$\alpha$ emission. Low blueshifted $V_{\mathrm{centroid}}$ for the {[}\ion{Fe}{2}{]} line implies that the NE main outflow from MWC 1080A is the blueshifted outflow with a low inclination angle with respect to the sky plane. In addition, from the H$\alpha$ line image, we interpreted a faint H$\alpha$ feature located in the opposite direction from MWC 1080A as the SW cavity edge formed by the counter outflow corresponding to the NE main outflow.

4. The H$_{2}$ lines with a blueshifted $V_{\mathrm{centroid}}$ of $-$31 to $-$10 km s$^{-1}$ and a broad $\Delta V$ of 18--43 km s$^{-1}$ were detected near/in the SE molecular region. From the line ratio analysis, they were identified as thermal H$_{2}$ lines. Some of their components have strong spatial and kinematic correlations with the shock-induced {[}\ion{Fe}{2}{]} lines with a highly-blueshifted $V_{\mathrm{centroid}}$ of $-$239 to $-$42 km s$^{-1}$ and a broad $\Delta V$ of 77--139 km s$^{-1}$, which can be explained by stellar outflow shocks and the ambient molecular gas entrained by the shocks. Out of them, the brightest H$_{2}$ and {[}\ion{Fe}{2}{]} emissions were detected near the peak position of a {[}\ion{S}{2}{]} HH object, ``Knot A.'' Comparing double-velocity {[}\ion{Fe}{2}{]} components detected here with the bow shock model of \citet{1987ApJ...316..323H}, we suggest that ``Knot A'' is a high-velocity (a shock velocity of 300--400 km s$^{-1}$) and highly-blueshifted (an inclination angle of $>$45$\arcdeg$) outflow moving towards the SE direction (referred to as ``Outflow A'' with PA $\simeq$ 135$\arcdeg$). We also propose that some other {[}\ion{Fe}{2}{]} and H$_{2}$ components are possibly associated with another {[}\ion{S}{2}{]} HH object, ``Knot B.'' We found that there is another faint H$\alpha$ feature in the direction of the SW counter outflow corresponding to ``Knot B'' (referred to as ``Outflow B'' with PA $\simeq$ 77$\arcdeg$) and those {[}\ion{Fe}{2}{]} and H$_{2}$ components lie near the axis of the outflow. In an H$\alpha$-to-continuum ratio map, we confirmed that ``Knot A'' and ``Knot B'' have inherent H$\alpha$ emissions excited by the outflows.

This study showed that atomic hydrogen line emissions observed in the MWC 1080 region mostly originate from the stellar emission of MWC 1080A scattered by dust. However, the extended components originating from ionized gas related to two stellar outflows and the compact components of two young stars were also found. In conclusion, the H$\alpha$ nebulosity in the MWC 1080 region is overall dominated by dust scattering of the stellar emission but locally by shock-excited ionized gases, depending on locations. Our results demonstrate the importance of high-resolution observations. The MWC 1080 region includes at least $\sim$50 young stars. Therefore, there could possibly be more outflows associated with young stars than we have found. In this study, we identified at least three outflows from MWC 1080A and other young stars. Further optical/NIR spectroscopic observations with a wider spatial coverage over this region would enable us to find more unknown outflows and their accurate driving sources.

\acknowledgments
We thank the anonymous referee for useful comments that helped to improve the manuscript. This work was supported by the National Research Foundation of Korea (NRF) grant funded by the Korea government (MSIT) (No. 2019R1F1A1064112). This work used the Immersion Grating Infrared Spectrograph (IGRINS) that was developed under a collaboration between the University of Texas at Austin and the Korea Astronomy and Space Science Institute (KASI) with the financial support of the US National Science Foundation under grant AST-1229522, of the University of Texas at Austin, and of the Korean GMT Project of KASI. These results made use of the Lowell Discovery Telescope (LDT) at Lowell Observatory. Lowell is a private, non-profit institution dedicated to astrophysical research and public appreciation of astronomy and operates the LDT in partnership with Boston University, the University of Maryland, the University of Toledo, Northern Arizona University and Yale University. This paper makes use of data obtained as part of the IPHAS (http://www.iphas.org/) carried out at the Isaac Newton Telescope (INT). This research has made use of the SIMBAD database, operated at CDS, Strasbourg, France.

\appendix
\section{IGRINS Observations of Additional Slit Positions}

\begin{figure}[ht!]
\centering
\includegraphics[scale=0.35]{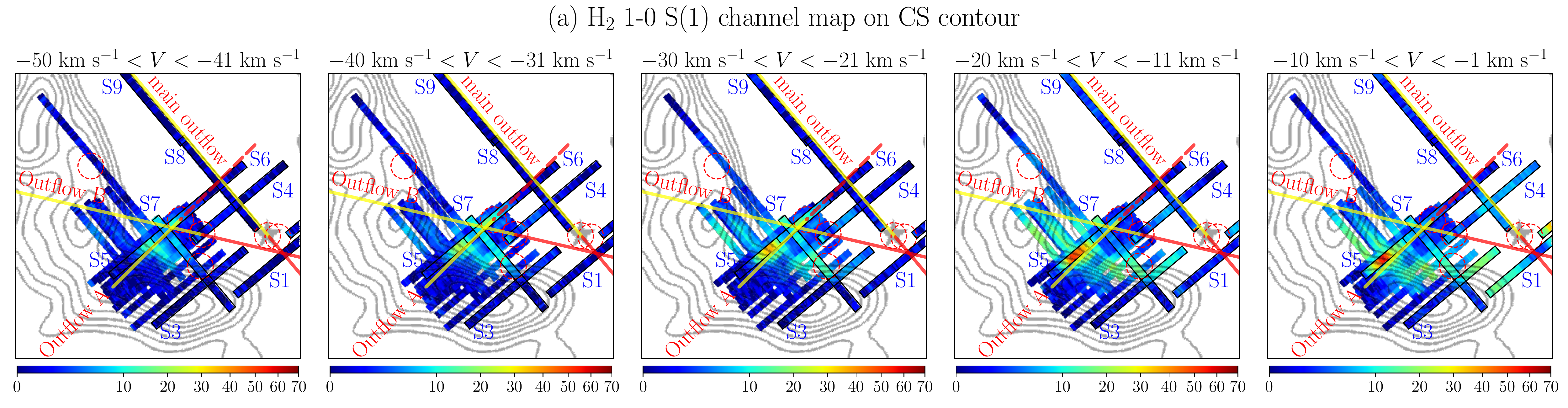}
\includegraphics[scale=0.35]{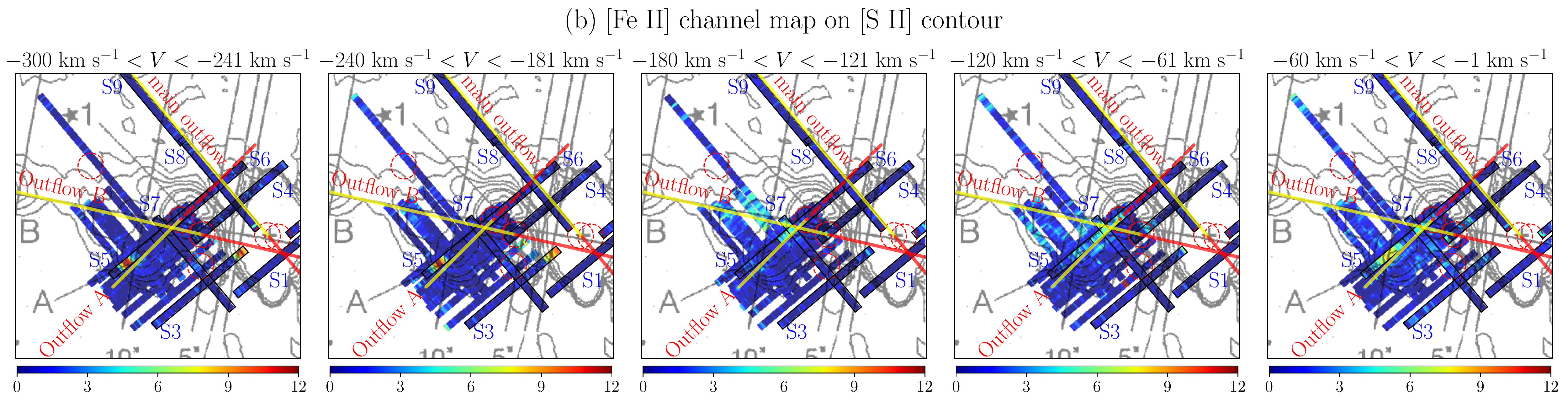}
\caption{(a) H$_{2}$ 1-0 S(1) and (b) {[}\ion{Fe}{2}{]} velocity channel maps around the SE molecular region, which are overlaid on the CS \citep{2008ApJ...673..315W} and {[}\ion{S}{2}{]} \citep{1992A&A...262..229P} contours, respectively. The color scales are all in units of the detection significance (signal/$\sigma$). The data obtained from additional observations for twelve slit positions were overplotted together with the results of the slit positions S1-S10 (denoted by the rectangles). For the overlapped regions, the S1-S10 data were used because they have better data quality. The integrated velocity range is indicated at the top of each channel map. Red dashed circles (young stars) and yellow+red lines (axes of the main outflow, ``Outflow A,'' and ``Outflow B'') are the same as in Figure \ref{fig:basic}. Here, the yellow and red lines represent the blueshifted and redshifted part of the outflows, respectively.\label{fig:add}}
\end{figure}

We also carried out additional observations for twelve slit positions with IGRINS to confirm the origins of the {[}\ion{Fe}{2}{]} and H$_{2}$ emissions detected near the SE molecular region. Although the results were not very useful because of bad weather conditions, we plotted the H$_{2}$ 1-0 S(1) and {[}\ion{Fe}{2}{]} velocity channel maps in Figure \ref{fig:add} to see rough shapes of distributions of the H$_{2}$ and {[}\ion{Fe}{2}{]} emissions. The H$_{2}$ 1-0 S(1) and {[}\ion{Fe}{2}{]} maps were overlaid on the CS molecular and {[}\ion{S}{2}{]} contours, respectively. The CS and {[}\ion{S}{2}{]} contours were adapted from Figure 3(b) of \citet{2008ApJ...673..315W} and Figure 8(d) of \citet{1992A&A...262..229P}, respectively.

\clearpage
\begin{landscape}
\LongTables
\begin{deluxetable*}{lcccccccccc}
\tablecaption{Line Profiles and Fluxes for Br$\gamma$ Sources and MWC 1080A\label{table:brglp}}
\tablewidth{700pt}
\tabletypesize{\footnotesize}
\tablehead{
&&& \multicolumn{5}{c}{Extended Sources} && \multicolumn{2}{c}{Compact Sources} \\
\cline{4-8} \cline{10-11}
\colhead{Line} & \colhead{$\lambda$ ($\micron$)} && \colhead{Br SW} & \colhead{Br NW} & \colhead{Br SE} & \colhead{Br NE 1} & \colhead{Br NE 2} & \colhead{MWC 1080A} &
\colhead{Br 1080E} & \colhead{Br YSO 1}
}
\startdata
&& $\mathnormal{V}_{\mathrm{centroid}}$\tablenotemark{a} (km s$^{-1}$) & $-$6 $\pm$ 1 & $-$2 $\pm$ 1 & $-$10 $\pm$ 1 & $-$10 $\pm$ 1 & $-$11 $\pm$ 1 & $-$8 $\pm$ 1 & 8 $\pm$ 1 & $-$11 $\pm$ 2 \\
Br$\gamma$ & 2.16612 & $\Delta\mathnormal{V}$\tablenotemark{b} (km s$^{-1}$) & 162 $\pm$ 1 & 171 $\pm$ 2 & 151 $\pm$ 2 & 165 $\pm$ 3 & 153 $\pm$ 2 & 163 $\pm$ 2 & 235 $\pm$ 2 & 234 $\pm$ 10 \\
&& $\mathnormal{F}/\mathnormal{F}_{0}$\tablenotemark{c} & 1.00 $\pm$ 0.01 & 1.00 $\pm$ 0.01 & 1.00 $\pm$ 0.01 & 1.00 $\pm$ 0.02 & 1.00 $\pm$ 0.01 & 1.00 $\pm$ 0.01 & 1.00 $\pm$ 0.01 & 1.00 $\pm$ 0.18 \\
\hline
&& $\mathnormal{V}_{\mathrm{centroid}}$\tablenotemark{a} (km s$^{-1}$) & $-$19 $\pm$ 1 & $-$6 $\pm$ 2 & $-$18 $\pm$ 1 & $-$30 $\pm$ 2 & $-$16 $\pm$ 1 & $-$14 $\pm$ 1 & $-$4 $\pm$ 2 & \nodata \\
Br 10 & 1.73669 & $\Delta\mathnormal{V}$\tablenotemark{b} (km s$^{-1}$) & 140 $\pm$ 2 & 160 $\pm$ 6 & 123 $\pm$ 3 & 161 $\pm$ 5 & 148 $\pm$ 3 & 130 $\pm$ 2 & 262 $\pm$ 7 & \nodata \\
&& $\mathnormal{F}/\mathnormal{F}_{0}$\tablenotemark{c} & 0.56 $\pm$ 0.01 & 0.45 $\pm$ 0.04 & 0.51 $\pm$ 0.01 & 0.59 $\pm$ 0.04 & 0.60 $\pm$ 0.01 & 0.45 $\pm$ 0.00 & 0.28 $\pm$ 0.02 & \nodata \\
\hline
&& $\mathnormal{V}_{\mathrm{centroid}}$\tablenotemark{a} (km s$^{-1}$) & $-$23 $\pm$ 1 & $-$21 $\pm$ 2 & $-$22 $\pm$ 1 & $-$29 $\pm$ 2 & $-$27 $\pm$ 1 & $-$14 $\pm$ 1 & $-$12 $\pm$ 2 & \nodata \\
Br 11 & 1.68111 & $\Delta\mathnormal{V}$\tablenotemark{b} (km s$^{-1}$) & 141 $\pm$ 2 & 178 $\pm$ 6 & 146 $\pm$ 3 & 194 $\pm$ 6 & 149 $\pm$ 2 & 146 $\pm$ 1 & 216 $\pm$ 6 & \nodata \\
&& $\mathnormal{F}/\mathnormal{F}_{0}$\tablenotemark{c} & 0.44 $\pm$ 0.01 & 0.46 $\pm$ 0.04 & 0.63 $\pm$ 0.01 & 0.59 $\pm$ 0.05 & 0.67 $\pm$ 0.01 & 0.43 $\pm$ 0.00 & 0.17 $\pm$ 0.01 & \nodata \\
\hline
&& $\mathnormal{V}_{\mathrm{centroid}}$\tablenotemark{a} (km s$^{-1}$) & $-$10 $\pm$ 1 & \nodata & $-$20 $\pm$ 3 & $-$31 $\pm$ 4 & $-$10 $\pm$ 2 & $-$14 $\pm$ 1 & $-$14 $\pm$ 4 & \nodata \\
Br 12 & 1.64117 & $\Delta\mathnormal{V}$\tablenotemark{b} (km s$^{-1}$) & 140 $\pm$ 3 & \nodata & 115 $\pm$ 8 & 227 $\pm$ 11 & 177 $\pm$ 5 & 145 $\pm$ 2 & 292 $\pm$ 10 & \nodata \\
&& $\mathnormal{F}/\mathnormal{F}_{0}$\tablenotemark{c} & 0.37 $\pm$ 0.01 & \nodata & 0.23 $\pm$ 0.05 & 0.61 $\pm$ 0.14 & 0.54 $\pm$ 0.03 & 0.37 $\pm$ 0.00 & 0.22 $\pm$ 0.03 & \nodata \\
\hline
&& $\mathnormal{V}_{\mathrm{centroid}}$\tablenotemark{a} (km s$^{-1}$) & $-$13 $\pm$ 2 & 1 $\pm$ 4 & $-$7 $\pm$ 3 & \nodata & $-$13 $\pm$ 3 & $-$15 $\pm$ 1 & \nodata & \nodata \\
Br 13 & 1.61137 & $\Delta\mathnormal{V}$\tablenotemark{b} (km s$^{-1}$) & 161 $\pm$ 5 & 159 $\pm$ 11 & 134 $\pm$ 8 & \nodata & 107 $\pm$ 8 & 131 $\pm$ 2 & \nodata & \nodata \\
&& $\mathnormal{F}/\mathnormal{F}_{0}$\tablenotemark{c} & 0.45 $\pm$ 0.02 & 0.30 $\pm$ 0.09 & 0.30 $\pm$ 0.06 & \nodata & 0.25 $\pm$ 0.07 & 0.29 $\pm$ 0.00 & \nodata & \nodata \\
\hline
&& $\mathnormal{V}_{\mathrm{centroid}}$\tablenotemark{a} (km s$^{-1}$) & $-$13 $\pm$ 1 & $-$21 $\pm$ 3 & $-$15 $\pm$ 2 & $-$12 $\pm$ 2 & $-$13 $\pm$ 2 & $-$16 $\pm$ 1 & 8 $\pm$ 4 & \nodata \\
Br 14 & 1.58849 & $\Delta\mathnormal{V}$\tablenotemark{b} (km s$^{-1}$) & 134 $\pm$ 3 & 201 $\pm$ 7 & 141 $\pm$ 5 & 172 $\pm$ 6 & 161 $\pm$ 5 & 158 $\pm$ 2 & 218 $\pm$ 12 & \nodata \\
&& $\mathnormal{F}/\mathnormal{F}_{0}$\tablenotemark{c} & 0.28 $\pm$ 0.01 & 0.42 $\pm$ 0.05 & 0.37 $\pm$ 0.02 & 0.43 $\pm$ 0.04 & 0.35 $\pm$ 0.02 & 0.34 $\pm$ 0.00 & 0.07 $\pm$ 0.02 & \nodata \\
\enddata
\tablenotetext{a}{The centroid velocity $V_{\mathrm{centroid}}$ obtained from the Gaussian fitting.}
\tablenotetext{b}{The FWHM velocity width $\Delta V$ obtained from the Gaussian fitting.}
\tablenotetext{c}{The line flux $F$ was obtained from the Gaussian fitting and then normalized to the Br$\gamma$ line flux $F_{0}$ that is displayed in Figure \ref{fig:brglp}.}
\tablecomments{The line profiles and fluxes of Br 10--14 were calculated by the same method as used for Br$\gamma$ (see Figure \ref{fig:brglp}), and the lines measured with $\ge$3$\sigma$ significance (based on the line flux) are presented. For Br YSO 1, only the Br$\gamma$ emission component was detected at a significance level of $\ge$3$\sigma$. All of the line fluxes were not corrected for the reddening effect.}
\end{deluxetable*}

\clearpage
\LongTables
\begin{deluxetable*}{lcccccccccccc}
\tablecaption{Line Profiles and Fluxes for H$_{2}$ Sources: Narrow Components\label{table:h2lp1}}
\tablewidth{700pt}
\tabletypesize{\footnotesize}
\tablehead{
\colhead{Line} & \colhead{$\lambda$ ($\micron$)} && \colhead{H$_{2}$ SW-E} & \colhead{H$_{2}$ SW-W} & \colhead{H$_{2}$ NW} & \colhead{H$_{2}$ SE} & \colhead{H$_{2}$ NE 1} &
\colhead{H$_{2}$ knot A} & \colhead{H$_{2}$ knot B} & \colhead{H$_{2}$ YSO 2} & \colhead{H$_{2}$ NE 2} & \colhead{H$_{2}$ NE cavity axis}
}
\startdata
&& $V_{\mathrm{centroid}}$\tablenotemark{a} (km s$^{-1}$) & $-$5 $\pm$ 0 & $-$2 $\pm$ 0 & $-$2 $\pm$ 0 & $-$5 $\pm$ 0 & 0 $\pm$ 0 & $-$9 $\pm$ 0 & 5 $\pm$ 0 & 1 $\pm$ 0 & $-$5 $\pm$ 1 & 8 $\pm$ 1 \\
H$_{2}$ 1-0 S(2) & 2.03376 & $\Delta V$\tablenotemark{b} (km s$^{-1}$) & 12 $\pm$ 1 & 9 $\pm$ 0 & 11 $\pm$ 0 & 13 $\pm$ 1 & 19 $\pm$ 1 & 17 $\pm$ 1 & 10 $\pm$ 0 & 13 $\pm$ 1 & 20 $\pm$ 2 & 15 $\pm$ 1 \\
&& $F/F_{0}$\tablenotemark{c} & 0.44 $\pm$ 0.02 & 0.40 $\pm$ 0.00 & 0.32 $\pm$ 0.00 & 0.51 $\pm$ 0.02 & 0.41 $\pm$ 0.01 & 0.36 $\pm$ 0.01 & 0.49 $\pm$ 0.00 & 0.58 $\pm$ 0.02 & 0.37 $\pm$ 0.04 & 0.64 $\pm$ 0.04 \\
\hline
&& $V_{\mathrm{centroid}}$\tablenotemark{a} (km s$^{-1}$) & $-$5 $\pm$ 0 & $-$2 $\pm$ 0 & $-$2 $\pm$ 0 & $-$6 $\pm$ 0 & $-$2 $\pm$ 0 & $-$10 $\pm$ 0 & 3 $\pm$ 0 & 1 $\pm$ 0 & $-$6 $\pm$ 0 & 9 $\pm$ 0 \\
H$_{2}$ 1-0 S(1) & 2.12183 & $\Delta V$\tablenotemark{b} (km s$^{-1}$) & 9 $\pm$ 0 & 10 $\pm$ 0 & 11 $\pm$ 0 & 11 $\pm$ 0 & 19 $\pm$ 1 & 18 $\pm$ 0 & 11 $\pm$ 0 & 11 $\pm$ 0 & 19 $\pm$ 1 & 14 $\pm$ 1 \\
&& $F/F_{0}$\tablenotemark{c} & 1.00 $\pm$ 0.00 & 1.00 $\pm$ 0.00 & 1.00 $\pm$ 0.00 & 1.00 $\pm$ 0.00 & 1.00 $\pm$ 0.01 & 1.00 $\pm$ 0.00 & 1.00 $\pm$ 0.00 & 1.00 $\pm$ 0.01 & 1.00 $\pm$ 0.01 & 1.00 $\pm$ 0.02 \\
\hline
&& $V_{\mathrm{centroid}}$\tablenotemark{a} (km s$^{-1}$) & $-$5 $\pm$ 0 & $-$2 $\pm$ 0 & $-$1 $\pm$ 0 & $-$5 $\pm$ 0 & $-$2 $\pm$ 1 & $-$9 $\pm$ 0 & 4 $\pm$ 0 & 1 $\pm$ 0 & $-$7 $\pm$ 0 & 8 $\pm$ 1 \\
H$_{2}$ 1-0 S(0) & 2.22329 & $\Delta V$\tablenotemark{b} (km s$^{-1}$) & 11 $\pm$ 1 & 10 $\pm$ 1 & 10 $\pm$ 1 & 12 $\pm$ 1 & 18 $\pm$ 2 & 20 $\pm$ 1 & 12 $\pm$ 0 & 11 $\pm$ 1 & 10 $\pm$ 1 & 12 $\pm$ 3 \\
&& $F/F_{0}$\tablenotemark{c} & 0.55 $\pm$ 0.01 & 0.53 $\pm$ 0.01 & 0.45 $\pm$ 0.02 & 0.71 $\pm$ 0.01 & 0.59 $\pm$ 0.06 & 0.26 $\pm$ 0.01 & 0.58 $\pm$ 0.00 & 0.90 $\pm$ 0.01 & 0.37 $\pm$ 0.01 & 0.32 $\pm$ 0.09 \\
\hline
&& $V_{\mathrm{centroid}}$\tablenotemark{a} (km s$^{-1}$) & $-$4 $\pm$ 0 & $-$3 $\pm$ 0 & $-$2 $\pm$ 0 & $-$4 $\pm$ 1 & 1 $\pm$ 1 & $-$12 $\pm$ 0 & 4 $\pm$ 0 & 2 $\pm$ 1 & \nodata & \nodata \\
H$_{2}$ 2-1 S(1) & 2.24772 & $\Delta V$\tablenotemark{b} (km s$^{-1}$) & 10 $\pm$ 1 & 10 $\pm$ 2 & 12 $\pm$ 1 & 10 $\pm$ 2 & 23 $\pm$ 2 & 20 $\pm$ 1 & 11 $\pm$ 1 & 14 $\pm$ 1 & \nodata & \nodata \\
&& $F/F_{0}$\tablenotemark{c} & 0.45 $\pm$ 0.01 & 0.31 $\pm$ 0.03 & 0.24 $\pm$ 0.00 & 0.27 $\pm$ 0.03 & 0.41 $\pm$ 0.03 & 0.15 $\pm$ 0.00 & 0.37 $\pm$ 0.01 & 0.41 $\pm$ 0.02 & \nodata & \nodata \\
\hline
&& $V_{\mathrm{centroid}}$\tablenotemark{a} (km s$^{-1}$) & $-$4 $\pm$ 0 & $-$1 $\pm$ 0 & $-$1 $\pm$ 0 & $-$4 $\pm$ 0 & $-$3 $\pm$ 0 & $-$11 $\pm$ 0 & 4 $\pm$ 0 & 0 $\pm$ 0 & $-$7 $\pm$ 0 & 8 $\pm$ 0 \\
H$_{2}$ 1-0 Q(1) & 2.40659 & $\Delta V$\tablenotemark{b} (km s$^{-1}$) & 10 $\pm$ 1 & 9 $\pm$ 0 & 10 $\pm$ 0 & 9 $\pm$ 0 & 16 $\pm$ 1 & 21 $\pm$ 1 & 12 $\pm$ 0 & 10 $\pm$ 0 & 10 $\pm$ 1 & 9 $\pm$ 1 \\
&& $F/F_{0}$\tablenotemark{c} & 1.62 $\pm$ 0.02 & 2.08 $\pm$ 0.00 & 2.13 $\pm$ 0.00 & 1.57 $\pm$ 0.01 & 1.41 $\pm$ 0.01 & 1.06 $\pm$ 0.01 & 1.43 $\pm$ 0.00 & 1.82 $\pm$ 0.01 & 1.21 $\pm$ 0.02 & 0.75 $\pm$ 0.05 \\
\hline
&& $V_{\mathrm{centroid}}$\tablenotemark{a} (km s$^{-1}$) & $-$6 $\pm$ 0 & $-$2 $\pm$ 0 & $-$2 $\pm$ 0 & $-$3 $\pm$ 0 & $-$5 $\pm$ 0 & $-$10 $\pm$ 1 & 4 $\pm$ 0 & 0 $\pm$ 0 & \nodata & \nodata \\
H$_{2}$ 1-0 Q(2) & 2.41344 & $\Delta V$\tablenotemark{b} (km s$^{-1}$) & 11 $\pm$ 1 & 9 $\pm$ 1 & 10 $\pm$ 0 & 10 $\pm$ 2 & 13 $\pm$ 1 & 20 $\pm$ 2 & 11 $\pm$ 1 & 13 $\pm$ 1 & \nodata & \nodata \\
&& $F/F_{0}$\tablenotemark{c} & 0.52 $\pm$ 0.04 & 0.61 $\pm$ 0.01 & 0.58 $\pm$ 0.01 & 0.60 $\pm$ 0.06 & 0.66 $\pm$ 0.03 & 0.39 $\pm$ 0.05 & 0.63 $\pm$ 0.02 & 0.99 $\pm$ 0.05 & \nodata & \nodata \\
\hline
&& $V_{\mathrm{centroid}}$\tablenotemark{a} (km s$^{-1}$) & $-$4 $\pm$ 0 & $-$2 $\pm$ 0 & $-$2 $\pm$ 0 & $-$5 $\pm$ 1 & $-$3 $\pm$ 1 & $-$10 $\pm$ 0 & 3 $\pm$ 0 & 0 $\pm$ 0 & $-$9 $\pm$ 1 & 9 $\pm$ 1 \\
H$_{2}$ 1-0 Q(3) & 2.42373 & $\Delta V$\tablenotemark{b} (km s$^{-1}$) & 11 $\pm$ 1 & 13 $\pm$ 1 & 12 $\pm$ 0 & 13 $\pm$ 3 & 17 $\pm$ 1 & 19 $\pm$ 1 & 14 $\pm$ 1 & 7 $\pm$ 1 & 19 $\pm$ 2 & 10 $\pm$ 1 \\
&& $F/F_{0}$\tablenotemark{c} & 0.87 $\pm$ 0.02 & 0.92 $\pm$ 0.01 & 0.99 $\pm$ 0.01 & 0.97 $\pm$ 0.23 & 0.82 $\pm$ 0.03 & 0.80 $\pm$ 0.02 & 0.81 $\pm$ 0.02 & 0.35 $\pm$ 0.02 & 0.75 $\pm$ 0.08 & 0.81 $\pm$ 0.11 \\
\enddata
\tablenotetext{a}{The centroid velocity $V_{\mathrm{centroid}}$ obtained from the Gaussian fitting.}
\tablenotetext{b}{The FWHM velocity width $\Delta V$ obtained from the Gaussian fitting.}
\tablenotetext{c}{The line flux $F$ was obtained from the Gaussian fitting and then normalized to the H$_{2}$ 1-0 S(1) line flux $F_{0}$ that is displayed in Figure \ref{fig:h2lp}.}
\tablecomments{The same method as for H$_{2}$ 1-0 S(1) (see Figure \ref{fig:h2lp}) was adopted to calculate the profiles and fluxes of the other lines, and the components corresponding to the individual H$_{2}$ 1-0 S(1) narrow components are listed. Only the lines measured with $\ge$3$\sigma$ significance (based on the line flux) are presented. For H$_{2}$ 1080E and H$_{2}$ NE cavity edge, no narrow components were detected with $\ge$3$\sigma$ significance. All of the line fluxes were not corrected for the reddening effect.}
\end{deluxetable*}

\clearpage
\LongTables
\begin{deluxetable*}{lccccccccc}
\tablecaption{Line Profiles and Fluxes for H$_{2}$ Sources: Broad Components\label{table:h2lp2}}
\tablewidth{700pt}
\tabletypesize{\footnotesize}
\tablehead{
\colhead{Line} & \colhead{$\lambda$ ($\micron$)} && \colhead{H$_{2}$ SE} & \colhead{H$_{2}$ 1080E} & \colhead{H$_{2}$ knot A} &
\colhead{H$_{2}$ knot B} & \colhead{H$_{2}$ YSO 2} & \colhead{H$_{2}$ NE cavity axis} & \colhead{H$_{2}$ NE cavity edge}
}
\startdata
&& $V_{\mathrm{centroid}}$\tablenotemark{a} (km s$^{-1}$) & $-$12 $\pm$ 2 & $-$11 $\pm$ 2 & $-$23 $\pm$ 2 & $-$29 $\pm$ 0 & $-$21 $\pm$ 1 & $-$10 $\pm$ 1 & $-$6 $\pm$ 1 \\
H$_{2}$ 1-0 S(2) & 2.03376 & $\Delta V$\tablenotemark{b} (km s$^{-1}$) & 34 $\pm$ 3 & 44 $\pm$ 4 & 35 $\pm$ 2 & 22 $\pm$ 1 & 32 $\pm$ 3 & 12 $\pm$ 2 & 28 $\pm$ 3 \\
&& $F/F_{0}$\tablenotemark{c} & 0.29 $\pm$ 0.03 & 0.30 $\pm$ 0.04 & 0.33 $\pm$ 0.02 & 0.26 $\pm$ 0.00 & 0.27 $\pm$ 0.02 & 0.31 $\pm$ 0.03 & 0.40 $\pm$ 0.06 \\
\hline
&& $V_{\mathrm{centroid}}$\tablenotemark{a} (km s$^{-1}$) & $-$12 $\pm$ 0 & $-$14 $\pm$ 0 & $-$23 $\pm$ 1 & $-$31 $\pm$ 0 & $-$21 $\pm$ 0 & $-$12 $\pm$ 1 & $-$4 $\pm$ 0 \\
H$_{2}$ 1-0 S(1) & 2.12183 & $\Delta V$\tablenotemark{b} (km s$^{-1}$) & 31 $\pm$ 1 & 43 $\pm$ 1 & 32 $\pm$ 1 & 29 $\pm$ 0 & 41 $\pm$ 1 & 22 $\pm$ 2 & 24 $\pm$ 1 \\
&& $F/F_{0}$\tablenotemark{c} & 1.00 $\pm$ 0.00 & 1.00 $\pm$ 0.01 & 1.00 $\pm$ 0.02 & 1.00 $\pm$ 0.00 & 1.00 $\pm$ 0.00 & 1.00 $\pm$ 0.04 & 1.00 $\pm$ 0.01 \\
\hline
&& $V_{\mathrm{centroid}}$\tablenotemark{a} (km s$^{-1}$) & \nodata & $-$12 $\pm$ 1 & $-$25 $\pm$ 3 & $-$33 $\pm$ 1 & $-$23 $\pm$ 1 & $-$8 $\pm$ 1 & $-$1 $\pm$ 1 \\
H$_{2}$ 1-0 S(0) & 2.22329 & $\Delta V$\tablenotemark{b} (km s$^{-1}$) & \nodata & 28 $\pm$ 3 & 47 $\pm$ 4 & 39 $\pm$ 2 & 51 $\pm$ 3 & 13 $\pm$ 3 & 14 $\pm$ 3 \\
&& $F/F_{0}$\tablenotemark{c} & \nodata & 0.26 $\pm$ 0.03 & 0.27 $\pm$ 0.03 & 0.32 $\pm$ 0.01 & 0.32 $\pm$ 0.02 & 0.27 $\pm$ 0.07 & 0.18 $\pm$ 0.04 \\
\hline
&& $V_{\mathrm{centroid}}$\tablenotemark{a} (km s$^{-1}$) & \nodata & $-$13 $\pm$ 1 & \nodata & \nodata & \nodata & \nodata & \nodata \\
H$_{2}$ 2-1 S(1) & 2.24772 & $\Delta V$\tablenotemark{b} (km s$^{-1}$) & \nodata & 14 $\pm$ 3 & \nodata & \nodata & \nodata & \nodata & \nodata \\
&& $F/F_{0}$\tablenotemark{c} & \nodata & 0.11 $\pm$ 0.02 & \nodata & \nodata & \nodata & \nodata & \nodata \\
\hline
&& $V_{\mathrm{centroid}}$\tablenotemark{a} (km s$^{-1}$) & $-$10 $\pm$ 1 & $-$10 $\pm$ 0 & $-$31 $\pm$ 4 & $-$31 $\pm$ 1 & $-$19 $\pm$ 1 & \nodata & $-$2 $\pm$ 1 \\
H$_{2}$ 1-0 Q(1) & 2.40659 & $\Delta V$\tablenotemark{b} (km s$^{-1}$) & 27 $\pm$ 1 & 28 $\pm$ 1 & 46 $\pm$ 5 & 43 $\pm$ 1 & 45 $\pm$ 1 & \nodata & 24 $\pm$ 2 \\
&& $F/F_{0}$\tablenotemark{c} & 1.15 $\pm$ 0.02 & 0.96 $\pm$ 0.02 & 0.47 $\pm$ 0.11 & 0.87 $\pm$ 0.02 & 0.92 $\pm$ 0.01 & \nodata & 0.97 $\pm$ 0.06 \\
\hline
&& $V_{\mathrm{centroid}}$\tablenotemark{a} (km s$^{-1}$) & $-$12 $\pm$ 3 & \nodata & \nodata & $-$30 $\pm$ 1 & $-$21 $\pm$ 3 & \nodata & \nodata \\
H$_{2}$ 1-0 Q(2) & 2.41344 & $\Delta V$\tablenotemark{b} (km s$^{-1}$) & 22 $\pm$ 4 & \nodata & \nodata & 44 $\pm$ 4 & 52 $\pm$ 4 & \nodata & \nodata \\
&& $F/F_{0}$\tablenotemark{c} & 0.34 $\pm$ 0.10 & \nodata & \nodata & 0.50 $\pm$ 0.06 & 0.37 $\pm$ 0.06 & \nodata & \nodata \\
\hline
&& $V_{\mathrm{centroid}}$\tablenotemark{a} (km s$^{-1}$) & $-$12 $\pm$ 5 & $-$14 $\pm$ 1 & $-$24 $\pm$ 4 & $-$32 $\pm$ 0 & $-$19 $\pm$ 1 & $-$14 $\pm$ 1 & \nodata \\
H$_{2}$ 1-0 Q(3) & 2.42373 & $\Delta V$\tablenotemark{b} (km s$^{-1}$) & 21 $\pm$ 3 & 26 $\pm$ 2 & 35 $\pm$ 4 & 30 $\pm$ 1 & 45 $\pm$ 1 & 17 $\pm$ 2 & \nodata \\
&& $F/F_{0}$\tablenotemark{c} & 0.48 $\pm$ 0.11 & 0.55 $\pm$ 0.05 & 0.64 $\pm$ 0.11 & 0.74 $\pm$ 0.02 & 0.84 $\pm$ 0.02 & 0.81 $\pm$ 0.11 & \nodata \\
\enddata
\tablenotetext{a}{The centroid velocity $V_{\mathrm{centroid}}$ obtained from the Gaussian fitting.}
\tablenotetext{b}{The FWHM velocity width $\Delta V$ obtained from the Gaussian fitting.}
\tablenotetext{c}{The line flux $F$ was obtained from the Gaussian fitting and then normalized to the H$_{2}$ 1-0 S(1) line flux $F_{0}$ that is displayed in Figure \ref{fig:h2lp}.}
\tablecomments{The same method as for H$_{2}$ 1-0 S(1) (see Figure \ref{fig:h2lp}) was adopted to calculate the profiles and fluxes of the other lines, and the components corresponding to the individual H$_{2}$ 1-0 S(1) broad components are listed. Only the lines measured with $\ge$3$\sigma$ significance (based on the line flux) are presented. For H$_{2}$ NE 1 and H$_{2}$ NE 2, no broad components were detected with $\ge$3$\sigma$ significance. H$_{2}$ SW-E, H$_{2}$ SW-W, and H$_{2}$ NW with only negligible broad components of H$_{2}$ 1-0 S(1) were also excluded in the table. All of the line fluxes were not corrected for the reddening effect.}
\end{deluxetable*}

\LongTables
\begin{deluxetable*}{lcccccccc}
\tablecaption{Line Profiles and Fluxes for {[}\ion{Fe}{2}{]} Sources\label{table:felp}}
\tablewidth{700pt}
\tabletypesize{\footnotesize}
\tablehead{
\colhead{Line} & \colhead{$\lambda$ ($\micron$)} && \colhead{Fe SE} & \colhead{Fe 1080E} & \multicolumn{2}{c}{Fe knot A} & \colhead{Fe knot B} & \colhead{Fe NE cavity edge}
}
\startdata
&& $V_{\mathrm{centroid}}$\tablenotemark{a} (km s$^{-1}$) & $-$239 $\pm$ 1 & $-$42 $\pm$ 3 & $-$233 $\pm$ 2 & $-$48 $\pm$ 1 & $-$88 $\pm$ 2 & $-$10 $\pm$ 1 \\
{[}\ion{Fe}{2}{]} & 1.64400 & $\Delta V$\tablenotemark{b} (km s$^{-1}$) & 77 $\pm$ 3 & 99 $\pm$ 7 & 118 $\pm$ 5 & 80 $\pm$ 3 & 139 $\pm$ 5 & 43 $\pm$ 3 \\
&& $F_{0}/\sigma_{F_{0}}$\tablenotemark{c} & 20.35 & 4.31 & 10.72 & 27.14 & 14.30 & 12.03 \\
\enddata
\tablenotetext{a}{The centroid velocity $V_{\mathrm{centroid}}$ obtained from the Gaussian fitting in Figure \ref{fig:felp}.}
\tablenotetext{b}{The FWHM velocity width $\Delta V$ obtained from the Gaussian fitting in Figure \ref{fig:felp}.}
\tablenotetext{c}{The detection significance $F_{0}/\sigma_{F_{0}}$ obtained from the Gaussian fitting in Figure \ref{fig:felp}.}
\end{deluxetable*}
\clearpage
\end{landscape}


\clearpage
\begin{thebibliography}{}
\bibitem[Alonso-Albi et al.(2009)]{2009A&A...497..117A} Alonso-Albi, T., Fuente, A., Bachiller, R., et al. 2009, \aap, 497, 117
\bibitem[Barentsen et al.(2014)]{2014MNRAS.444.3230B} Barentsen, G., Farnhill, H. J., Drew, J. E., et al. 2014, \mnras, 444, 3230
\bibitem[Berriman et al.(2003)]{2003ASPC..295..343B} Berriman, G. B., Good, J. C., Curkendall, D. W., et al. 2003, in ASP Conf. Ser. 295, Astronomical Data Analysis Software and Systems XII, ed. H. E. Payne, R. I. Jedrzejewski, \& R. N. Hook (San Francisco, CA: ASP), 343
\bibitem[Black \& van Dishoeck(1987)]{1987ApJ...322..412B} Black, J. H., \& van Dishoeck, E. F. 1987, \apj, 322, 412
\bibitem[Cant\'{o} et al.(1984)]{1984ApJ...282..631C} Cant\'{o}, J., Rodr\'{i}guez, L. F., Calvet, N., \& Levreault, R. M. 1984, \apj, 282, 631
\bibitem[Cardelli et al.(1989)]{1989ApJ...345..245C} Cardelli, J. A., Clayton, G. C., \& Mathis, J. S. 1989, \apj, 345, 245
\bibitem[Cohen \& Kuhi(1979)]{1979ApJS...41..743C} Cohen M., \& Kuhi L., 1979, \apjs, 41, 743
\bibitem[Curiel et al.(1989)]{1989ApL&C..27..299C} Curiel, S., Rodr\'{i}guez, L. F., Cant\'{o}, J., et al. 1989, Astrophysical Letters and Communications, 27, 299
\bibitem[Davis et al.(2001)]{2001MNRAS.326..524D} Davis, C. J., Ray, T. P., Desroches, L., \& Aspin, C. 2001, \mnras, 326, 524
\bibitem[Davis et al.(2003)]{2003A&A...397..693D} Davis, C. J., Whelan, E., Ray, T. P., \& Chrysostomou, A. 2003, \aap, 397, 693
\bibitem[Drew et al.(2005)]{2005MNRAS.362..753D} Drew, J. E., Greimel, R., Irwin, M. J., et al. 2005, \mnras, 362, 753
\bibitem[Folha \& Emerson(2001)]{2001A&A...365...90F} Folha, D. F., \& Emerson, J. P. 2001, \aap, 365, 90
\bibitem[Fuente et al.(2003)]{2003ApJ...598L..39F} Fuente, A., Rodr\'{i}guez-Franco, A., Testi, L., et al. 2003, \apjl, 598, L39
\bibitem[Gaia Collaboration(2018)]{2018yCat.1345....0G} Gaia Collaboration 2018, yCat, 1345, 0
\bibitem[Garcia Lopez et al.(2016)]{2016MNRAS.456..156G} Garcia Lopez, R., Kurosawa, R., Caratti o Garatti, A., et al. 2016, \mnras, 456, 156
\bibitem[Girart et al.(2002)]{2002RMxAA..38..169G} Girart, J. M., Curiel, S., Rodr\'{i}guez, L. F., \& Cant\'{o}, J. 2002, RevMexAA, 38, 169
\bibitem[Gonz\'{a}lez-Solares et al.(2008)]{2008MNRAS.388...89G} Gonz\'{a}lez-Solares, E. A., Walton, N. A., Greimel, R., et al. 2008, \mnras, 388, 89
\bibitem[Hartigan et al.(1987)]{1987ApJ...316..323H} Hartigan, P., Raymond, J., \& Hartmann, L. 1987, \apj, 316, 323
\bibitem[Harvey et al.(1979)]{1979ApJ...231..115H} Harvey, P. M., Thronson, H. A., \& Gatley, I. 1979, \apj, 231, 115
\bibitem[Herbig(1960)]{1960ApJS....4..337H} Herbig, G. H. 1960, \apjs, 4, 337
\bibitem[Hillenbrand et al.(1992)]{1992ApJ...397..613H} Hillenbrand L. A., Strom S. E., Vrba F. J., \& Keene J., 1992, \apj, 397, 613
\bibitem[Jacob et al.(2010)]{2010ascl.soft10036J} Jacob, J. C., Katz, D. S., Berriman, G. B., et al. 2010, Montage: An Astronomical Image Mosaicking Toolkit, Astrophysics Source Code Library, ascl:1010.036
\bibitem[Kaplan et al.(2017)]{2017ApJ...838..152K} Kaplan, K. F., Dinerstein, H. L., Oh, H., et al. 2017, \apj, 838, 152
\bibitem[Kim et al.(2018)]{2018ApJS..238...28K} Kim, I.-J., Pyo, J., Jeong, W.-S., et al. 2018, \apjs, 238, 28
\bibitem[Kraus et al.(2008)]{2008A&A...489.1157K} Kraus, S., Hofmann, K.-H., Benisty, M., et al. 2008, \aap, 489, 1157
\bibitem[Kurosawa et al.(2006)]{2006MNRAS.370..580K} Kurosawa, R., Harries, T. J., \& Symington, N. H. 2006, \mnras, 370, 580
\bibitem[Lee et al.(2017)]{2017zndo....845059L} Lee, J.-J., Gullikson, K., \& Kaplan, K. 2017, igrins/plp 2.2.0. Zenodo. http://doi.org/10.5281/zenodo.845059
\bibitem[Leinert et al.(1997)]{1997A&A...318..472L} Leinert, C., Richichi, A., \& Haas, M. 1997, \aap, 318, 472
\bibitem[Li et al.(2014)]{2014ApJ...796...74L} Li, D., Mari\~{n}as, N., \& Telesco, C. M. 2014, \apj, 796, 74
\bibitem[Mace et al.(2018)]{2018SPIE10702E..0QM} Mace, G., Sokal, K., Lee, J.-J., et al. 2018, Proc. SPIE, 10702, 107020Q
\bibitem[Marston \& McCollum(2008)]{2008A&A...477..193M} Marston, A. P., \& McCollum, B. 2008, \aap, 193, 202
\bibitem[Najita et al.(1996)]{1996ApJ...456..292N} Najita, J., Carr, J. S., \& Tokunaga, A. T. 1996, \apj, 456, 292
\bibitem[Oh et al.(2016a)]{2016ApJ...833..275O} Oh, H., Pyo, T.-S., Kaplan, K., et al. 2016a, \apj, 833, 275
\bibitem[Oh et al.(2016b)]{2016ApJ...817..148O} Oh, H., Pyo, T.-S., Yuk, I.-S., et al. 2016b, \apj, 817, 148
\bibitem[Oudmaijer et al.(1997)]{1997MNRAS.291..797O} Oudmaijer, R. D., Busfield, G., \& Drew, J. E. 1997, \mnras, 291, 797
\bibitem[Park et al.(2014)]{2014SPIE.9147E..1DP} Park, C., Jaffe, D. T., Yuk, I.-S., et al. 2014, Proc. SPIE, 9147, 91471D
\bibitem[Poetzel et al.(1992)]{1992A&A...262..229P} Poetzel, R., Mundt, R., \& Ray, T. P. 1992, \aap, 262, 229
\bibitem[Polomski et al.(2002)]{2002AJ....124.2207P} Polomski, E. F., Telesco, C. M., Pi\~{n}a, R., \& Schulz, B. 2002, \aj, 124, 2207
\bibitem[Pyo et al.(2002)]{2002ApJ...570..724P} Pyo, T.-S., Hayashi, M., Kobayashi, N., et al. 2002, \apj, 570, 724
\bibitem[Rousselot et al.(2000)]{2000A&A...354.1134R} Rousselot, P., Lidman, C., Cuby, J.-G., Moreels, G., \& Monnet, G. 2000, \aap, 354, 1134
\bibitem[Sandell et al.(2011)]{2011ApJ...727...26S} Sandell, G., Weintraub, D. A., \& Hamidouche, M. 2011, \apj, 727, 26
\bibitem[Storey \& Hummer(1995)]{1995MNRAS.272...41S} Storey, P. J., \& Hummer, D. G. 1995, \mnras, 272, 41
\bibitem[Takami et al.(2006)]{2006ApJ...641..357T} Takami, M., Chrysostomou, A., Ray, T. P., et al. 2006, \apj, 641, 357
\bibitem[Tambovtseva et al.(2014)]{2014A&A...562A.104T} Tambovtseva, L. V., Grinin, V. P., \& Weigelt, G., 2014, \aap, 562, A104
\bibitem[Tambovtseva et al.(2016)]{2016A&A...590A..97T} Tambovtseva, L. V., Grinin, V. P., \& Weigelt, G., 2016, \aap, 590, A97
\bibitem[Tanaka et al.(1989)]{1989ApJ...336..207T} Tanaka, M., Hasegawa, T., Hayashi, S. S., Brand, P. W. J. L., \& Gatley, I. 1989, \apj, 336, 207
\bibitem[Wang et al.(2008)]{2008ApJ...673..315W} Wang, S., Looney, L. W., Brandner, W., \& Close, L. M. 2008, \apj, 673, 315
\bibitem[Wolniewicz et al.(1998)]{1998ApJS..115..293W} Wolniewicz, L., Simbotin, I., \& Dalgarno, A. 1998, \apjs, 115, 293
\bibitem[Wouterloot \& Brand(1989)]{1989A&AS...80..149W} Wouterloot, J. G. A., \& Brand, J. 1989, \aaps, 80, 149
\bibitem[Yoshida et al.(1992)]{1992PASJ...44...77Y} Yoshida, S., Kogure, T., Nakano, M., Tatematsu, K., \& Wiramihardja, S. D. 1992, \pasj, 44, 77
\bibitem[Yuk et al.(2010)]{2010SPIE.7735E..1MY} Yuk, I.-S., Jaffe, D. T., Barnes, S., et al. 2010, Proc. SPIE, 7735, 77351M
\end{thebibliography}
\end{document}